\pdfoutput=1
\documentclass[11pt,twoside,a4paper,cmspaper,final,collab]{cms-tdr}
\def\svnVersion{3ab50dd}\def\svnDate{2024/11/22}\def\cmsCernNoTag{CERN-EP-2024-259}\def\cmsCernDate{\today}\def\cmsMessage{Published in the Journal of Instrumentation as \href{http://dx.doi.org/10.1088/1748-0221/19/11/P11021}{\doi{10.1088/1748-0221/19/11/P11021}.}}
\begin{document}\cmsNoteHeader{TRG-19-001}

\newcommand\instL{\ensuremath{\,\text{cm}^{-2}\text{s}^{-1}}\xspace}
\newcommand{\Linst}{\ensuremath{\mathcal{L}_{\text{inst}}}\xspace}
\newcommand{\Runone}{\text{\mbox{Run\,1}}\xspace}
\newcommand{\Runtwo}{\text{\mbox{Run\,2}}\xspace}
\newcommand{\Runthree}{\text{\mbox{Run\,3}}\xspace}
\newcommand\Tstrut{\rule{0pt}{2.6ex}}
\newcommand\Bstrut{\rule[-0.9ex]{0pt}{0pt}}
\newcommand{\PShpm}{{\HepParticle{\PSh}{}{\pm}}\Xspace}
\newcommand{\DeepCSV}{{DeepCSV}\xspace}
\newcommand{\cmsTable}[1]{\resizebox{\textwidth}{!}{#1}}

\cmsNoteHeader{TRG-19-001} 
\title{Performance of the CMS high-level trigger during LHC \Runtwo}

\date{\today}

\abstract{
The CERN LHC provided proton and heavy ion collisions during its \Runtwo operation period from 2015 to 2018. Proton-proton collisions reached a peak instantaneous luminosity of $2.1\times 10^{34}$\instL, twice the initial design value, at $\sqrt{s}=13\TeV$. The CMS experiment records a subset of the collisions for further processing as part of its online selection of data for physics analyses, using a two-level trigger system: the Level-1 trigger, implemented in custom-designed electronics, and the high-level trigger, a streamlined version of the offline reconstruction software running on a large computer farm. This paper presents the performance of the CMS high-level trigger system during LHC \Runtwo for physics objects, such as leptons, jets, and missing transverse momentum, which meet the broad needs of the CMS physics program and the challenge of the evolving LHC and detector conditions.   Sophisticated algorithms that were originally used in offline reconstruction were deployed online. Highlights include a machine-learning \PQb tagging algorithm and a reconstruction algorithm for tau leptons that decay hadronically.
}

\hypersetup{
pdfauthor={CMS Collaboration},
pdftitle={Performance of the CMS high-level trigger during LHC Run 2},
pdfsubject={CMS},
pdfkeywords={CMS,  trigger}}

\maketitle 

\tableofcontents

\section{Introduction}

The CERN LHC collides bunches of particles at a maximum rate of about
40\unit{MHz} at several experimental sites including
CMS. During LHC \Runtwo in 2015--2018, the maximum instantaneous luminosity \Linst reached
$2.1\times 10^{34}$\instL, twice the initial design value, at $\sqrt{s}=13\TeV$.
The mean pileup (PU), or simultaneous inelastic proton-proton ($\Pp\Pp$) collisions per bunch
crossing, was about 50.
To select collision events of potential interest to physics analyses, the CMS trigger system
divides the processing into two levels: a first level (L1) that is
implemented in custom-designed electronics, and a high-level trigger (HLT) implemented
in software and executed on commodity computers. The HLT further refines the purity of the collection of physics objects
that are selected at L1, with an input event rate to the HLT limited to about
100\unit{kHz} by the detector read-out electronics, and targets an average output rate of
about 1\unit{kHz} during \Runtwo for
standard $\Pp\Pp$ collision events for offline storage and prompt
reconstruction.
The HLT also accommodates combinations of objects, such as leptons and
jets or final states in \PB physics, to target specific
physics analyses,  although the focus of this paper is on the reconstruction and
object identification tools used at the HLT.
Novel techniques introduced during \Runone, such as
a high rate storage of events with a reduced data content
(``data scouting'')
and storage of additional full events for delayed processing
(``data parking''),
both described in Ref.~\cite{CMS:EXO-23-007},
continued during \Runtwo to further extend the physics program.

The performance of the CMS L1 and HLT trigger systems during
\Runone of the LHC, with $\sqrt{s}=7$ and 8\TeV for $\Pp\Pp$ collisions,
is described in Ref.~\cite{CMS:trigger-run1}. The L1 trigger was
subsequently upgraded during \Runtwo of the LHC as part of the Phase-1 upgrades of
CMS~\cite{CMS-L1T-TDR}, and its performance is
described in Ref.~\cite{TRG-17-001}. \Runtwo of the LHC delivered
challenges to the \Runone L1 and HLT algorithms,
including a higher $\sqrt{s}$ of 13\TeV, higher
luminosity, larger PU, and further detector aging (primarily from radiation damage). The
algorithms deployed at the HLT were revised to address these challenges
and were made flexible enough to adapt to various changing detector conditions
that occurred during \Runtwo, together with the installation of
detector components and electronics that were part of the CMS Phase-1
upgrades~\cite{CMS:phase-1-upgrade}.
Further adaptations of the HLT for \Runthree of the LHC, such as the
inclusion of graphical processing units for computation, are described
in Ref.~\cite{CMS:2023gfb}.

This paper is organized as follows. Section~\ref{sec:expt} describes
the CMS experiment and the LHC operating conditions during \Runtwo, and
Section~\ref{sec:online} describes the architecture of the HLT, as well
as a breakdown of its rate and processing
time. Section~\ref{sec:objectperf} describes the reconstruction and
performance of the physics objects in the HLT that are broadly
applicable to a wide range of physics analyses, such as lepton,
jet, and energy sum triggers.
Finally, a summary is given in Section~\ref{sec:summary}.

\section{Experimental conditions during \texorpdfstring{\Runtwo}{Run 2}}
\label{sec:expt}

\subsection{The CMS detector}

The central feature of the CMS apparatus is a superconducting solenoid
of 6\unit{m} internal diameter, providing a magnetic field of
3.8\unit{T}. Within the solenoid volume are a silicon pixel and strip
tracker, a lead tungstate crystal electromagnetic calorimeter (ECAL),
and a brass and scintillator hadron calorimeter (HCAL), each composed
of a barrel and two endcap sections. Forward calorimeters extend the
pseudorapidity coverage provided by the barrel and endcap
detectors. Muons are detected in gas-ionization chambers embedded in
the steel flux-return yoke outside the solenoid.
A more detailed description of the CMS detector, together with a
definition of the coordinate system used and the relevant kinematic
variables, can be found in Refs.~\cite{CMS-Detector,CMS:2023gfb}.
The silicon tracker used in 2016 measured charged particles within the
range $\abs{\eta} < 2.5$. For nonisolated particles with $1 < \pt < 10\GeV$ and $\abs{\eta} < 1.4$, the track resolutions were typically 1.5\% in \pt and 25--90 (45--150)\mum in the transverse (longitudinal) impact parameter $d_{xy}$ ($d_z$)~\cite{CMS:2014pgm}. At the start of 2017, a new pixel detector was installed~\cite{Phase1Pixel}; the upgraded tracker measured particles up to $\abs{\eta} = 3.0$ with typical resolutions of 1.5\% in \pt and 20--75\mum in $d_{xy}$~\cite{DP-2020-049} for nonisolated particles of $1 < \pt < 10\GeV$. According to simulation studies~\cite{DP-2017-015}, similar improvements are expected in the longitudinal direction.

\subsection{The LHC operations}

The operational period of LHC \Runtwo covered the years 2015 to 2018.
The year 2015 was dominated by commissioning activities in the wake of LHC Long Shutdown 1,
with a correspondingly small amount of integrated luminosity delivered
to the experiments.  Additionally, the electronics for the
CMS L1 trigger was upgraded and installed for data taking at the
beginning of 2016. Hence we will focus mainly on the 2016--2018 period.

Traditionally, each year starts with an interleaved commissioning--production period, 
where the \Linst is progressively increased by the insertion of additional proton bunches in
opposite directions to make two oppositely running proton beams in a single LHC fill.
After the beams achieve their maximum occupancy projected for the year, 
the LHC conditions are optimized throughout the year to achieve the
needs of the 
physics programs of the experiments. Within a given LHC fill,
the \Linst generally decreases in tandem with the
natural depletion of the beams, which
allows for the activation of looser trigger algorithms that
can increase the physics reach of an experiment by keeping the data
bandwidth to storage effectively constant. For some
\Runtwo fills, the beam focusing was periodically adjusted to
achieve approximately constant \Linst for an extended amount of
time (luminosity leveling).

\begin{table}[htbp]
  \renewcommand{\arraystretch}{1.2}
  \centering
  \topcaption{The LHC operations parameters during \Runtwo. The maximum
    PU is for standard physics fills with more than 600 bunches,
    and the average PU is calculated assuming an inelastic
    cross section of 80\unit{mb}.
    The 8b4e bunch-filling configuration values are given in brackets,
    and they refer to an LHC configuration used in 2017 to mitigate beam losses.
    The \sqrtsNN is the center-of-mass energy per nucleon.
  }
  \begin{tabular}{@{}lccc @{}} 
    \hline
    &	{2016} & {2017} & {2018} \\
    \Bstrut {Proton-proton}\\
    \cline{1-1}
    Max. colliding bunches  			& 2220	& 2556 [1868]	& 2556 \\
    \,\,\, in CMS                  			& 2208	& 2544 [1866] & 2544 \\
    Max. \Linst ($\times10^{34}\percms$) 		& 1.5 		& 1.7 [2.1]		& 2.1 \\
    Max. PU  			& 46.9 		& 47.5 [78.8]		& 64.7 \\
    Avg. PU 	& 27		& 31 [42]		& 37 \\
    \hline
    \Tstrut \Bstrut {Heavy ions}\\
    \cline{1-1}
    Collisions species										& pPb			& XeXe, pp	& PbPb  	\\
    $\sqrtsNN (\TeVns{})$
    & 8.16 			& 5.44, 5.02 	& 5.02  \\
    \hline
  \end{tabular}
  \label{tab:lhcparams}
\end{table}

There are also special LHC runs that occur throughout the running period with a variety of purposes.
Low-PU runs, with much less than 0.3 $\Pp\Pp$ interactions per bunch crossing, are important for particular standard model (SM) measurements,
such as the measurement of the \PW boson mass.
On the other hand, high-PU runs, with fewer filled bunches, are also available for experimental performance measurements
to prepare for the steady luminosity increase during
\Runtwo, as well as for that
expected for LHC \Runthree and for the High-Luminosity LHC era.
Other special run setups include:
reference $\Pp\Pp$ collisions for the heavy ion program;
luminosity studies during van der Meer scans~\cite{GRAFSTROM201597},
both at the initial proton injection energy and at the maximum energy;
as well as special machine development runs for LHC beam studies.

The experience acquired through \Runtwo led to progressively
smoother LHC operations throughout the years~\cite{Bruce:2019ohe, Wenninger:2668326}.
In 2016, the maximum \Linst had a rapid initial increase to the nominal value of $1.0\times10^{34}\percms$,
followed by a gradual increase during the year to ${\approx}1.5\times10^{34}\percms$.
In 2017, this pre-shutdown performance was easily achieved by the start of the run,
and adjustments made throughout the year allowed the crossing of the
$2.0\times10^{34}\percms$ milestone by October.
Amongst these adjustments, we highlight the deployment
of the ``8b4e'' filling scheme (8 filled bunches followed by 4 empty buckets) accompanied
by the reduction of $\beta^*$ (a parameter related to the transverse beam size at the
interaction point) from 40 to 30\unit{cm}
to mitigate beam losses around the region of an LHC magnet interconnect~\cite{Wenninger:2668326,Mirarchi:2019mre}.
Finally, in 2018 the initial ramp-up period to routinely deliver maximum instantaneous luminosity was very quick, and
collision data were acquired at a steady pace during the year. A
summary of the LHC parameters for $\Pp\Pp$ and heavy ion
collisions for 2016--2018 is shown in Table~\ref{tab:lhcparams}.

\section{Online data selection}
\label{sec:online}

\subsection{The HLT architecture}
\label{sec:daq}

The HLT hardware consists of a large cluster of multi-core servers,
the event filter farm, that runs a Linux operating system.
The processing capacity of the Filter Farm was expanded gradually
throughout \Runtwo to cope with the evolving LHC and detector conditions.
By the end of \Runtwo,
the processing power was about $7.2 \times 10^5$ in HEPSPEC
2006~\cite{hs06} units. This was distributed across 360 nodes of dual
Intel Haswell E5-2680v3 processors (8640 cores), 324 nodes of dual Intel Broadwell
E5-2680v4 processors (9072 cores), and 400 nodes of dual Intel Skylake Gold 6130
processors ($12\,800$ cores).

As noted earlier, the HLT refines the purity and reduces the rate of events 
that are selected by L1, targeting a rate of about 1\unit{kHz} averaged over an LHC fill for
standard $\Pp\Pp$ collision events for offline storage and prompt
reconstruction. Additional storage beyond the 1\unit{kHz} rate is
allowed for data to be ``parked'', whereby the offline reconstruction
is postponed until a later, non-data-taking period (\eg, during a long
shutdown of the LHC). In 2018, the storage rate for parking was an
additional 3\unit{kHz}~\cite{CMS:EXO-23-007}.  The average raw data event size for these
standard $\Pp\Pp$ collision events at the average \Runtwo PU (37) is about
0.65\unit{MB} after compression, with a peak size near 1\unit{MB} at
the highest PU conditions.
A higher rate of reduced-size events also can be acquired, a technique
referred to as ``data scouting,'' where only the high-level physics objects, such as jets or leptons,
reconstructed at the HLT are stored on disk. No raw data from detector channels are stored for later offline analysis. For example, 5\unit{kHz}
of scouting events with an average event size of 8\unit{kB} were also
recorded in 2018~\cite{CMS:EXO-23-007}.

\subsection{Algorithms}

The data processing of the HLT is structured around the concept of an
HLT ``path'', which is a set of algorithmic processing steps run in a
predefined order that both reconstructs physics objects and makes
selections on these objects based on the physics requirements.
Each
HLT path is implemented as a sequence of steps generally of increasing
complexity, reconstruction refinement, and physics sophistication. For
example, the processing of intensive track reconstruction is usually
performed only after some initial reconstruction and selection based on the
calorimeters and muon detectors.
Each path also requires the selections in specific L1 triggers (``L1 seeds'')
to have been satisfied before execution would begin.
The reconstruction modules and
selection filters of the HLT use the same software framework used
for offline reconstruction and analyses (CMSSW~\cite{Jones:2015soc}). The framework
supports multi-threaded event processing, which optimizes memory
usage and is utilized for the HLT software.

The HLT paths selecting similar physics object topologies are grouped into primary data sets,
which are then grouped into streams.
Primary data sets define the samples used for offline processing,
and their trigger content is chosen such that the overlap is minimized
to avoid reconstructing offline the same event in multiple primary data sets.
Streams define the outputs of the HLT processes, which are transferred from CMS to offline computing facilities during data taking. 
The grouping of primary data sets into streams further reduces the overlap across HLT outputs,
allowing for a more efficient handling of these data transfers.

\subsection{Menus}

The HLT selects data for storage through the application of a trigger
``menu'', which is a collection of individual HLT paths. The trigger
path definitions, physics object thresholds (\eg, the transverse energy \ET threshold and the \pt threshold), and
rate allocations are set to meet the physics objectives of the
experiment. For \Runtwo, the HLT menus typically had around 600
paths for $\Pp\Pp$ data taking. This included the primary triggers for
analyses, as well as triggers for calibration, efficiency
measurements, control region measurements, \etc{} that were typically
looser than the primary triggers. These latter triggers were often
``prescaled'', meaning that they selected only a fraction of the
events that satisfied their conditions to limit their storage
rate. Approximately a dozen menus were deployed each year during 2016--2018 for operations
with $\Pp\Pp$ collisions.
Different sets of trigger menus were used for special LHC runs,
including heavy ion collision runs. A representative listing of the
primary triggers used in 2018 for physics analyses, along with their
thresholds and corresponding rates, is given in
Table~\ref{tab:HLT2-simplifiedMenu}. These triggers accounted for approximately
50\% of the overall  menu rate used to select data that were reconstructed
promptly. Note that the listed rates of each trigger are inclusive and not necessarily
unique. For example objects selected by a trigger applying isolation
would also be selected by a trigger not requiring isolation on the
same type of objects, provided that the other conditions (\eg, energy)
are met. 
Comprehensive details of the algorithms used and
their performances are discussed in Section~\ref{sec:objectperf}.

\begin{table}[htbp]
  \centering
  \topcaption{Representative set of HLT paths
    based on the basic HLT physics objects used during data taking in
    2018, the associated thresholds at the L1 and HLT, and the corresponding HLT output rates.
    The total menu rate at $\Linst=1.8\times 10^{34}\percms$, representative near the start of an LHC fill,  is 1.6\unit{kHz}.
  }
  \begin{tabular}{lccc}
    \hline
    HLT path & L1 thresholds [{\GeVns}] & HLT thresholds [{\GeVns}] & Rate [Hz] \\
    \hline
    Single muon & 22 & 50 & 49 \\
    Single muon (isolated) & 22 & 24 & 230 \\
    Double muon & 22 & 37, 27 & 16 \\
    Double muon (isolated) & 15, 7 & 17, 8 & 32 \\
    Single electron (isolated) & 30 & 32 & 180 \\
    Double electron & 25, 12 & 25, 25 & 16 \\
    Double electron (isolated) & 22, 12 & 23, 12 & 32 \\
    Single photon & 30 & 200 & 16 \\
    Single photon (isolated), & 30 & 110 & 16 \\
    \,\,\, barrel only ($\abs{\eta}<1.48$) &  &  &  \\
    Double photon & 25, 12 & 30, 18 & 32 \\
    Single tau & 120 & 180 & 16 \\
    Double tau & 32 & 35, 35 & 49 \\
    Single jet & 180 & 500 & 16 \\
    Single jet with substructure & 180 & 400 & 32 \\
    Multijets with \PQb tagging & $\HT > 320$ & $\HT >  330$ & 16 \\
    & jets $> 70, 55, 40, 40$ & jets $> 75, 60, 45, 40$ &  \\
    Total transverse momentum & 360 & 1050 & 16 \\
    Missing transverse momentum & 100 & 120 & 49 \\
    \hline
  \end{tabular}
  \label{tab:HLT2-simplifiedMenu}
\end{table}

The rest of the trigger menu not included in
Table~\ref{tab:HLT2-simplifiedMenu} consists of slightly more specialized
trigger paths that enhance the acceptance of events for targeted
analysis areas of the
CMS physics program. This includes ``cross-object'' triggers, such as
mixed double-lepton (\Pe+\PGm) triggers that target, \eg, $\PH\to \PW\PW$, $\PH\to \PGt\PGt$, and top quark
pair production in the dilepton final state, where \PH indicates the Higgs boson. Top quark
acceptance is further enhanced in other decay topologies, such as
lepton+jets or the all-hadronic channel. For the former, the \ET threshold on an
electron can be reduced from that used in the inclusive single-electron trigger,
with manageable rate increase, when used in coincidence with jets
with a total scalar \pt sum (\HT) above a given threshold. Likewise, for
the latter, the hadronic top quark decays can be selected via
requirements on the number of jets, \HT, and one or more \PQb-tagged jets.
The \PB physics program of CMS generally targets soft dimuon final states,
and thus uses additional requirements (\eg, invariant mass) in its HLT paths
to keep trigger rates manageable.
Since the overall list of these more specialized trigger paths numbers in the
hundreds, their performances are not described here. However, the
description and performance of the algorithms used for most of the
individual objects forming these paths are discussed in this article. 

\subsection{Rates and processing time}

\begin{figure}
  \centering
  \includegraphics[width=1.0\textwidth]{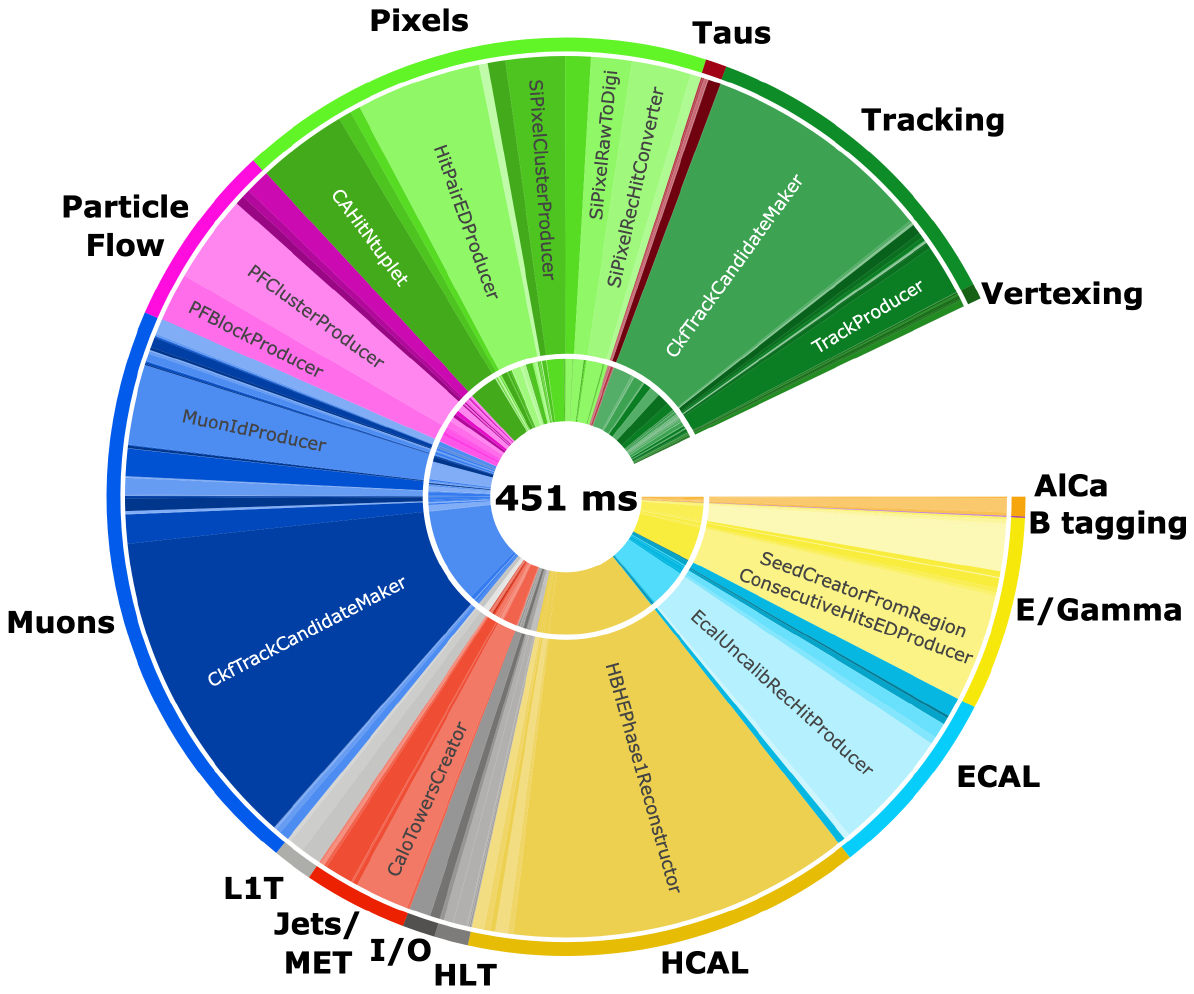}
  \caption{Pie chart distribution of the CPU time per event spent by the HLT for
    different parts of the event reconstruction.
    Reconstruction modules and filters are grouped by detector and physics object (outermost ring and similarly colored groupings).
    The middle ring reports the names of specific C++ classes in CMSSW used in the HLT reconstruction, and the various slices in the innermost ring refer
    to different instances (modules) of that given C++ class in the HLT menu.
    The empty slice indicates the time spent outside of the individual algorithms.
    \label{fig:CPUtiming}
  }
\end{figure}

The distribution of the CPU time spent
in processing the HLT menu by reconstruction category and by instances of
C++ classes within those categories is shown in
Fig.~\ref{fig:CPUtiming}. Overall, there are $\mathcal{O}(1200)$ instances stemming from $\mathcal{O}(200)$
algorithms that are run.
The HLT configuration is based on the one used in 2018, with only minimal
updates to the local reconstruction to reflect the ongoing
developments foreseen for LHC \Runthree.
The timing is measured for an average PU of 50 during a 2018
data-taking period on a full HLT
node (2x Intel Skylake Gold 6130) with hyper-threading enabled, running 16 jobs in
parallel with 4 threads each. The average processing time per event is 451\unit{ms};
scaling this performance to the full event filter farm capacity means that
it would be able to process an event input rate of approximately 130\unit{kHz},
above the nominal L1 rate target of 100\unit{kHz}.

\begin{figure}[!tbh]
  \centering
  \includegraphics[width=0.9\textwidth]{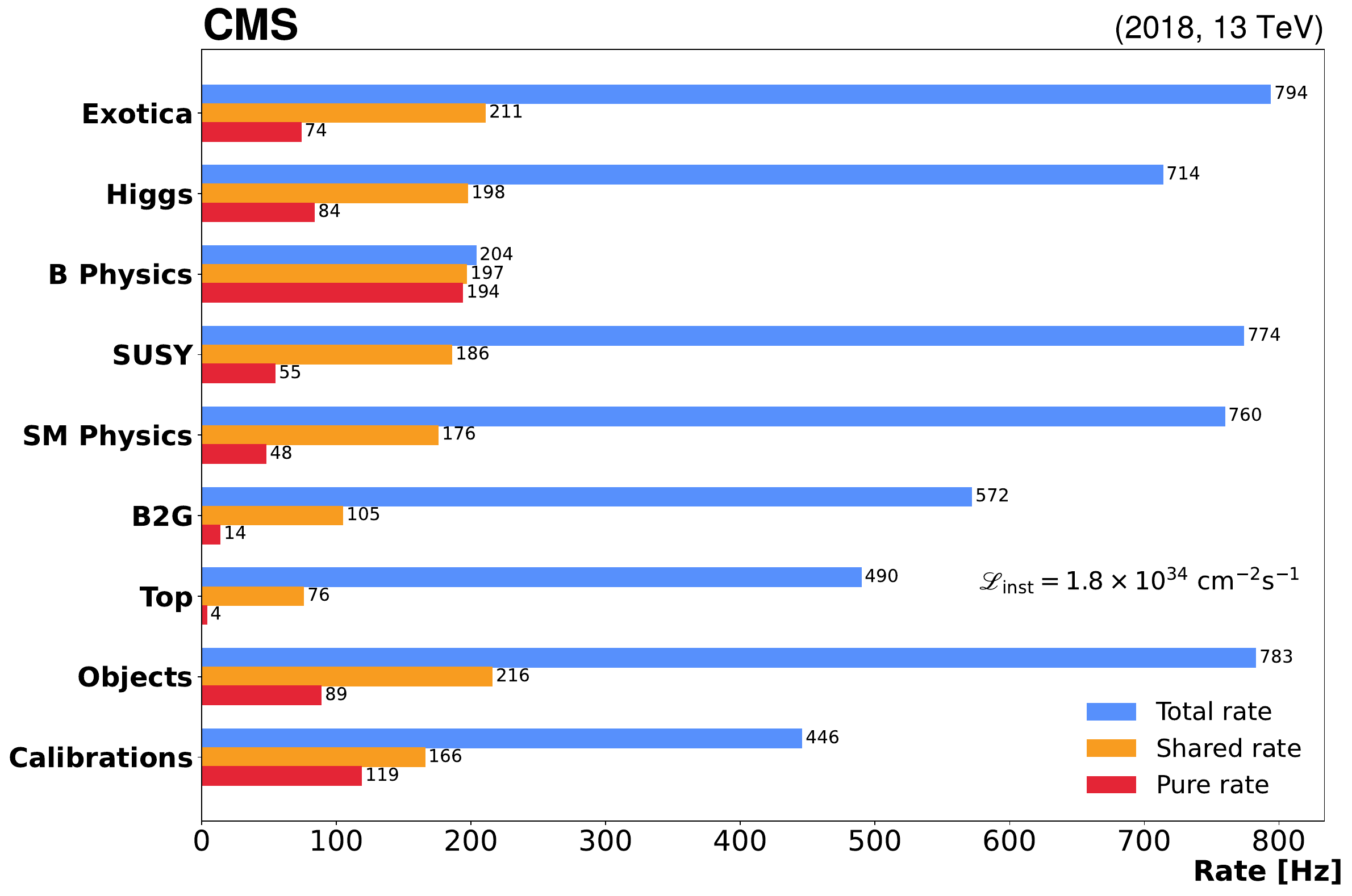}
  \caption{The HLT rate consumption by physics group in the standard physics
    streams for a \Runtwo menu deployed in September 2018. The ``Total
    Rate'' is the inclusive rate of all triggers owned by a group, and
    the ``Pure Rate'' is the exclusive rate of all triggers unique to
    that group. The ``Shared Rate'' is the rate calculated by dividing
    the rate of each trigger equally among all physics groups that use
    it, before summing the total group rate. It includes the Pure Rate
    of that physics group. The topic coverage of each group is discussed in the text.
    \label{fig:RatePerGroup}
  }
\end{figure}

Figure~\ref{fig:RatePerGroup} illustrates the HLT rates attributed to
each CMS physics group for the HLT menu deployed in September 2018,
which selects data for prompt reconstruction. The rates were
determined by running the HLT menu
on a special data set where events were selected that passed the L1 trigger without any additional HLT requirements.
The rates were normalized to an average \Linst of $1.8\times 10^{34}$\percms.
An event is attributed to a
given physics group if the latter requires (\ie, owns) at least one of the HLT
paths that triggered the event. For each physics group, three types of
rates are evaluated.
\begin{itemize}
\item Total: the inclusive rate arising from all HLT paths needed by that physics group.
\item Pure: the exclusive rate from all paths uniquely assigned to
  that physics group.
\item Shared: the sum of the pure rate and the fractional rate of the HLT
  paths shared with other groups, where the rate is split equally among all
  groups for a given path.
\end{itemize}
The sum of the Shared HLT rates is 1530\unit{Hz}. However, because the
selected data are recorded in separate data sets with some overlap (6.6\%)
for analysis and offline processing reasons,
the actual storage rate is 1640\unit{Hz} with the additional duplication. 

The CMS physics analysis groups focused on searches for physics beyond the SM are the
B2G (searches for new physics in boosted signatures),
SUSY (searches for new physics in final states with imbalanced \pt),
and Exotica (other topologies of new physics) groups.
The analysis groups focused on measurements are the
Higgs boson physics, top quark physics,
\PB physics, and other SM phenomena groups.
The ``Objects'' category in Fig.~\ref{fig:RatePerGroup} contains the HLT paths used by the physics object groups
(Tracking, Muon, Electron-Photon, Jet-MET, Tau, \PQb Tagging)
to characterize the performance of the online and offline reconstruction.
The ``Calibrations'' category includes all HLT paths
used for subdetector alignment and calibration purposes.

\begin{figure}[!tbh]
  \centering
  \includegraphics[width=0.75\textwidth]{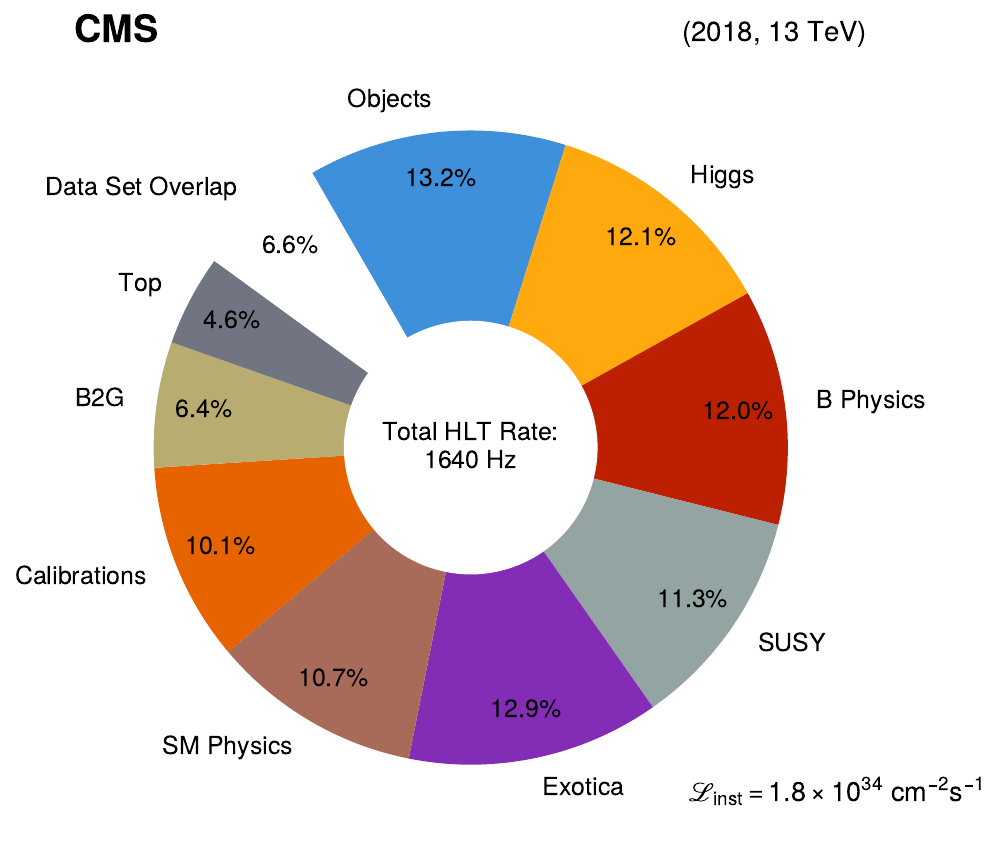}
  \caption{Share of the total HLT rate that each physics group
    contributes. ``Data Set Overlap'' refers to the events that are duplicated
    and saved into separate data sets for analysis and offline processing convenience, but which
    must be reconstructed separately offline.
    \label{fig:RateBudget}
  }
\end{figure}

The rate allocation
per physics group is also expressed as a 
pie chart in Fig.~\ref{fig:RateBudget}. 
Very roughly one-third of the HLT rate budget is devoted
to searches beyond the SM, one-third to measurements including the Higgs boson (but apart from \PB physics),
and one-third to \PB physics and physics object groups.

\section{The HLT reconstruction and performance}
\label{sec:objectperf}

The HLT paths in the menu are based on physics objects produced from reconstruction modules that use
information from the inner tracking system, calorimeters, and muon
detectors. Central to many of the object reconstructions
is the particle-flow (PF) algorithm~\cite{ParticleFlow}, which
aims to reconstruct all individual particles (electrons, muons, photons, and charged and neutral hadrons) in an event, combining
information provided by these systems.
The online PF reconstruction has a simplified version
of the offline reconstruction to fulfill
the timing limitation for online reconstruction.
The tracking has a
reduced number of iterations, down to three as discussed in Section~\ref{sec:tracking}.
Moreover, electron reconstruction is not integrated
into the online PF algorithm~\cite{ParticleFlow}. 
Brief descriptions of
the HLT reconstruction algorithms for physics objects and highlights
of their performance measured with data collected during \Runtwo are
described below.
The measured efficiencies of the lepton ($\Pell$) algorithms are typically obtained
using the ``tag-and-probe'' technique~\cite{CMS:2011aa}, which exploits
resonant dilepton production (\eg, $\PZ\to \Pell\Pell$ or $\PJGy \to \Pell\Pell$) events in data. One of the lepton candidates, called the
``tag,'' is required to satisfy a trigger requirement (\eg, a
single-lepton trigger) such that the
event is recorded irrespective of the other lepton, the ``probe.'' 
Offline selection requirements are applied to both tag and probe to
reduce the contribution of misidentified leptons. The trigger efficiency of the
probe can then be measured in an unbiased way as a function of various kinematic and
object quality parameters.
The measured efficiencies of jets and energy sums are
obtained using an unbiased data set, namely one triggered by a
lepton.

\subsection{Tracking}
\label{sec:tracking}

Charged particle tracks in the HLT are reconstructed from the hits in the pixel and strip tracker using a Kalman filtering technique~\cite{Fruhwirth:1987fm}, based on initial estimates of the track parameters obtained from hits in the pixel detectors (``seeds'').
The seed is propagated outwards and the track parameters are updated
with the information from compatible hits as they are found until no
more hits are found or the tracker boundary is reached. The track is
then propagated from the outermost hit inwards in search of additional
compatible hits, after which a fit to the resulting hit collection
determines the final track parameters. 
Similar to the offline track reconstruction~\cite{CMS:2014pgm}, the tracking is performed iteratively, starting with tight requirements on the \pt and displacement with respect to the beam spot of the seed, which become
looser for each subsequent iteration. Hits in the tracking detectors
already used in a track are removed at the beginning of the next
iteration. The general track reconstruction in the HLT consists of
three iterations. The first two require the maximum of four
consecutive pixel detector hits expected for one track from the detector geometry,
identified using a cellular automaton algorithm~\cite{Pantaleo:2293435}, to seed (\ie, initiate) the tracking. These
iterations first target high-\pt tracks before extending the coverage to low-\pt tracks, using the full
volume of the pixel detector. The third iteration relaxes the
requirement on the number of hits in the track seeds to three and is
restricted to the vicinity of jet candidates identified from
calorimeter information and the tracks reconstructed in the two
previous iterations. Tracks are clustered into vertices using the same deterministic annealing algorithm~\cite{726788} used in the offline reconstruction~\cite{CMS:2014pgm}. The vertex position is fitted using an adaptive vertex fitter~\cite{Fruhwirth:1027031}.

This configuration of the track reconstruction was deployed in 2017, after the Phase-1 pixel detector was installed. Reflecting the lower number of detector layers, fewer pixel hits were required to form track seeds in previous years. The higher quality of track seeds available in 2017 resulted in an increase in tracking efficiency by about 10\% for tracks with $\pt > 1.2\GeV$, with larger improvements present at lower \pt and high $\abs{\eta}$. At the same time, the track misidentification rate was reduced by a factor of 5--7.

During 2017, several issues with the installed Phase-1 pixel
detector were identified that led to a nonnegligible fraction of
inactive pixel detector modules in each event. Most notable was the failure
of some direct-current DC-DC converters (${\approx}5\%$) used to power the detector, which
resulted in an increasing fraction of inactive modules towards the
end of the data-taking period~\cite{CMS:2023gfb}.

During the year-end technical stop 2017--2018, the pixel detector was
equipped with new converters, and the initial performance was
restored. To safeguard against a possible recurrence of this
problem and other possible detector failures, an additional recovery
iteration was added. Track seeds consisting of just two pixel detector
hits (``doublets'') are created in regions of the detector where two inactive modules
overlap as seen from the interaction point. Because of the limited CPU
time available for the HLT reconstruction, this iteration is
restricted to tracks with $\pt > 1.2 \GeV$. 

The tracking efficiency and misidentification rate reported here are obtained from simulated
top quark pair (\ttbar)
events with a mean PU of 50. The efficiency and rate are
defined with respect to the Monte Carlo (MC) simulated objects, where the
tracks of the simulated particles are matched to reconstructed tracks
based on shared hits in the tracking detectors.
The tracking efficiency is defined as the fraction of simulated
particles from the signal interaction with $\pt > 0.9 \GeV$, $\abs{\eta} < 2.5$,
$d_{xy} < 35\unit{cm}$, and $d_z < 70\unit{cm}$ that are
matched to a reconstructed track.
The misidentification rate is defined as the fraction of reconstructed tracks that
could not be matched to a simulated particle.
To realistically model the effect of an imperfect pixel detector, a
map of inactive modules representing the status of the real detector
as of June 2018 is applied to the simulation.
The tracking performance is implicitly included in
the measured HLT object and algorithm performances that use tracking, reported in subsequent
sections of this article.

\begin{figure}
  \centering
  \includegraphics[width=0.48\textwidth]{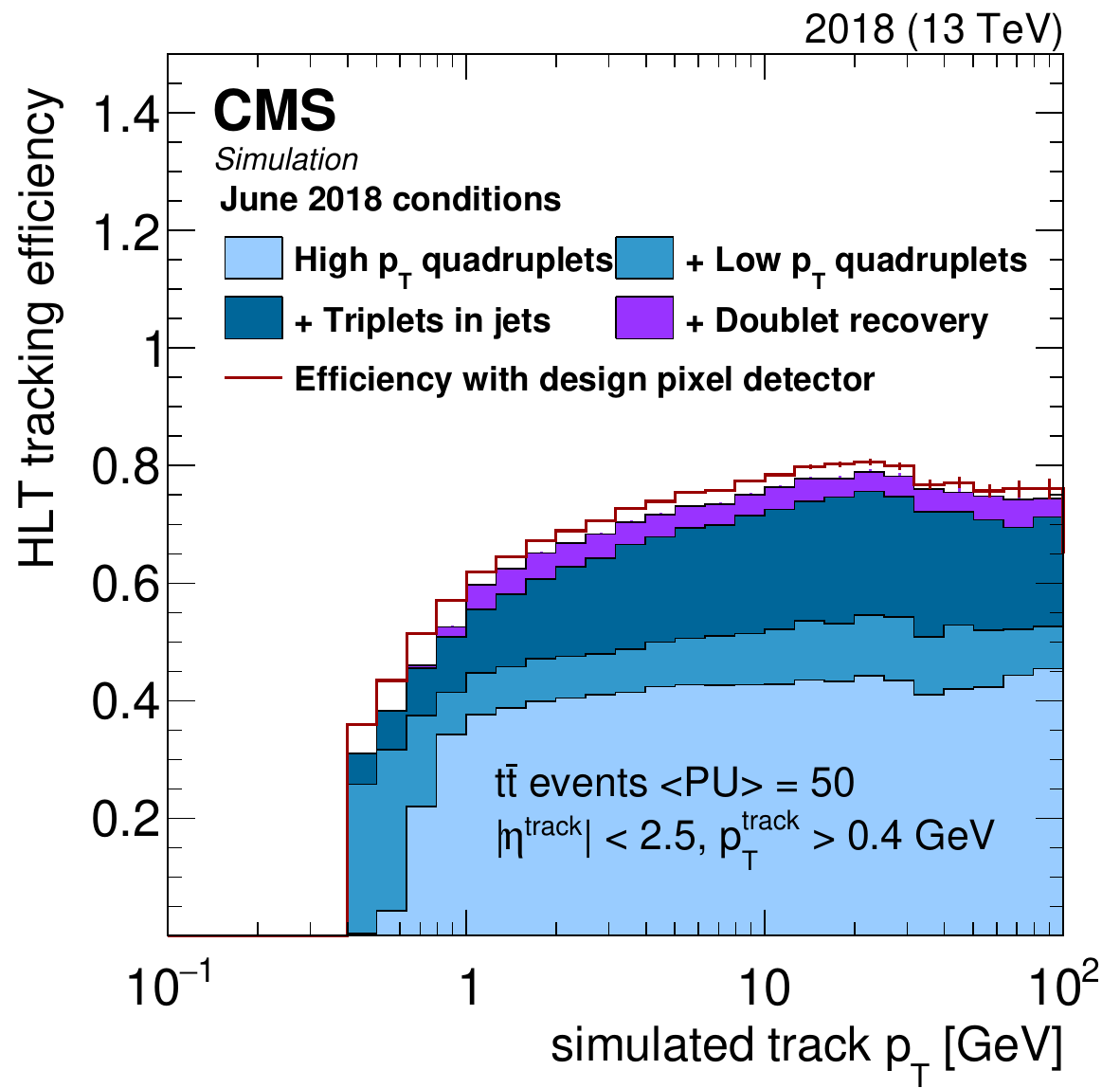}
  \includegraphics[width=0.48\textwidth]{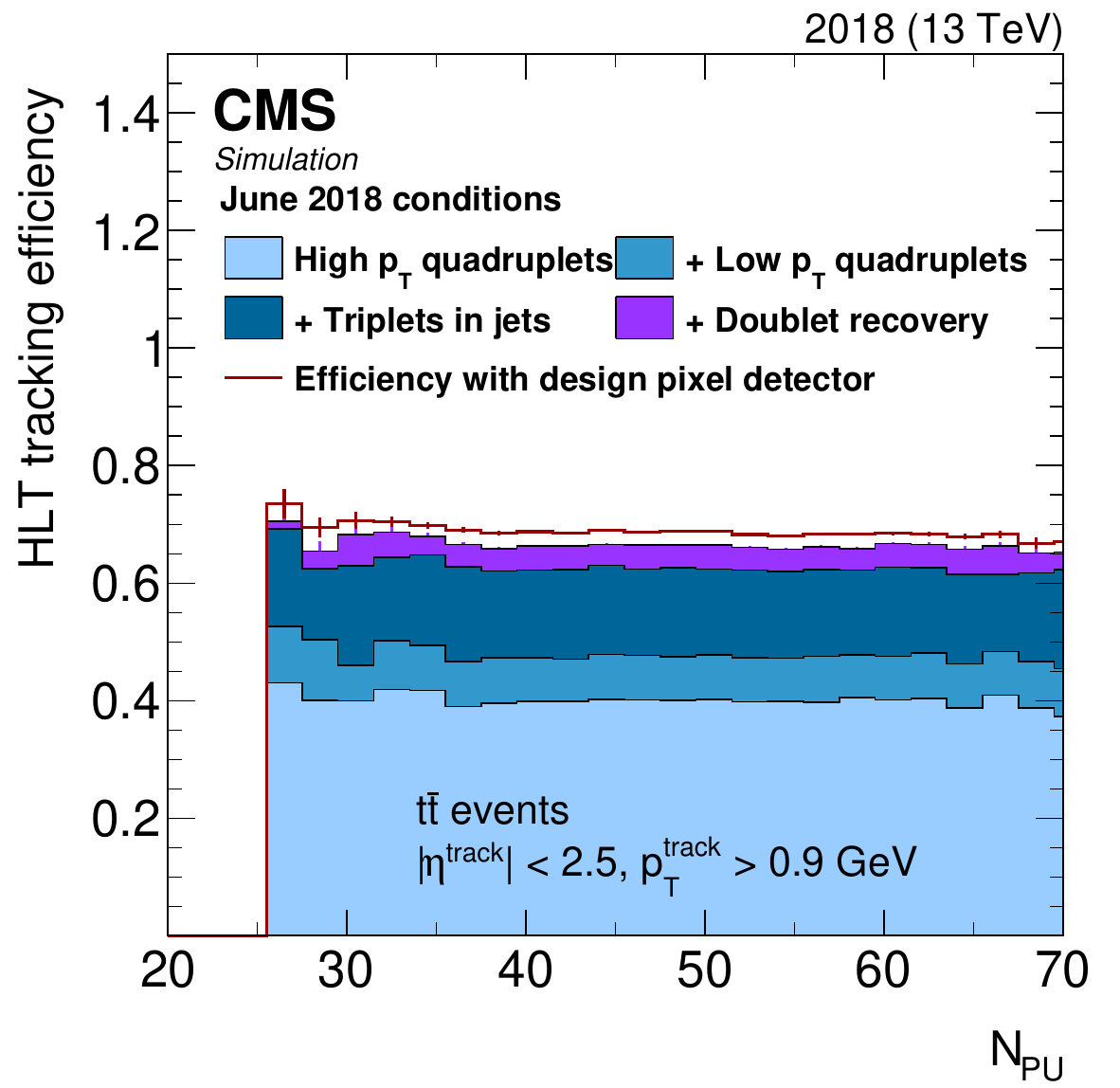}\\
  \includegraphics[width=0.48\textwidth]{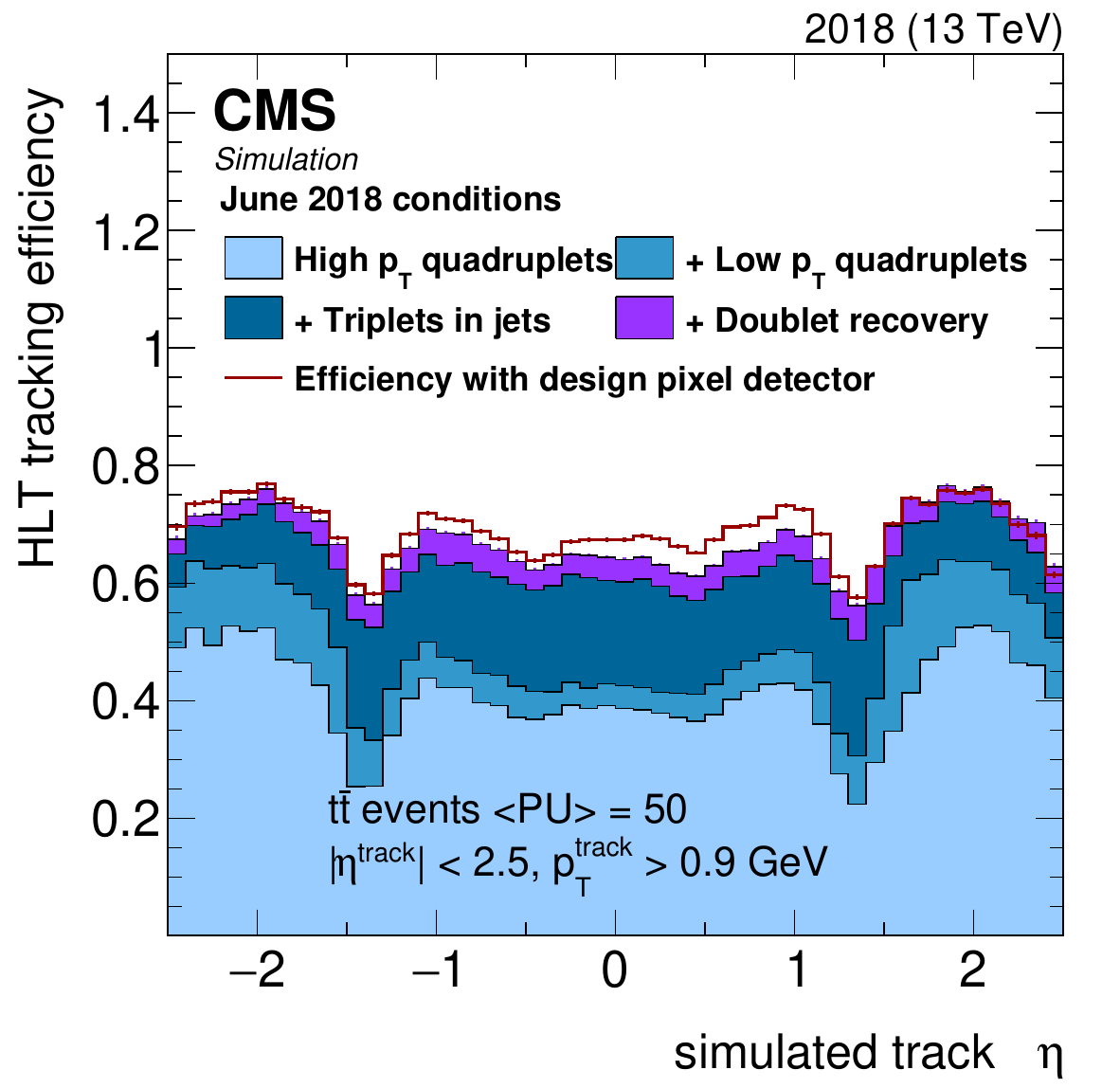}
  \includegraphics[width=0.48\textwidth]{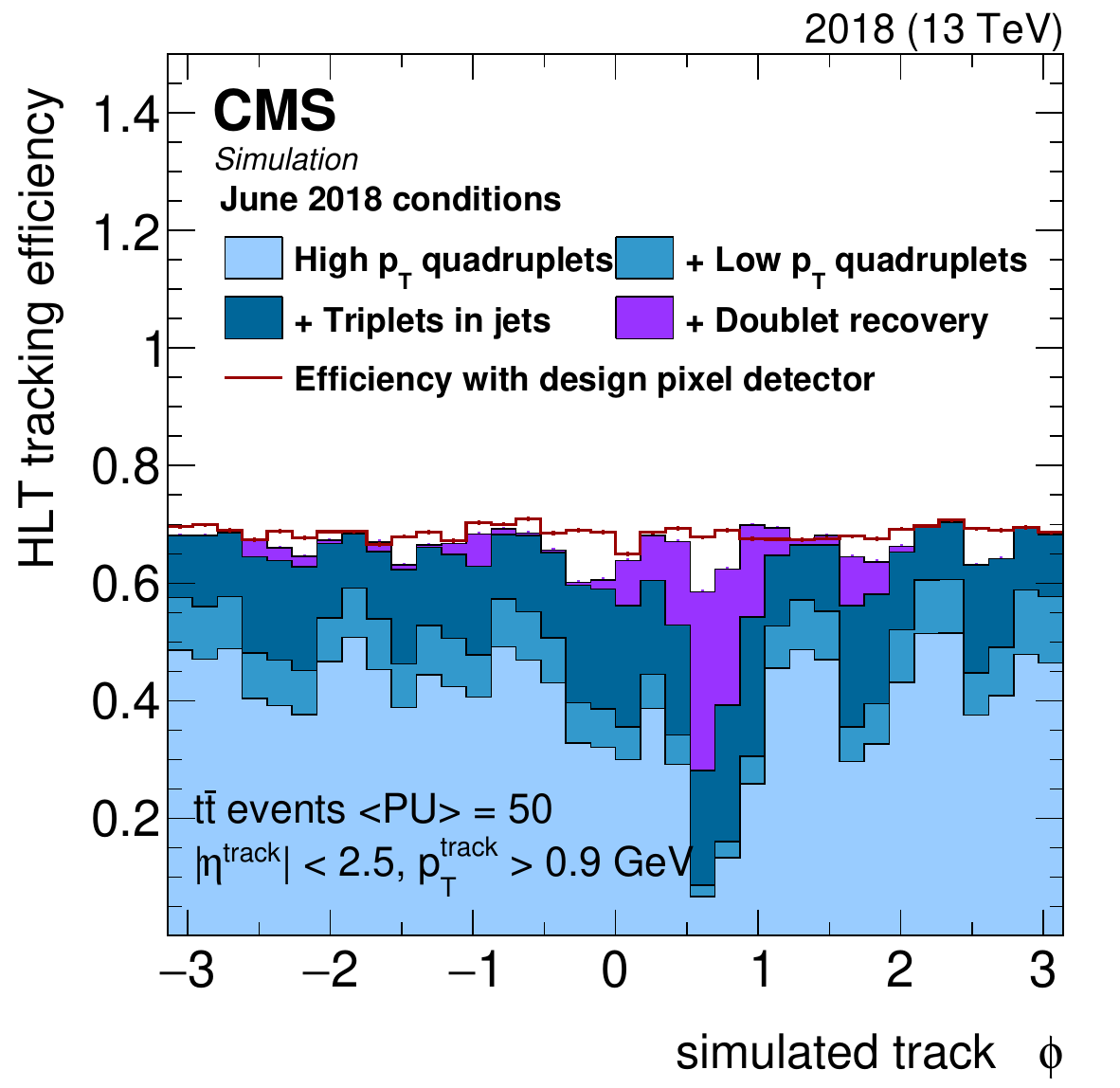}
  \caption{
    Tracking efficiency
    as a function of simulated track \pt (upper left), $N_{\text{PU}}$
    (upper right), $\eta$ (lower left), and $\phi$ (lower right).
    The contributions to
    the total efficiency from the different tracking iterations are shown
    in different colors. The initial three iterations are shown in
    shades of blue, and the contribution of the doublet recovery
    iteration is shown in violet. The simulation includes a map of inactive modules representing the
    status of the real detector as of June 2018.
    The performance that would be achieved
    with no inactive pixel detector modules and no doublet recovery iteration (design pixel detector) is shown as a red line. Details of the
    observed features are discussed in the text.
  }
  \label{fig:perf:trackingEffic}
\end{figure}

The tracking efficiency as a function of \pt, number of PU interactions ($N_{\text{PU}}$),
$\eta$, and $\phi$ is shown
in Fig.~\ref{fig:perf:trackingEffic}. 
The reduction of efficiency at large track \pt is characteristic of the sample used for the efficiency measurement. 
As the inactive modules
are not distributed uniformly throughout the detector, the tracking
performance is expected to be asymmetric in track $\eta$ and $\phi$. For
reference, the efficiency that would be achieved with the ``design pixel detector'',
namely with no inactive pixel detector modules,
is also shown in the figure. As the doublet-seeded iteration is not run in
the case of the design detector, in some cases the tracking
efficiency with the ``realistic detector'', which takes into account the pixel detector modules that have
become inactive, can be higher than with the
design detector. 
In the plateau region
around $\pt \approx 20\GeV$, the efficiency observed with the realistic detector
conditions is about 5\% lower than with the design detector.
This efficiency loss compared with the design detector is more
pronounced in the central part of the detector, and is concentrated
in the region around $\phi = 0.6$, where a significant number of inactive
modules is present.
The doublet-seeded tracking iteration is able to recover a significant
fraction of this efficiency loss above the \pt threshold of
1.2\GeV.
The performance of the
recovery procedure is not uniform across the $\eta$ and $\phi$ ranges since it
is invoked only if there are two
overlapping inactive modules, making it dependent on the
specific distribution of these modules.
The tracking efficiency is robust against the presence of PU,
decreasing only slightly with the number of additional
interactions. The performance of the doublet-seeded recovery is also
independent of PU.

\begin{figure}
  \centering
  \includegraphics[width=0.48\textwidth]{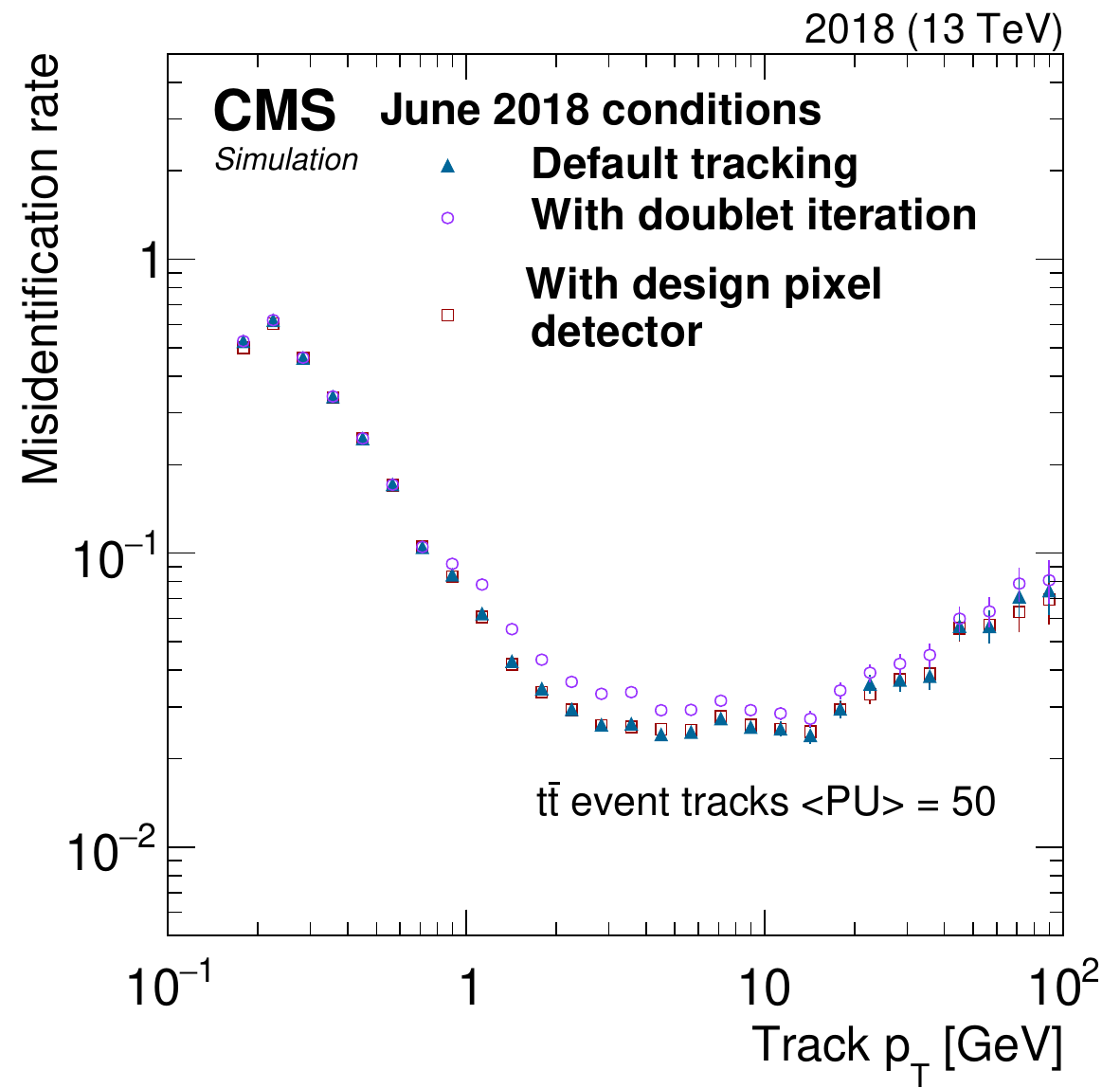}
  \includegraphics[width=0.48\textwidth]{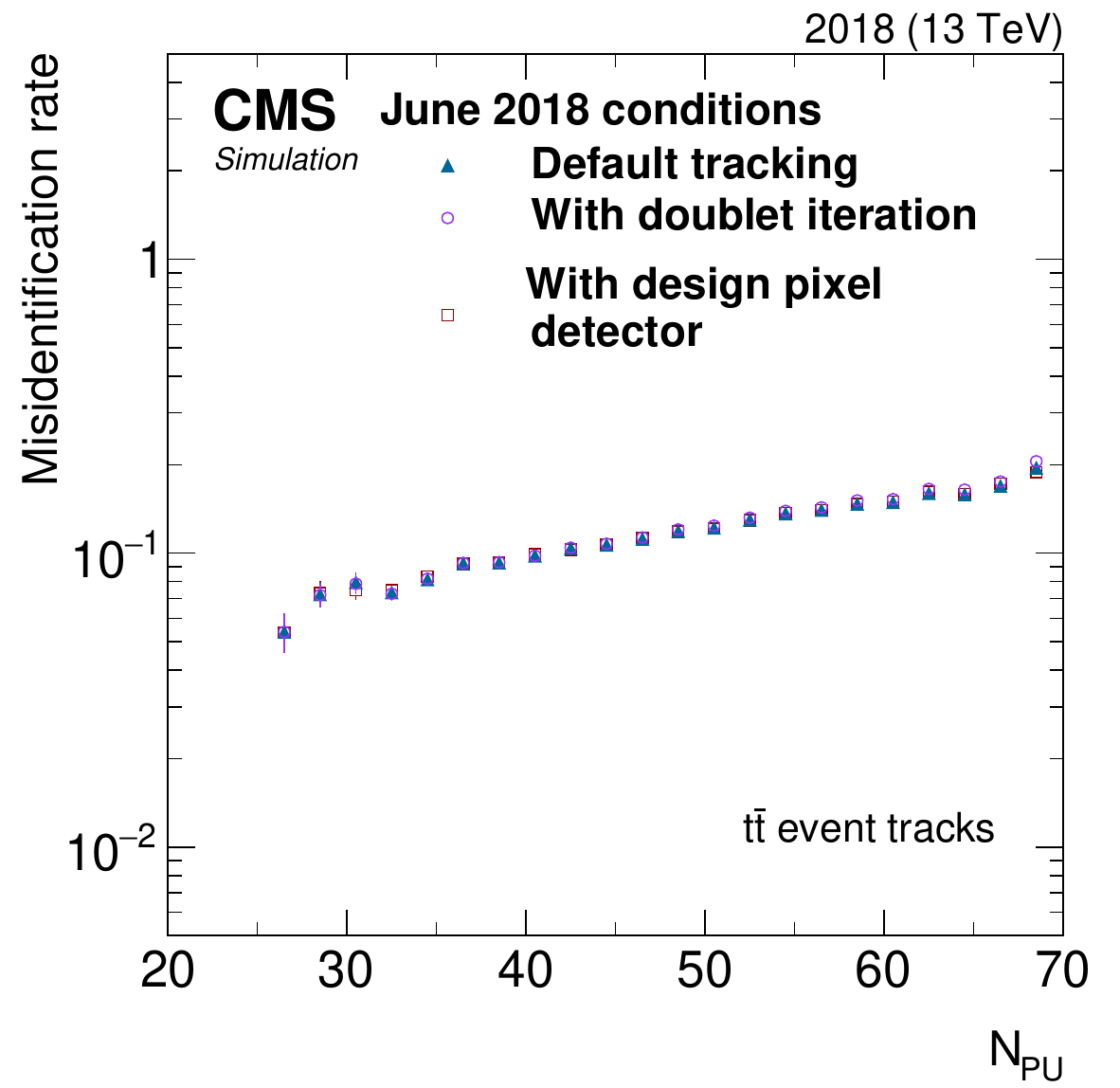}\\
  \includegraphics[width=0.48\textwidth]{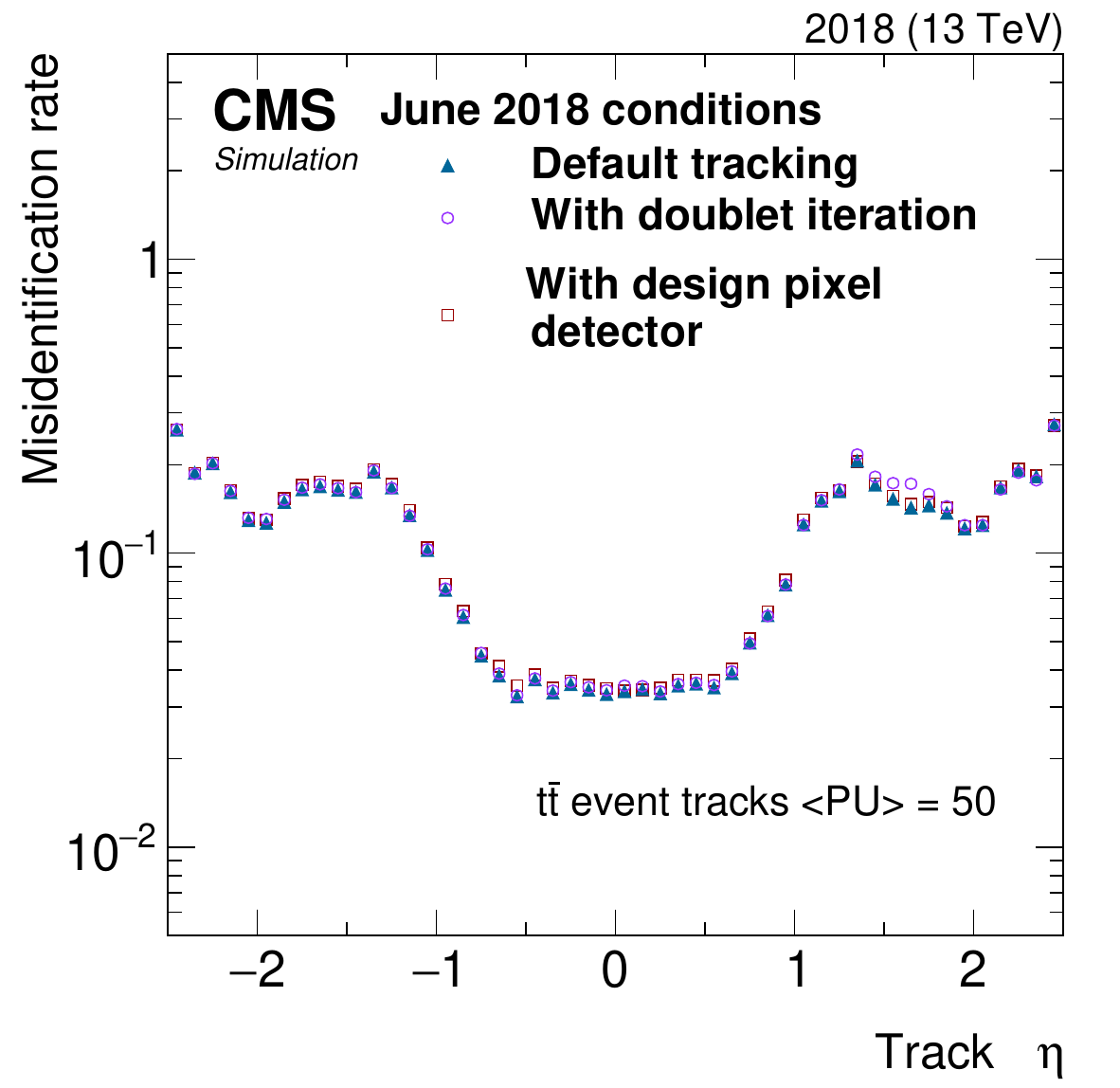}
  \includegraphics[width=0.48\textwidth]{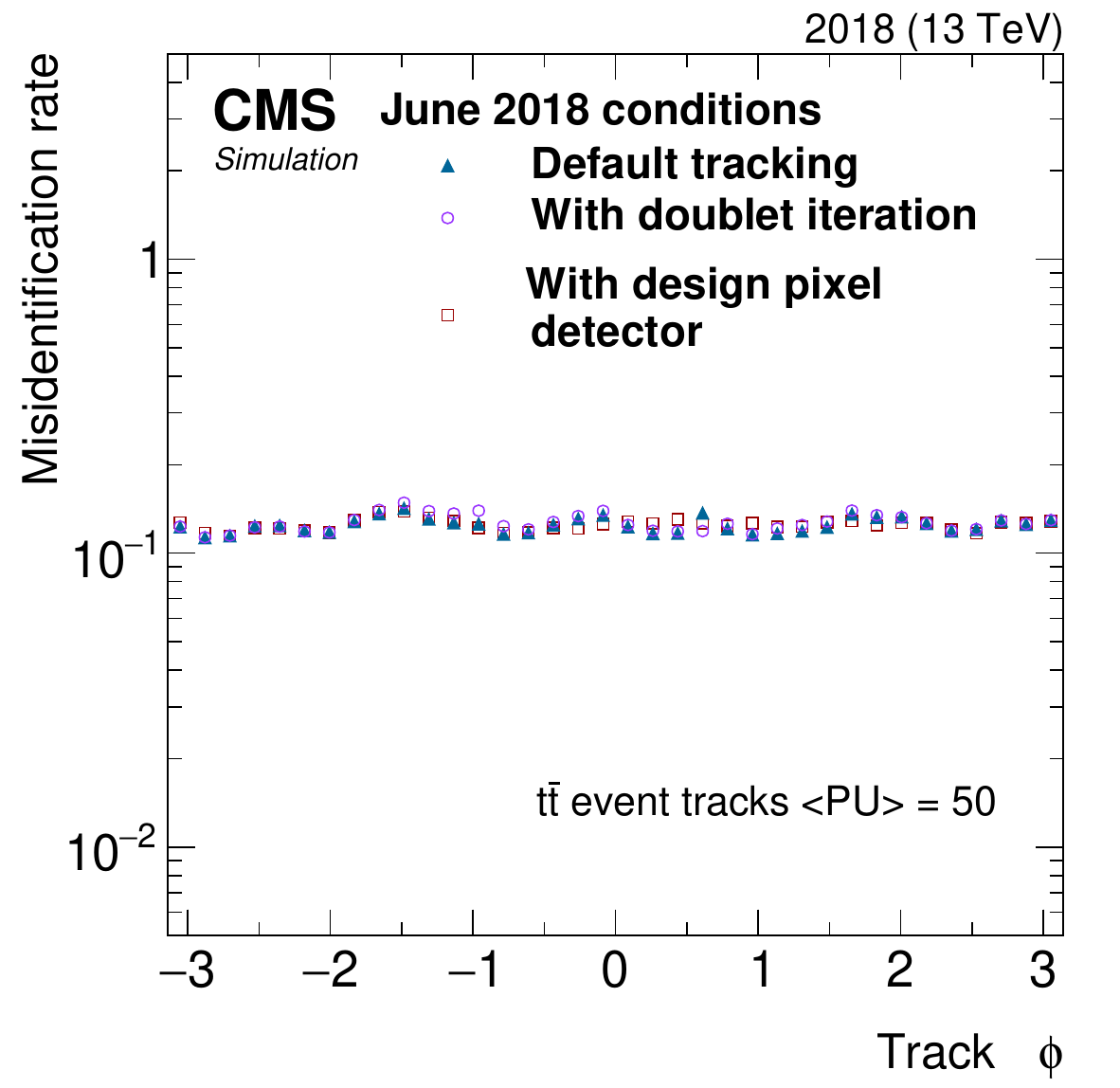}
  \caption{
    Tracking misidentification rate
    as a function of track \pt (upper left), $N_{\text{PU}}$
    (upper right), $\eta$ (lower left), and $\phi$ (lower right). No selection on track kinematics is applied. The misidentification rate for the first three iterations (default tracking) is
    shown in dark blue triangles, whereas the misidentification rate after including the doublet
    recovery iteration is shown in violet circles. The simulation includes a map of inactive modules representing the status of the real detector as of June 2018. The misidentification rate that would be
    observed with no inactive pixel detector modules and no doublet recovery iteration (design pixel detector) is shown in dark red squares.
  }
  \label{fig:perf:trackingFakeRate}
\end{figure}

The tracking misidentification rate as a function of \pt, $N_{\text{PU}}$, $\eta$, and $\phi$ is shown
in Fig.~\ref{fig:perf:trackingFakeRate}. 
There is no difference between the misidentification rates with the design and
realistic detector conditions. Taking into account the doublet-seeded
recovery iterations, a slight increase of the misidentification rate above the \pt
threshold of this iteration is observed.
When integrated over all \pt values, no significant increase in the
misidentification rate is observed for the doublet-seeded recovery iteration as
seen in the other plots.
The misidentification rate does increase with the number of additional PU
interactions for either tracking scenario and for the design
pixel detector. 

\subsection{Muons}
\label{sub:HLTRecoPerF_Muons}

Tracking algorithms are also deployed to identify and reconstruct
muons measured in the muon detectors in combination with the pixel
and strip trackers. Since the algorithms used during \Runtwo are
described in more detail in Ref.~\cite{MUO-19-001}, a brief summary
is given here.  

Muon track reconstruction at the HLT takes place in two steps: first using
hits only in the muon system (L2 reconstruction), followed by a
combination with hits in the inner tracking system (L3
reconstruction). The L2 reconstruction is equivalent to the 
standalone muon reconstruction performed offline.
The reconstruction at L3 is seeded by
an L2 muon and follows an iterative track reconstruction similar to
that described in the previous section in a region around the seed
starting from the outer tracking layers and working inward
(``outside-in'') or from the inner tracking layers working out
(``inside-out''). The latter inside-out approach also can be seeded
directly by muons reconstructed by the L1 trigger (``L1 muons'') using
muon detector information only.  
The L3 track reconstruction is essentially 100\% efficient with respect to
L1-identified muons, and it reduces the rate by more than an order of magnitude for the same \pt threshold because of the
improved momentum resolution of the inner tracking system. However,
the L3 reconstruction consumes approximately 20\% of the overall HLT CPU
time per event. After the track reconstruction, identification criteria are
applied as well as isolation criteria for the isolated muon
category. The isolation is based on the sum of \pt from additional
tracks associated with the primary vertex and calorimeter energy
deposits clustered using an algorithm based on the PF candidates in a cone of radius $\DR= \sqrt{(\Delta\phi)^2+(\Delta\eta)^2} = 0.3$ around the muon. The
estimated contribution from PU to the energy deposits in the
calorimeter is subtracted.

\begin{figure*}
  \centering
  \includegraphics[width=0.9\textwidth]{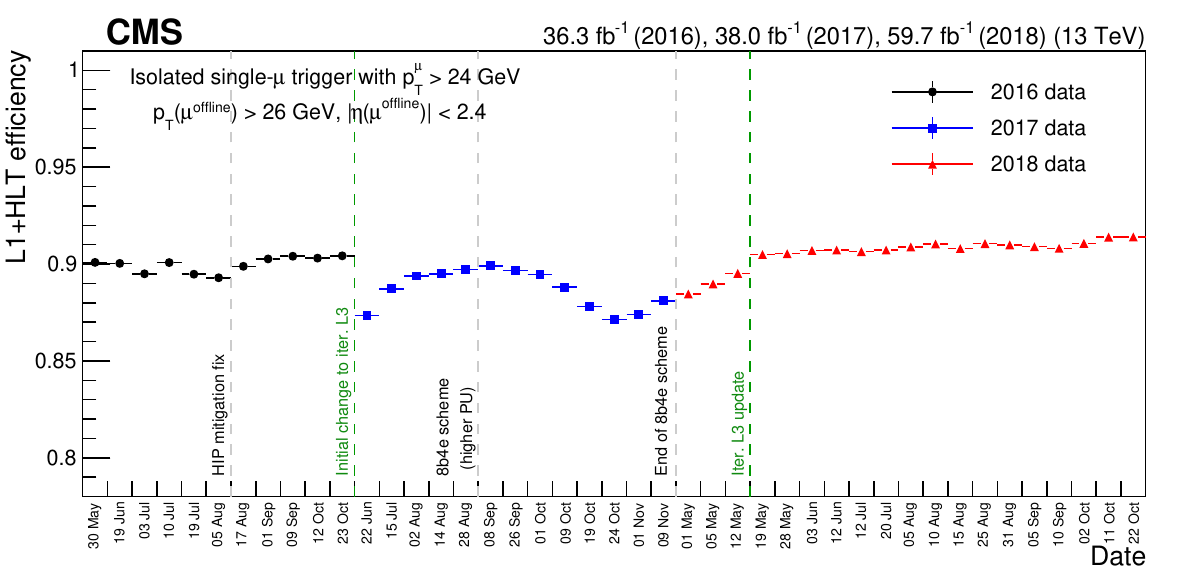}
  \caption{Evolution of the isolated single-muon trigger
    efficiency with $\pt > 24\GeV$ as a function of data-taking dates during
    the \Runtwo period from 2016 to 2018.
    Each point is
    the efficiency measured using the
    data with an integrated luminosity of about 3\fbinv.
    Dashed lines show the changes in the LHC or CMS
    conditions that could have an impact on the trigger performance,
    such as the fix for the degradation of the inner tracker as a result of
    heavily ionizing particles (``HIP mitigation fix'') or the change
    in the filling scheme for $\Pp\Pp$ collisions at the LHC (``8b4e scheme'')
    that led to higher PU in CMS until the end of 2017. The
    change of the reconstruction algorithm for L3 muons are
    presented as green dotted lines, including the replacement of
    cascade or tracker muon algorithm to the iterative algorithm
    (``Change to iter. L3'') and the update of the iterative algorithm
    to overcome the limitations observed in 2017 (``Iter. L3
    update'').
  }
  \label{fig:MuEffvsDateOffline}
\end{figure*}

The combined {L1}+HLT muon trigger efficiency of an isolated single-muon trigger
with $\pt > 24\GeV$ with respect to offline-reconstructed muons is presented in
Fig.~\ref{fig:MuEffvsDateOffline} as a function of the data-taking date,
which shows the effect of the evolution of the
muon reconstruction algorithm during \Runtwo, as well as the detector
and machine conditions. The maximum efficiency of about 90\% is
primarily set by the L1 trigger.
Offline-reconstructed muons with $\pt > 26\GeV$ are used.
In 2016, two different approaches were used to reconstruct
L3 muons. The first one (the ``cascade'' algorithm) starts from L2 muons as
seeds and consists of three different methods to reconstruct L3
muons. Each method uses the outside-in or inside-out approach with
different ways to find tracks in the inner tracker. The fastest method
in calculation time is used first, and then it proceeds to the next methods
only if an L3 muon is not found in the previous method.
The other approach (the ``tracker muon'' algorithm) starts from L1 muons as
seeds, which are used to define an inner tracker region to perform
the track reconstruction.
Reconstructed inner tracks matched to segments in the muon stations
are then tagged as muons.
By combining with the L3 muons from the cascade algorithm, it
improves the overall performance, especially when L2 muons are not
properly reconstructed. The performance was stable with an overall
efficiency of about 90\% during the whole of 2016 operations, showing
robustness in early 2016 during a period of degradation of the inner
strip tracking detectors before their operational parameters were retuned to
reduce their susceptibility to highly ionizing particles.

In 2017, a new algorithm for the L3 muon reconstruction (the ``iterative''
algorithm) was implemented. It combines the advantages of both the cascade
and tracker muon algorithms and replaced them. It starts
with the outside-in step seeded by L2 and continues to find more L3
muons by two inside-out steps seeded by L2 or L1 muons.
In early 2017, the performance of the initial version of the iterative
algorithm was not as good as that of the previous algorithm, since a
few technical weak points in the algorithm were not identified
during validation with simulated events. The performance was consistently improved by
implementing several patches during the data taking until the middle
of 2017. However, the efficiency decreased later, mainly as a result of pixel detector module
losses mentioned in the previous section and higher PU, as
indicated in Fig.~\ref{fig:MuEffvsDateOffline}.
The efficiency did rise nevertheless toward the end of 2017 and early 2018
because of a slight reduction in the amount of PU.
To improve the robustness of the
algorithm, significant changes were introduced in 2018. To recover the
efficiency, all L1 muons were used in the L1-seeded step by removing
the \pt requirement, and an
iterative tracking step was added in the inside-out steps using the
pixel doublet as seeds. In parallel, to improve the purity and rate,
identification criteria were imposed on L3 muons at the last step of
the algorithm. These improvements were implemented in May 2018 as
denoted in the figure, restoring the efficiency to be similar to the 2016
level, up to the end of the \Runtwo operation.

\begin{figure}
  \centering
  \includegraphics[width=0.48\textwidth]{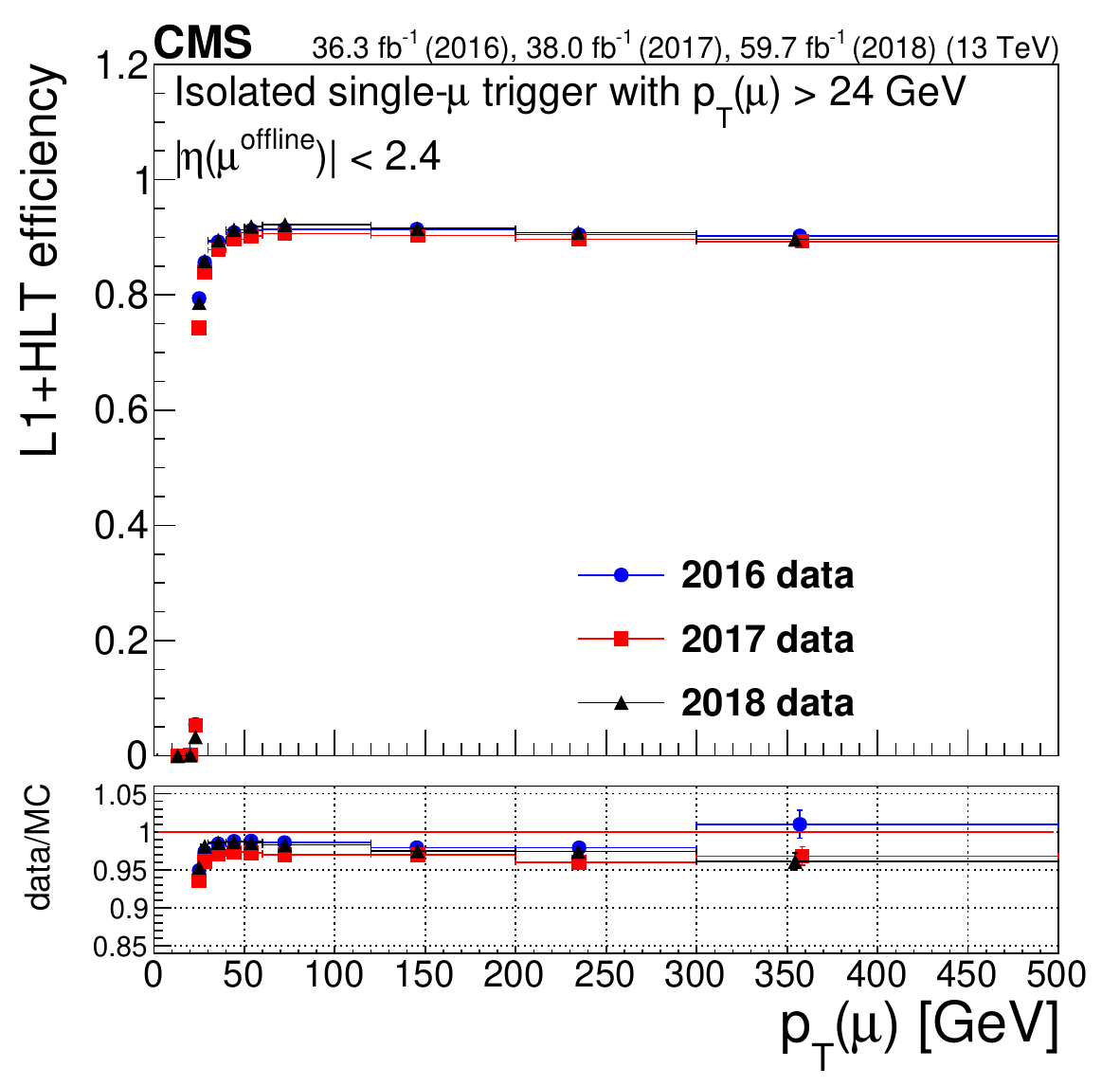}
  \includegraphics[width=0.48\textwidth]{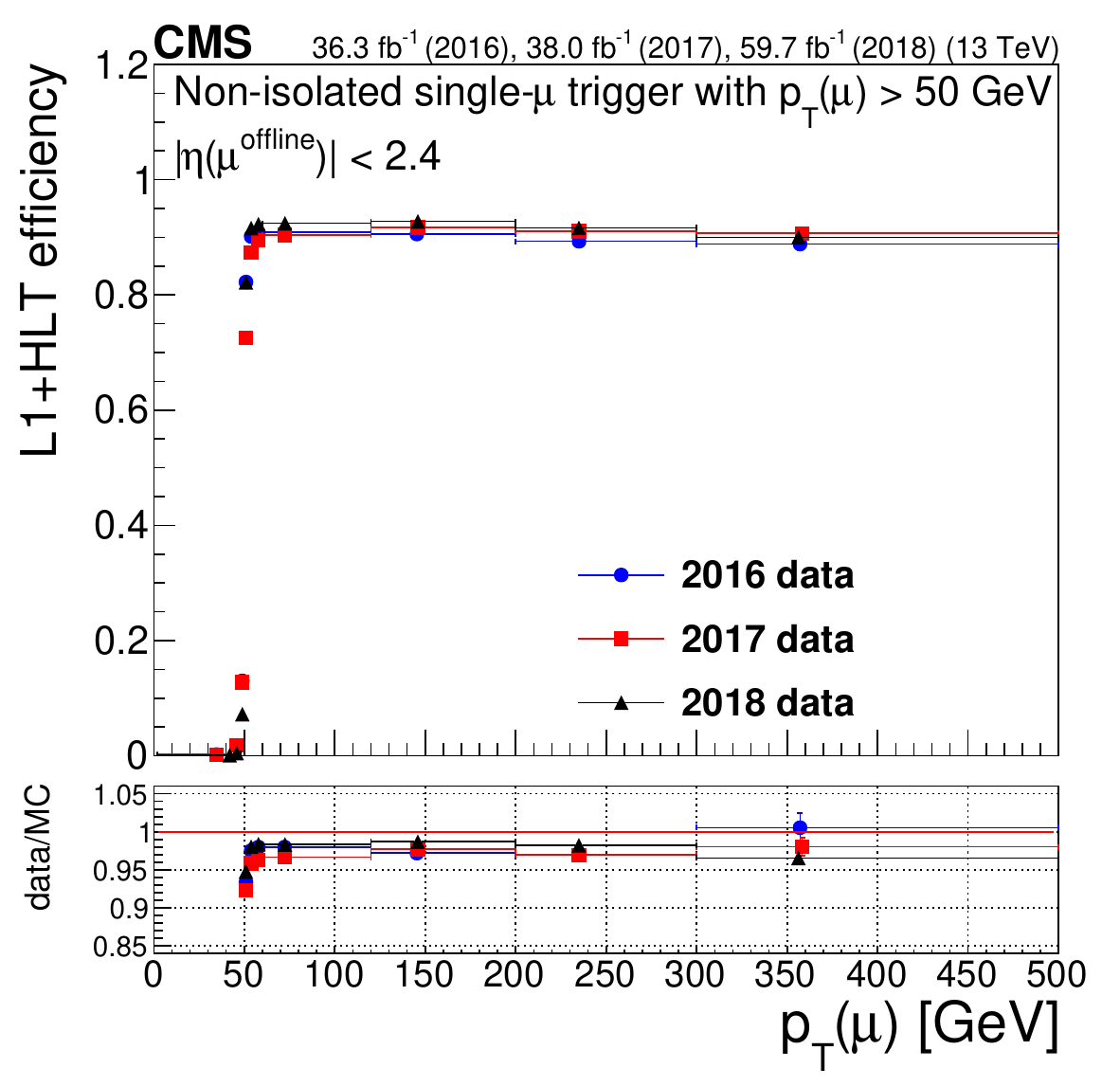} \\
  \includegraphics[width=0.48\textwidth]{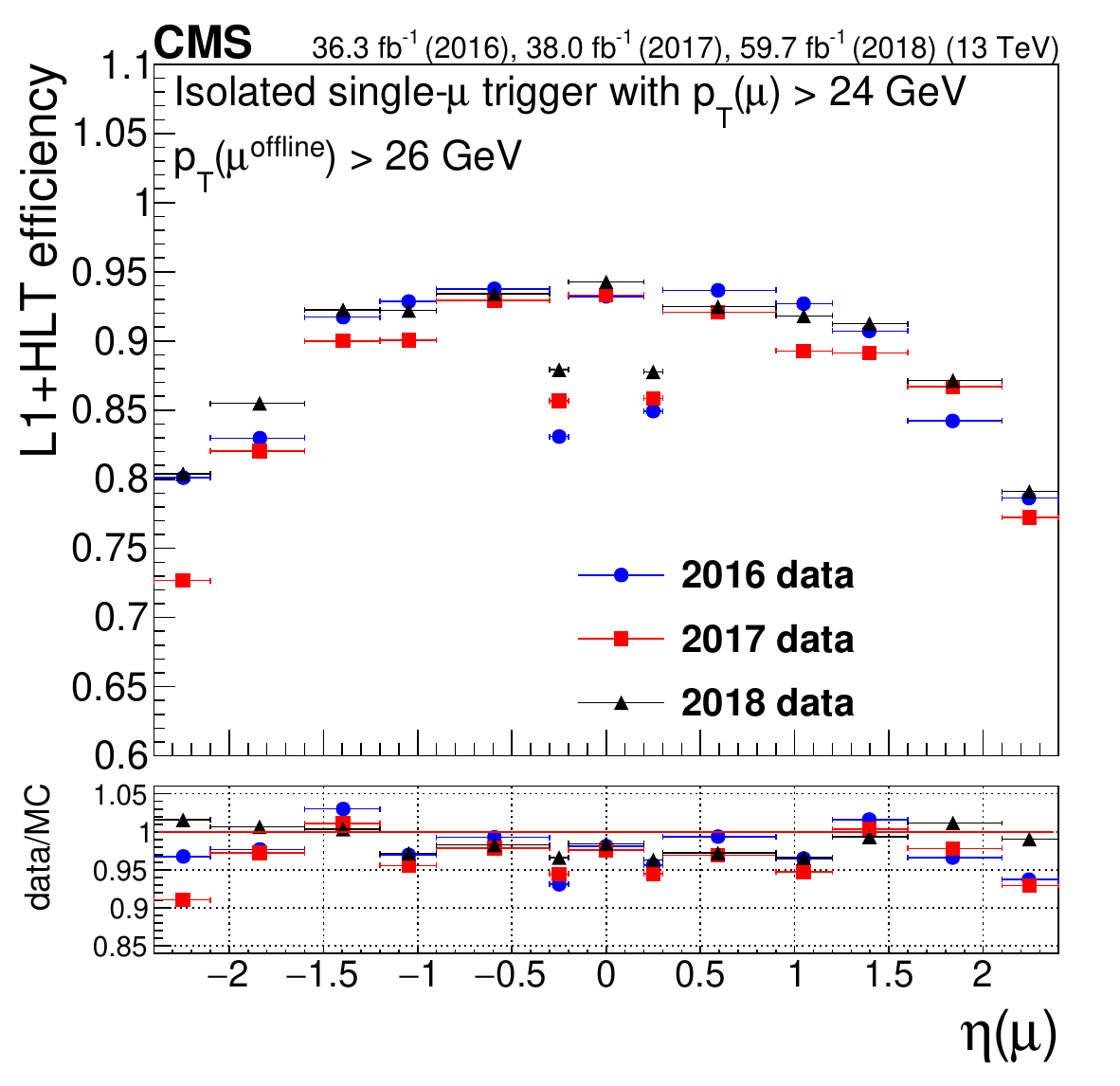}
  \includegraphics[width=0.48\textwidth]{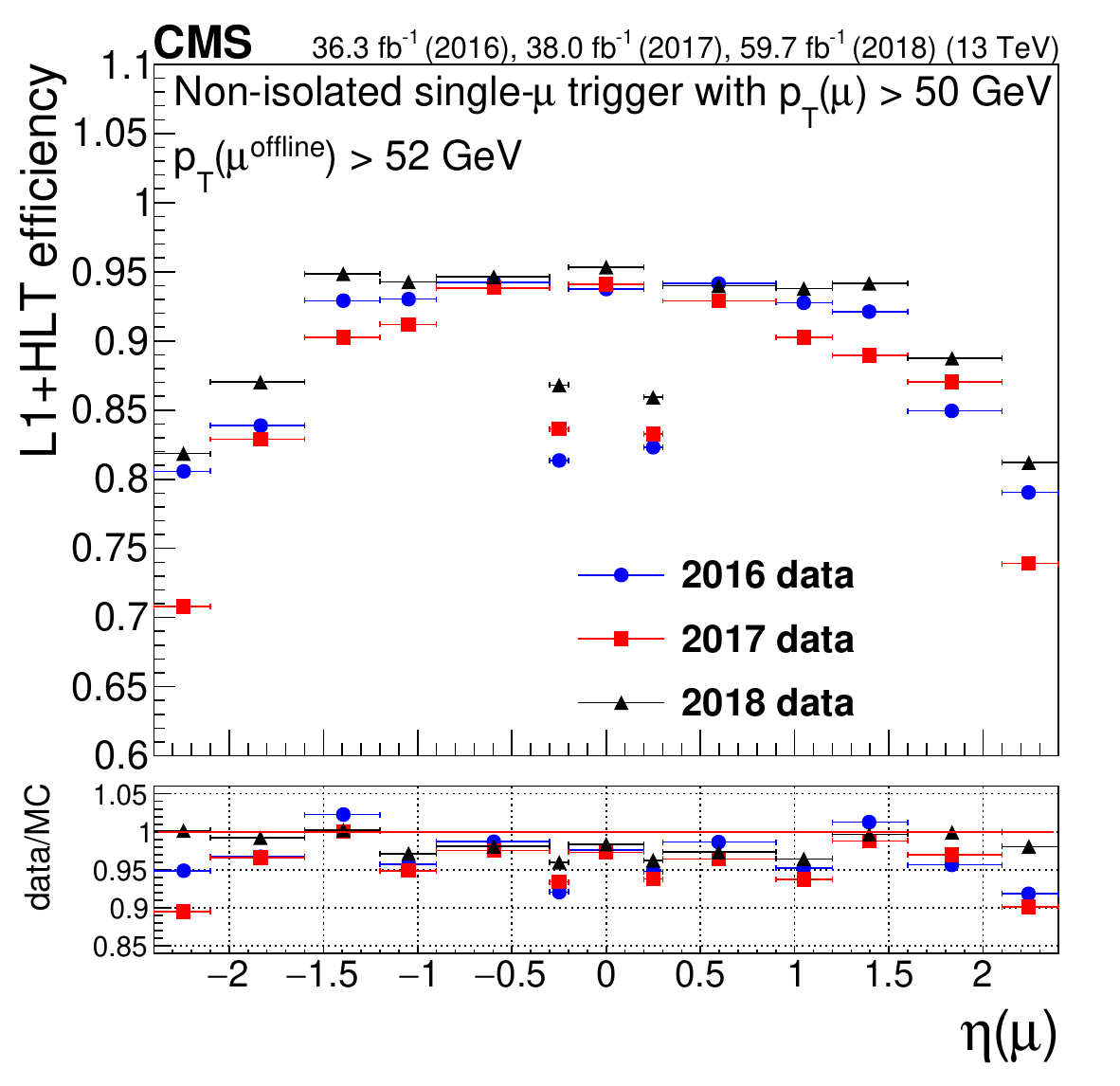} \\
  \includegraphics[width=0.48\textwidth]{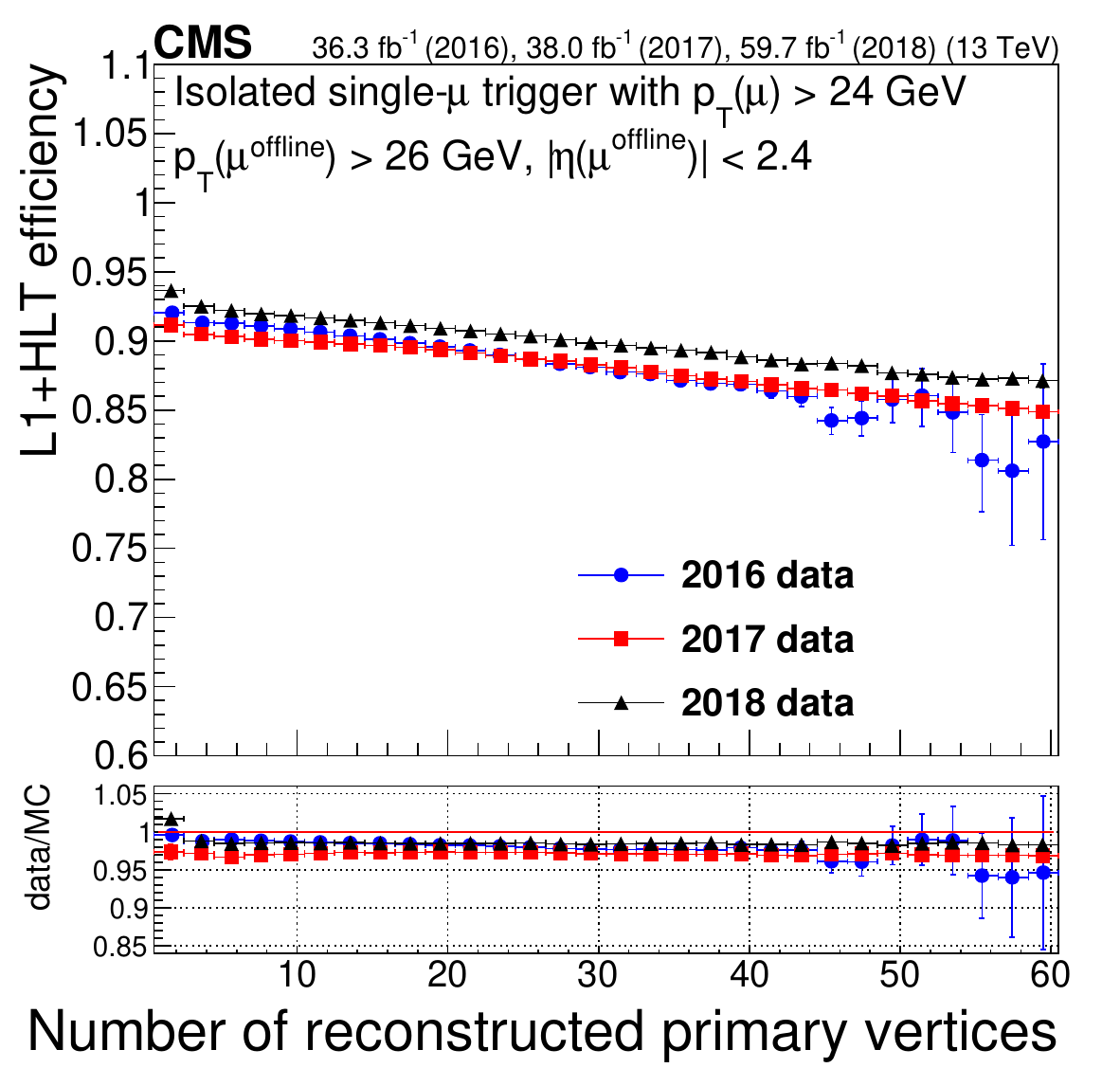}
  \includegraphics[width=0.48\textwidth]{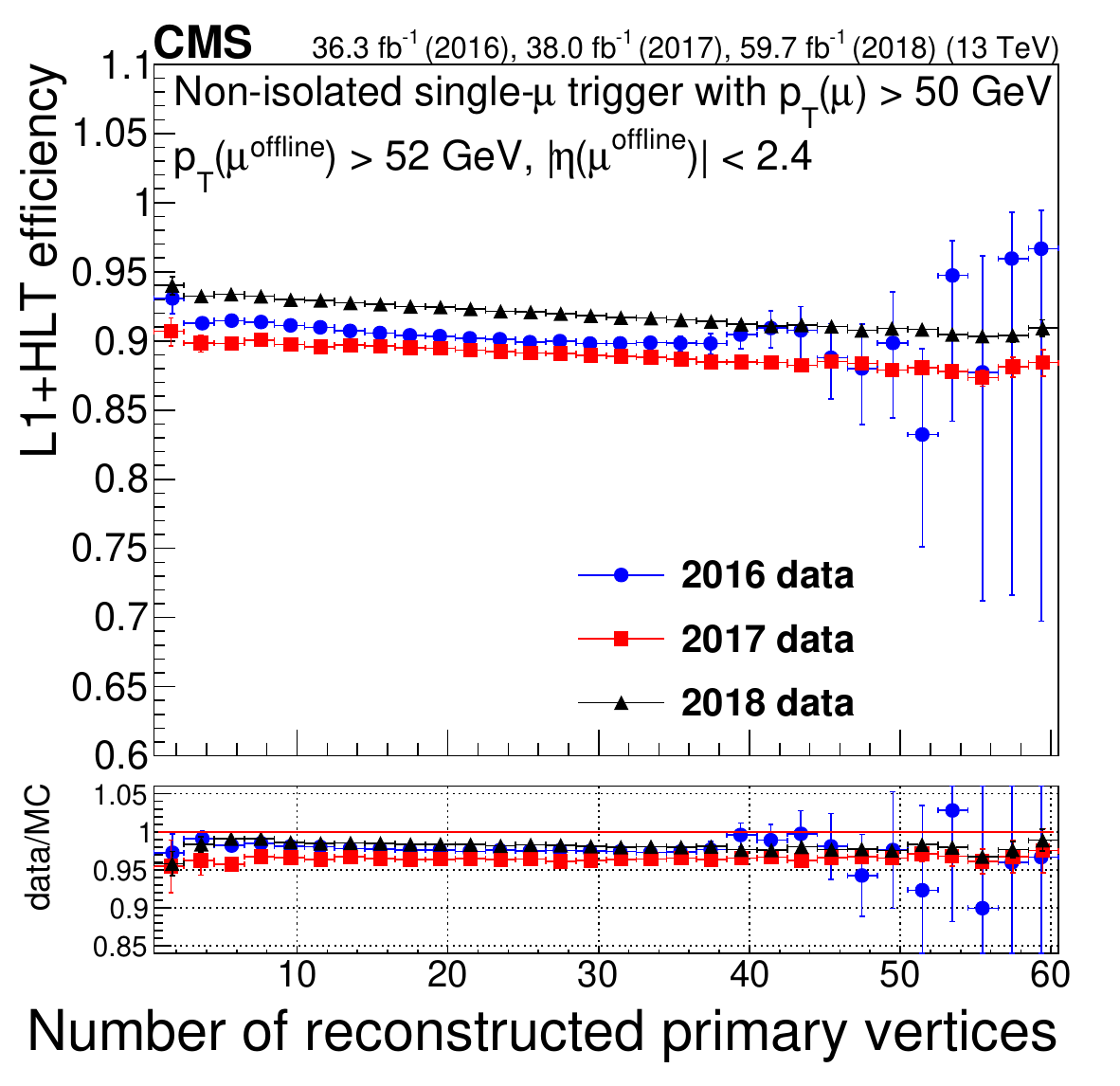}
  \caption{Trigger efficiencies for the isolated single-muon
    trigger with $\pt > 24\GeV$ (left column) and the
    nonisolated single-muon trigger with $\pt > 50\GeV$
    (right column), as functions of muon \pt (upper row), $\eta$ (middle
    row), and $N_{\text{vtx}}$ (lower row).
    The lower panel of each plot shows the ratio of data to MC simulation
The vertical bars on the markers represent statistical uncertainties.
  }
  \label{fig:perf:singleMuEff}
\end{figure}

Figure~\ref{fig:perf:singleMuEff} (left column) shows the efficiency of
the isolated single-muon trigger with $\pt > 24\GeV$
as a function of muon \pt, $\eta$, and the number
of reconstructed primary vertices ($N_{\text{vtx}}$) for three years of data taking:
2016, 2017, and 2018. The right column of
Fig.~\ref{fig:perf:singleMuEff} shows the corresponding
efficiency distributions for 
the nonisolated single-muon trigger with $\pt > 50\GeV$. The panel
below each figure shows the ratio of the efficiency measured for data
to that of simulation.

\begin{figure}
  \centering
  \includegraphics[width=0.48\textwidth]{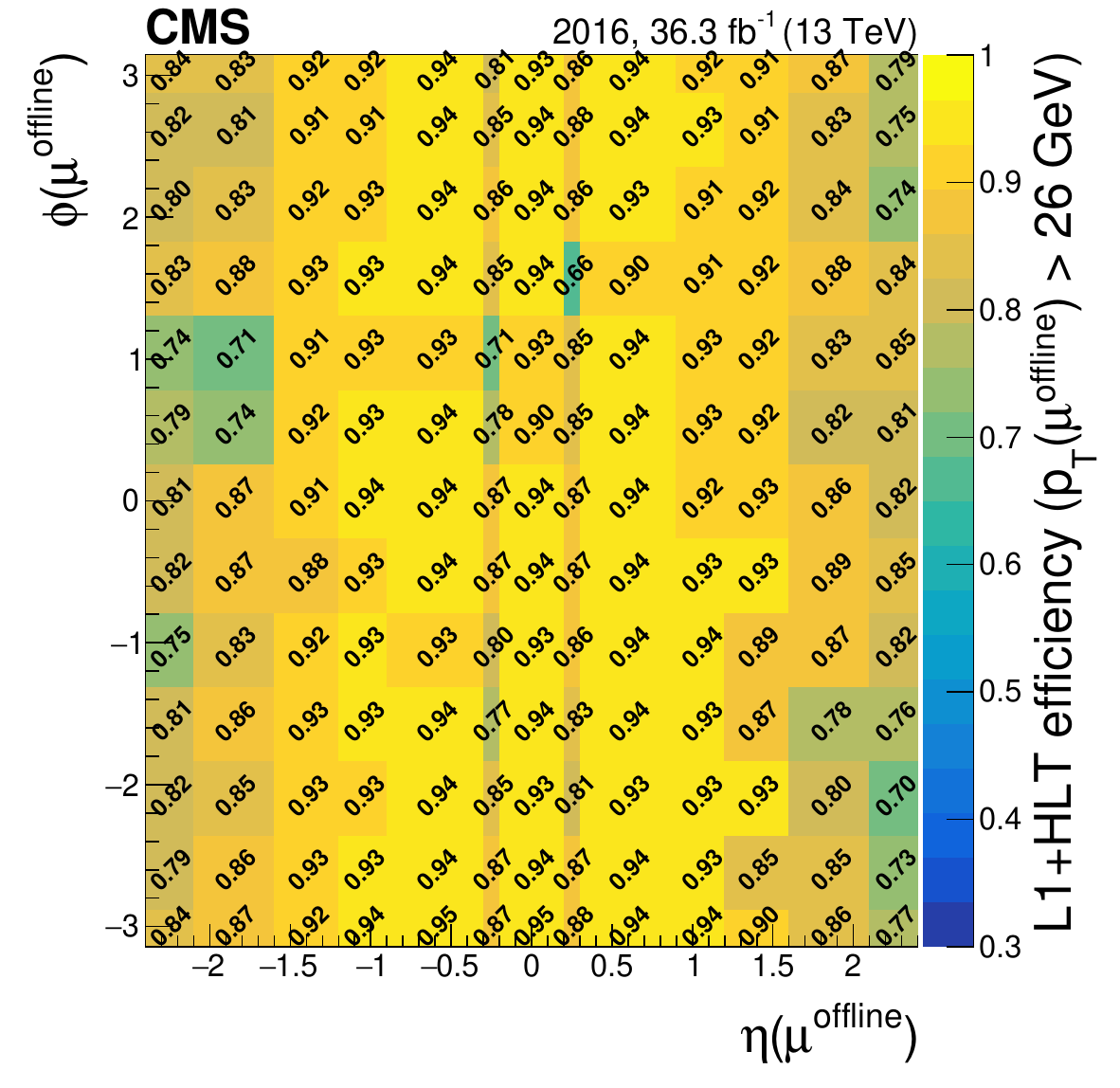}
  \includegraphics[width=0.48\textwidth]{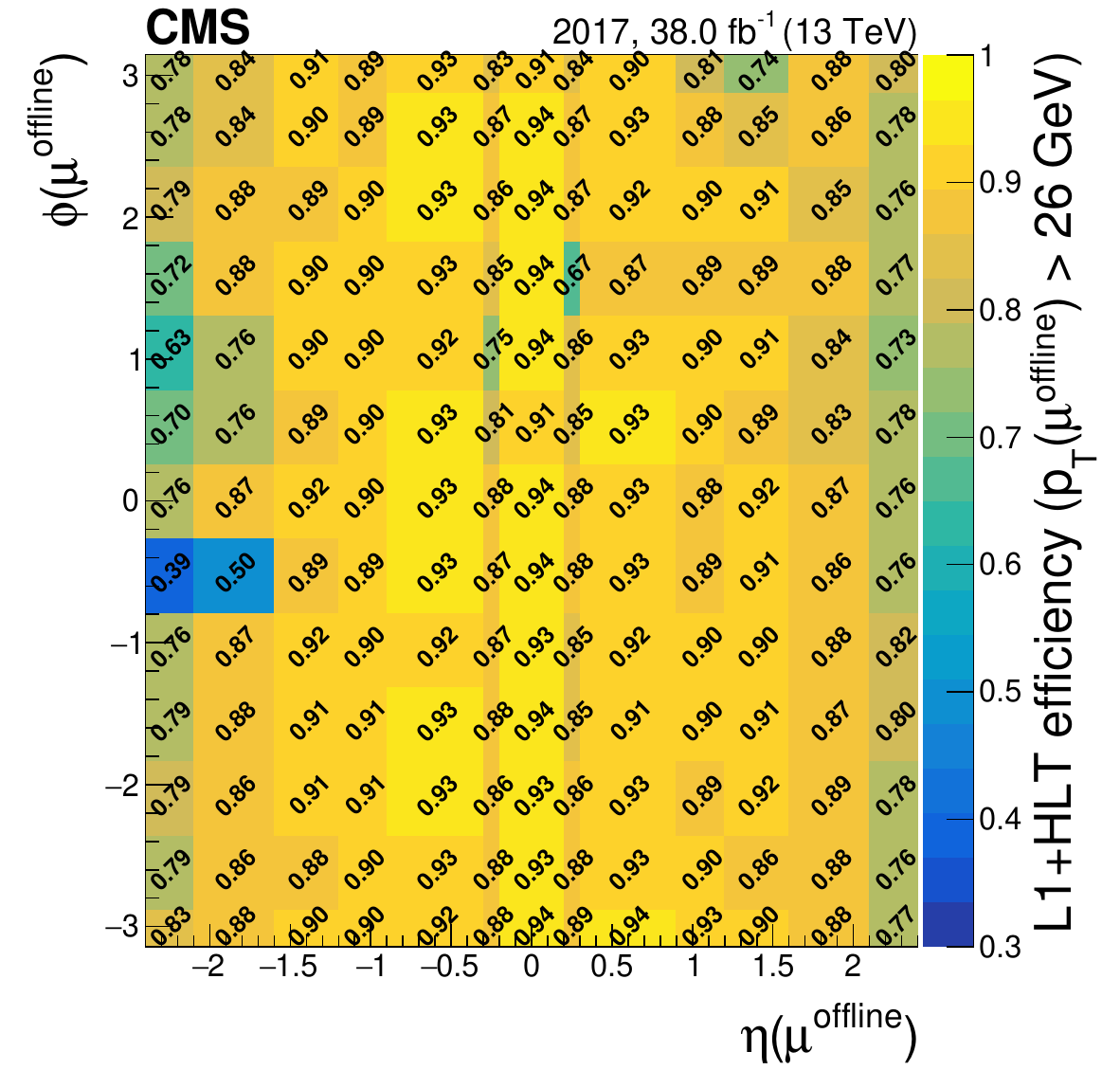}
  \includegraphics[width=0.48\textwidth]{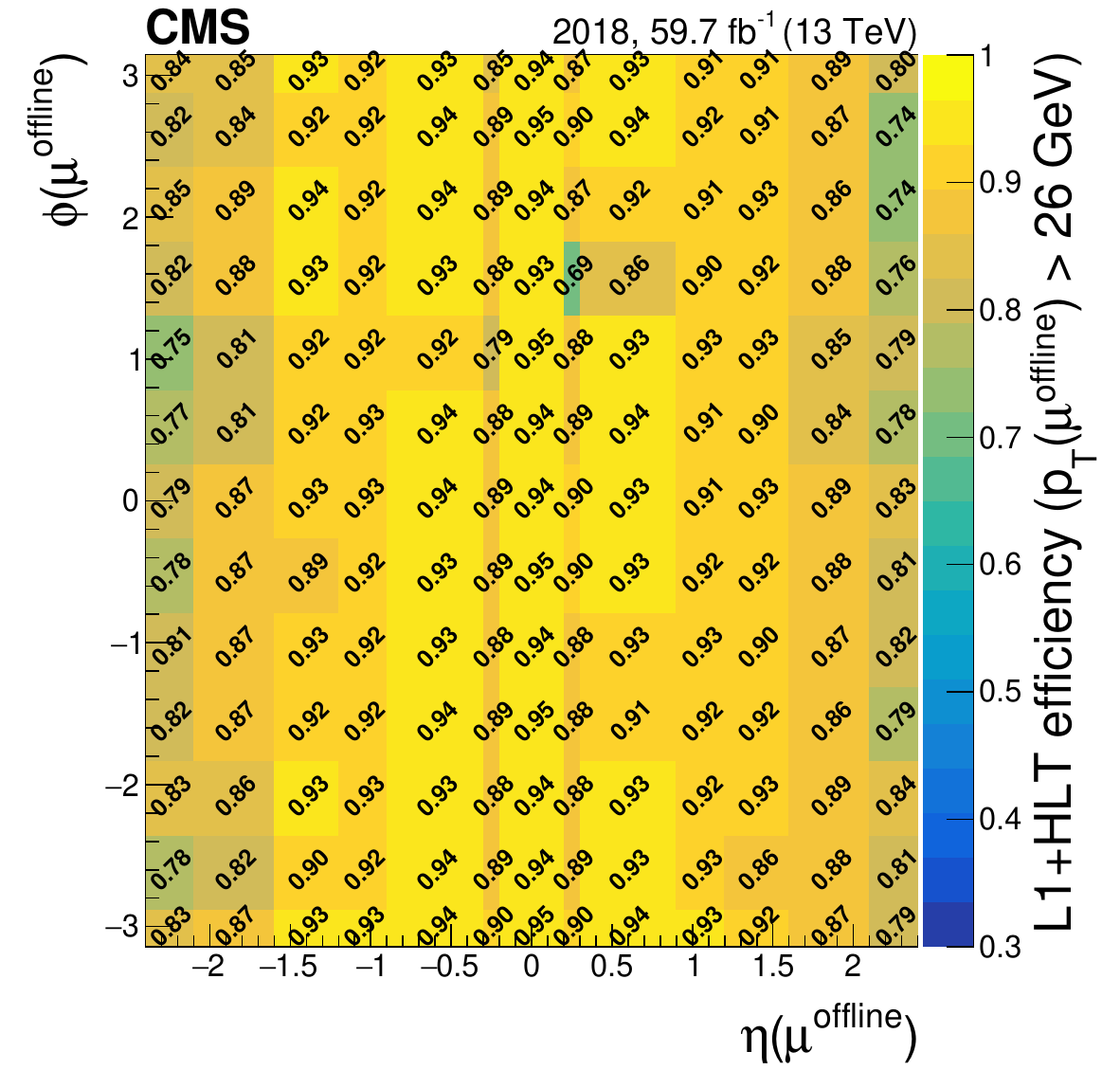}
  \caption{Efficiency of the isolated single-muon trigger with $\pt > 24\GeV$ as a function of the $\eta$ and $\phi$ of the muons in 2016 (upper left), 2017 (upper right), and 2018 (lower).
  }
  \label{fig:perf:MuEff2D}
\end{figure}

Figure~\ref{fig:perf:MuEff2D} shows the efficiency of the same isolated single-muon
trigger as in Fig.~\ref{fig:perf:singleMuEff}, plotted as a
function of the $\eta$ and $\phi$ of the muons. Different muon detector
technologies and different reconstruction algorithms in both the L1 and HLT were used
depending on the geometric region, but the overall efficiency is
generally stable across \Runtwo, except for a few specific regions related to issues in the muon detector.

The minimum \pt thresholds used for double-muon triggers are lower than those of single isolated muon triggers.
As Table~\ref{tab:HLT2-simplifiedMenu} shows, the lower-\pt threshold
of the isolated double-muon trigger used in the \Runtwo trigger menu
is 8\GeV. The efficiency of this
lower-\pt ``leg'' is reported in Ref.~\cite{MUO-19-001}. It exhibits a
sharp turn-on in the efficiency vs. \pt at the threshold, with a plateau efficiency of
${\approx}95\%$  that is stable across the years 2016--2018 to within about 1\%.

\subsection{Electrons and photons}
\label{sub:HLTRecoPerF_EGM}

The electron and photon candidates at L1 are based on trigger
towers defined by arrays of 5$\times$5 ECAL crystals along with
the HCAL tower directly behind them in the barrel, and
in the endcaps are formed from groups of 5--25 crystals depending on
their $\eta$-$\phi$ position~\cite{TRG-17-001}.
The trigger tower with the largest \ET is clustered
together with its adjacent \ET towers using a procedure
that also trims the energy deposits
to only include contiguous towers to match the electron or photon signature in the calorimeter.
To form an L1 candidate, energy clusters must satisfy additional identification criteria
and, optionally, isolation requirements.
The HLT electron and photon identification begins with a regional
reconstruction of the energy deposited in the ECAL crystals around the
L1 candidates. In \Runtwo, the signals in the ECAL crystals are
reconstructed by fitting the signal pulse with multiple template
functions, to mitigate out-of-time PU. The
signal amplitudes are then corrected by per-crystal correction factors
and per-channel calibration techniques, which, to deal with the
increasing ECAL crystal opacity from radiation damage, require
frequent updates to maintain performance. Clusters of ECAL deposits
within a certain geometric area around the seed cluster, called
``superclusters,'' are then built, using the same reconstruction
algorithm as used offline~\cite{CMS:2015xaf}.  
However, the energy correction applied to HLT superclusters is simpler
than the one used offline in that it employs ECAL information
only. This correction is needed to take into account possible energy
losses of the electrons and photons travelling through the detector
material. 
 After requesting a minimal threshold on the energy, requirements are
 applied based on properties of the energy deposits in the ECAL
 and HCAL subdetectors, according to the compactness and shape of
 electromagnetic showers. In the case of electrons, the ECAL
 supercluster is associated with a reconstructed track with a direction
 compatible with the cluster location.
 The first step is a match with pixel detector hits. Since 2017, the
 pixel matching algorithm requires three pixel detector hits rather than two, to maximize early
 background rejection, while a hit doublet is accepted only if the
 trajectory passes through a maximum of three active modules.
 Once the
 supercluster is associated with the pixel detector seeds, the
 electron track is reconstructed using a dedicated tracking algorithm,
 based on the Gaussian sum filter~\cite{GSF}. However, not all
 electron HLT paths run this algorithm: in some cases,
sufficient rate reduction is already achieved from pixel detector matching
 alone. 

Single- and double-electron triggers are the first selection step of
most analyses using electrons. In the following, their performance
is reported, using the full 2016, 2017, and 2018 data sets, corresponding to an
integrated luminosity of 136\fbinv~\cite{CMS-LUM-17-003, CMS-PAS-LUM-17-004, CMS-PAS-LUM-18-002}.
The performance of photon triggers, which are very similar to those of
electron triggers apart from the absence of the requirement on the
presence of matching tracks, is not reported here. This is because
photon triggers are typically designed for specific analyses and are not used
as extensively.

\begin{table}[!bthp]
  \centering
  \topcaption[]{Tag-and-probe selections used for the single- and double-electron trigger efficiency determination.}
  \begin{tabular}{ll} \hline
    Tag selection & Probe selection \\ \hline
    $\pt >  30$ (35)\GeV in 2016 (2017--2018)& $\pt >  5\GeV$ \\
    $\abs{\eta} < 2.1$ (except $1.44 <\abs{\eta}< 1.57$) & $\abs{\eta} < 2.5$\\
    Tight isolation and shower shape requirements & No extra identification criteria\\
    \multicolumn{1}{l}{Passing the single-electron HLT path } \\
    \multicolumn{1}{l}{with $\pt > 27\,(32)\GeV$ in 2016 (2017--2018)} \\ \hline
  \end{tabular}
  \label{tb:tp_sel}
\end{table}

The tag-and-probe selections~\cite{Khachatryan:2010xn} used to measure the efficiencies are listed in Table~\ref{tb:tp_sel}. 
Probes are then required to pass the HLT path under study. The analyzed triggers are the following, being those used by most of the physics analyses involving electrons:
\begin{itemize}
	\item Single-electron trigger with tight identification and isolation requirements: electron $\pt > 27\,(32)\GeV$ in 2016 (2017--2018).
	\item Double-electron trigger with loose identification and
          isolation requirements: highest-(lowest-)\pt
          electron $\pt > 23\,(12)\GeV$.
\end{itemize}

\begin{figure}[hbtp]
  \centering
    \includegraphics[width=0.48\textwidth]{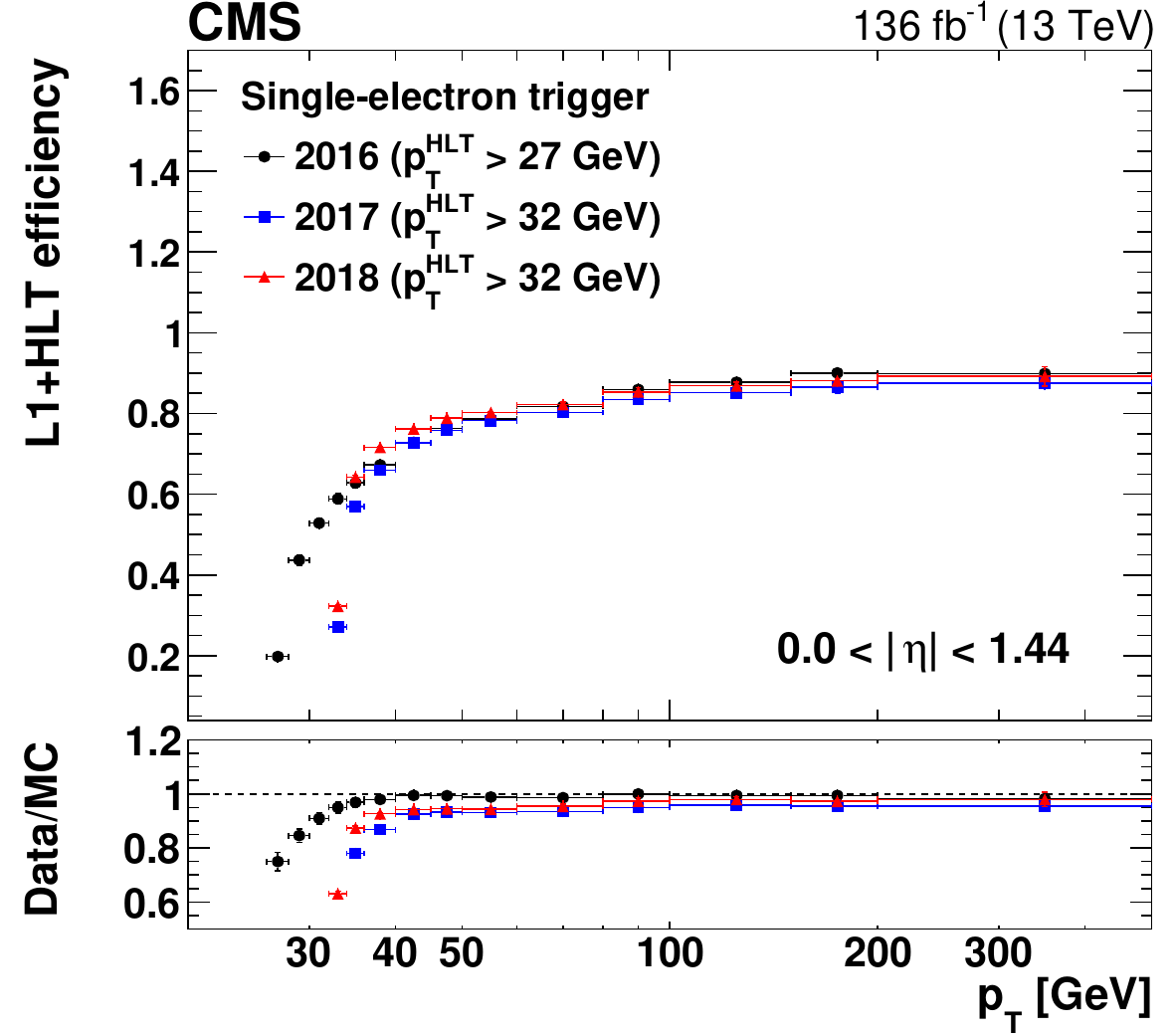}
    \includegraphics[width=0.48\textwidth]{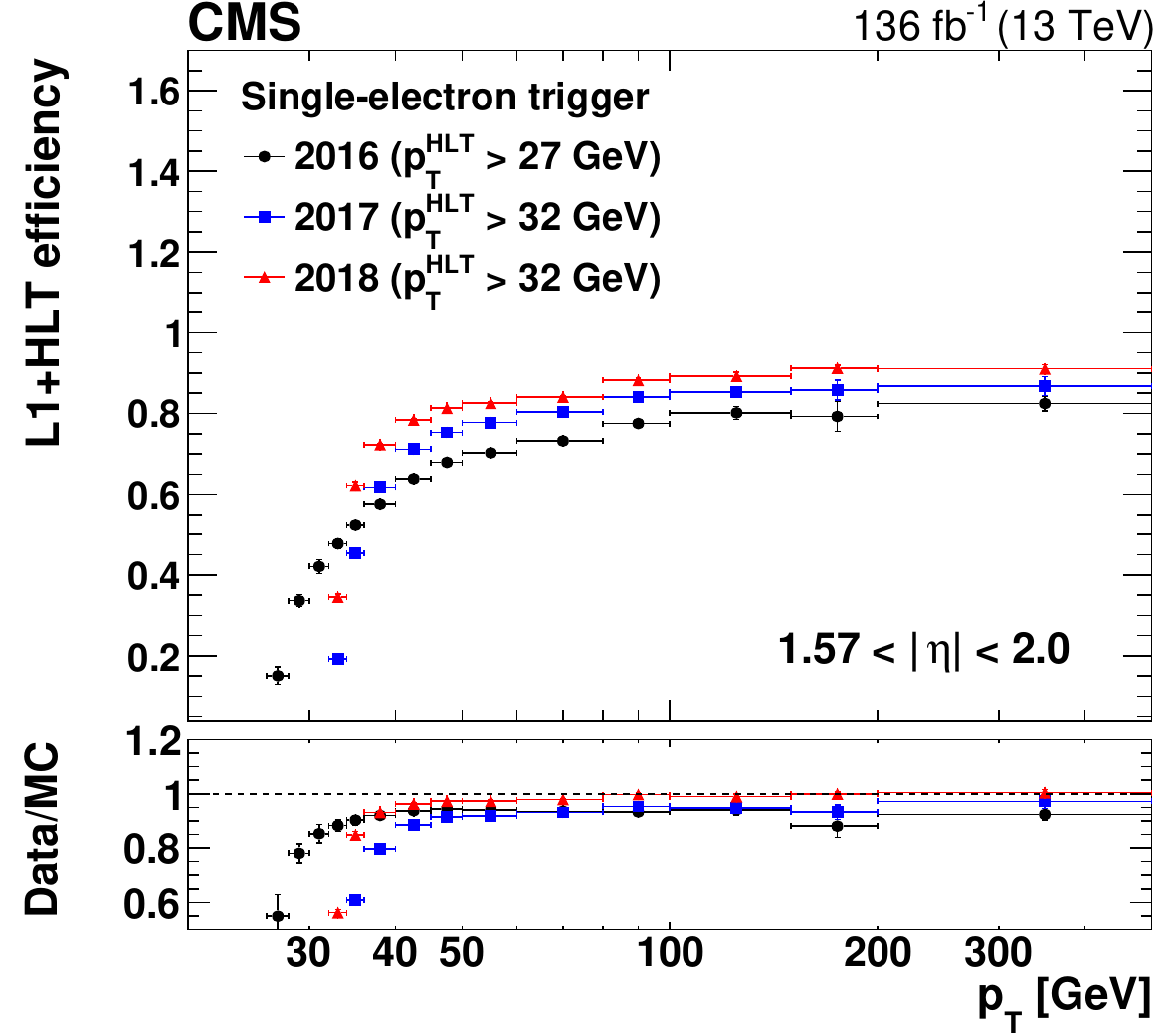}
    \includegraphics[width=0.48\textwidth]{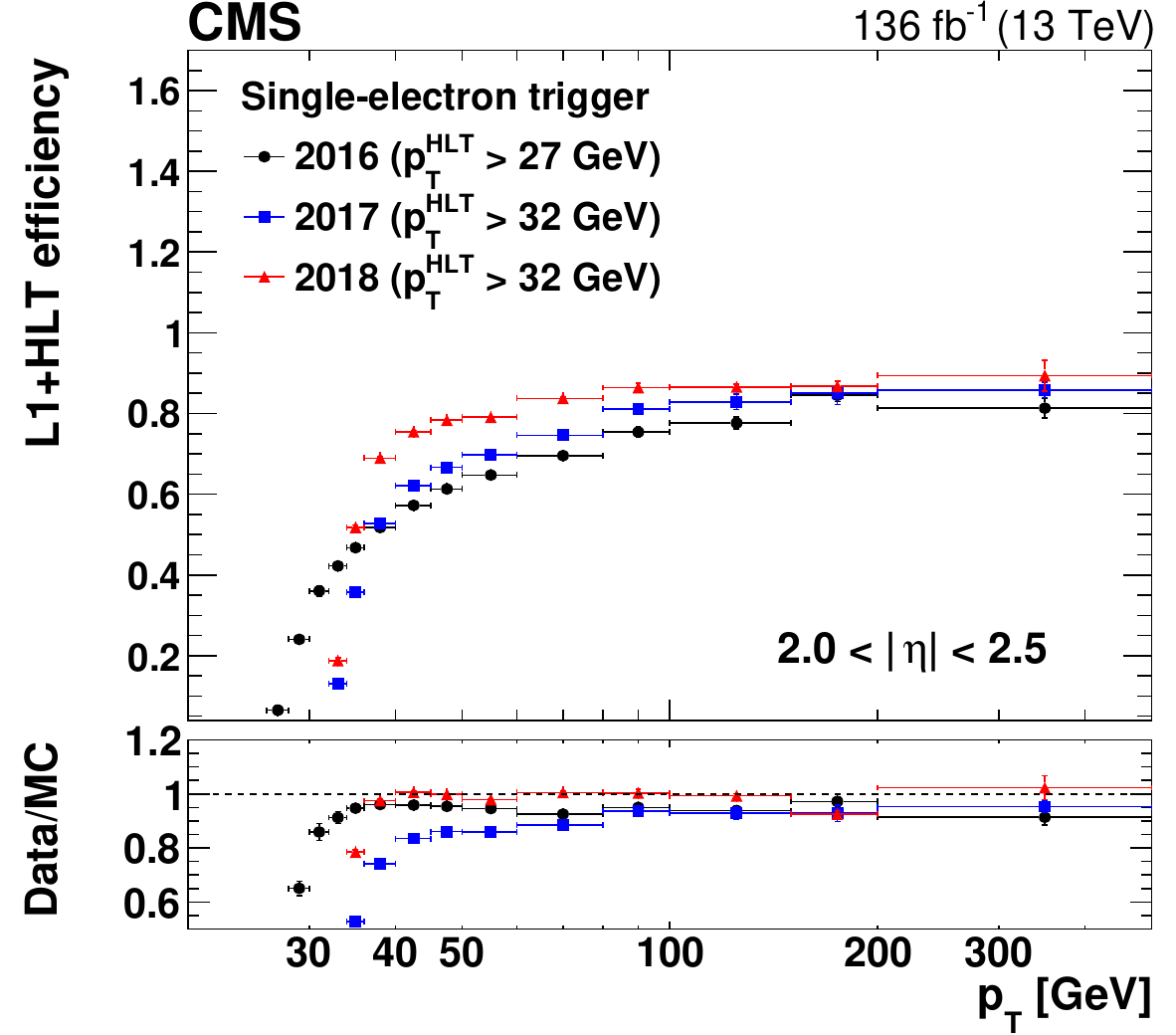}
  \caption{The L1+HLT efficiency of the single-electron HLT path
    with $\pt>27$ (32)\GeV in 2016 (2017 and 2018) with
    respect to an offline-reconstructed electron as a
    function of the electron \pt,
    obtained for
    $0 < \abs{\eta} <  1.44$ (upper left), 
    $1.57 < \abs{\eta} <  2.0$ (upper right),
    and $2.0 < \abs{\eta}  <  2.5$ (lower).
    The lower panel of each plot shows the ratio of data to MC simulation.
    The vertical bars on the markers represent combined statistical and systematic uncertainties.
  }
  \label{fig:SingleEle_pt}
\end{figure}

\begin{figure}[hbtp]
  \centering
    \includegraphics[width=0.48\textwidth]{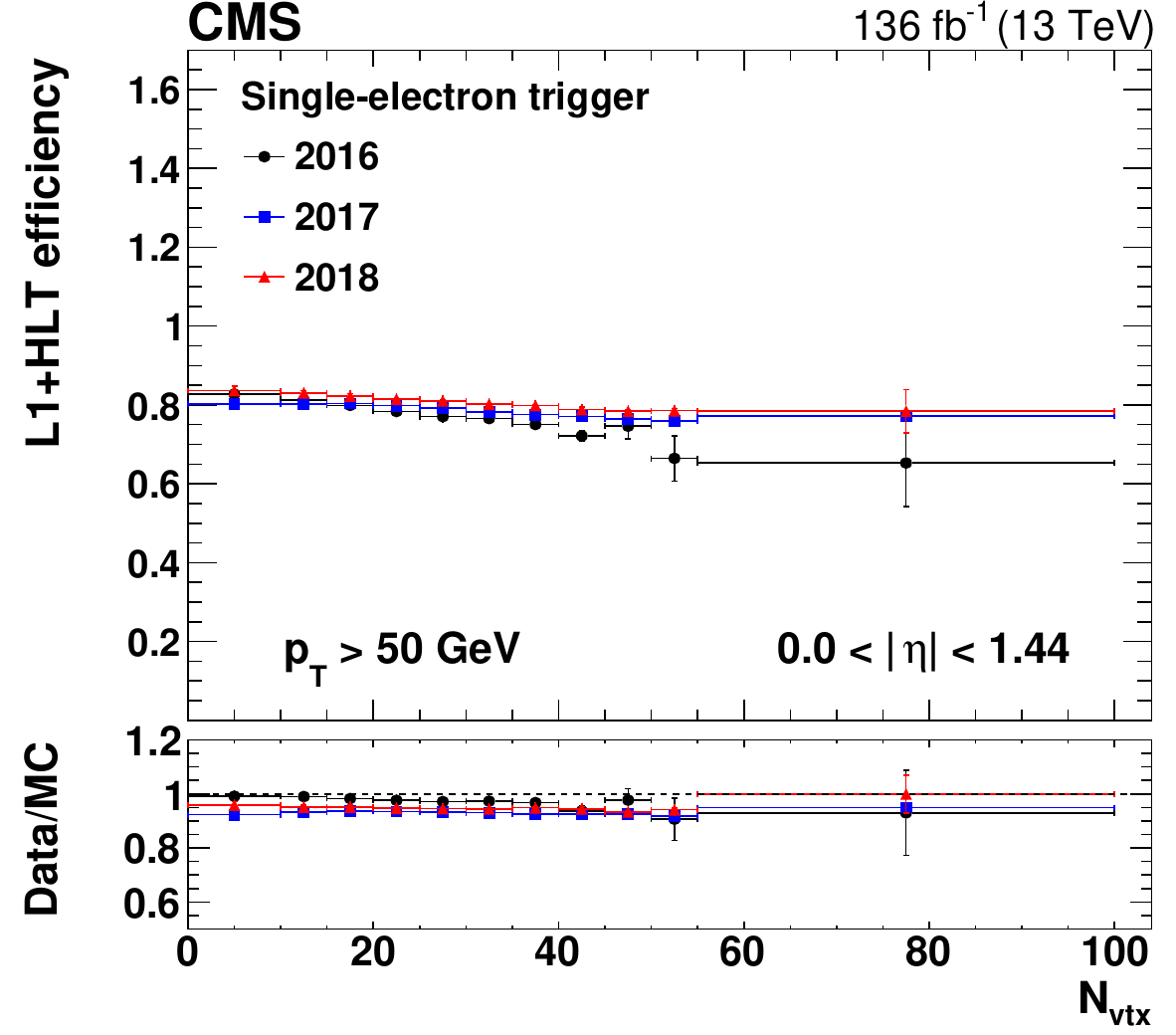}
    \includegraphics[width=0.48\textwidth]{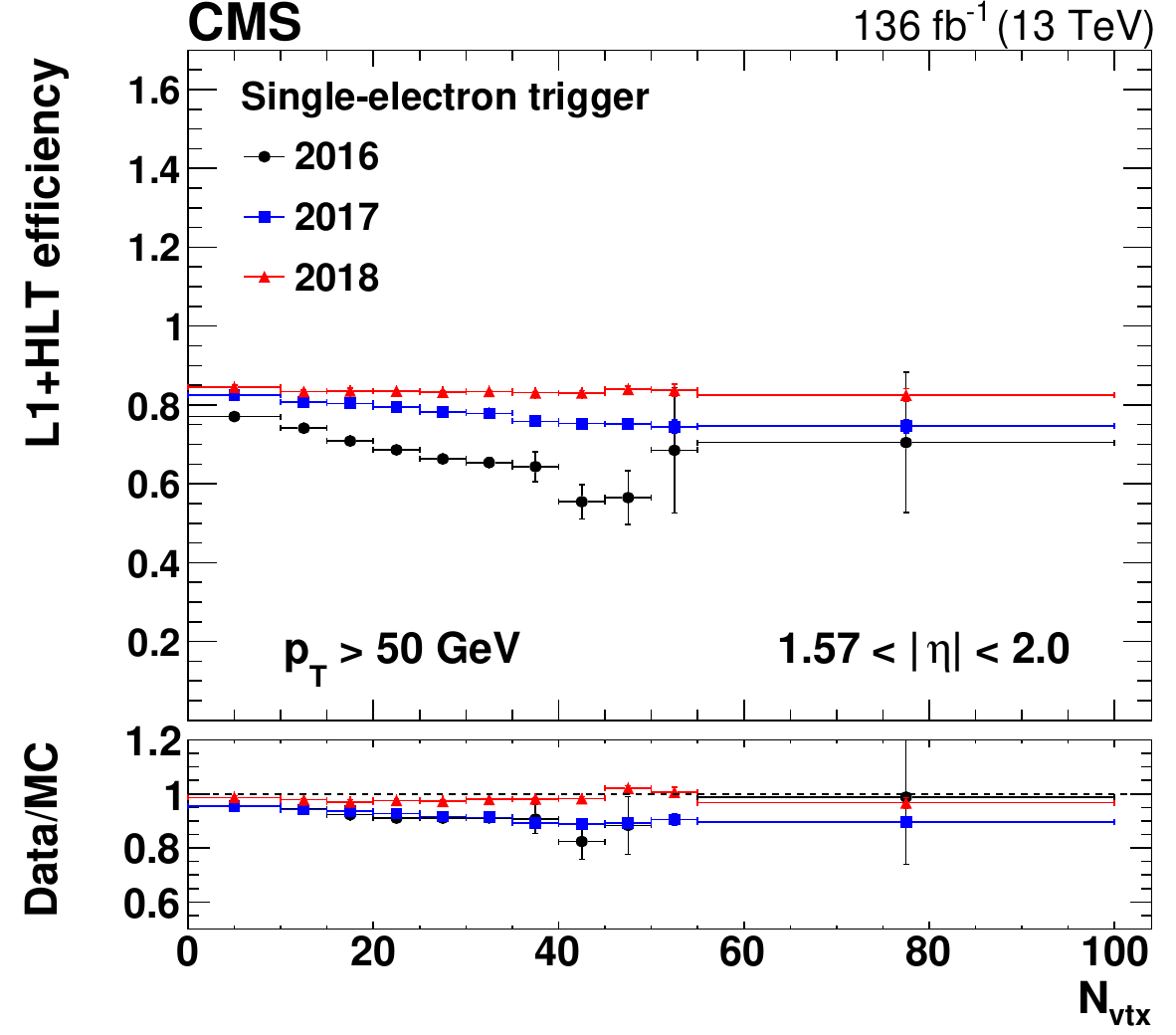}
    \includegraphics[width=0.48\textwidth]{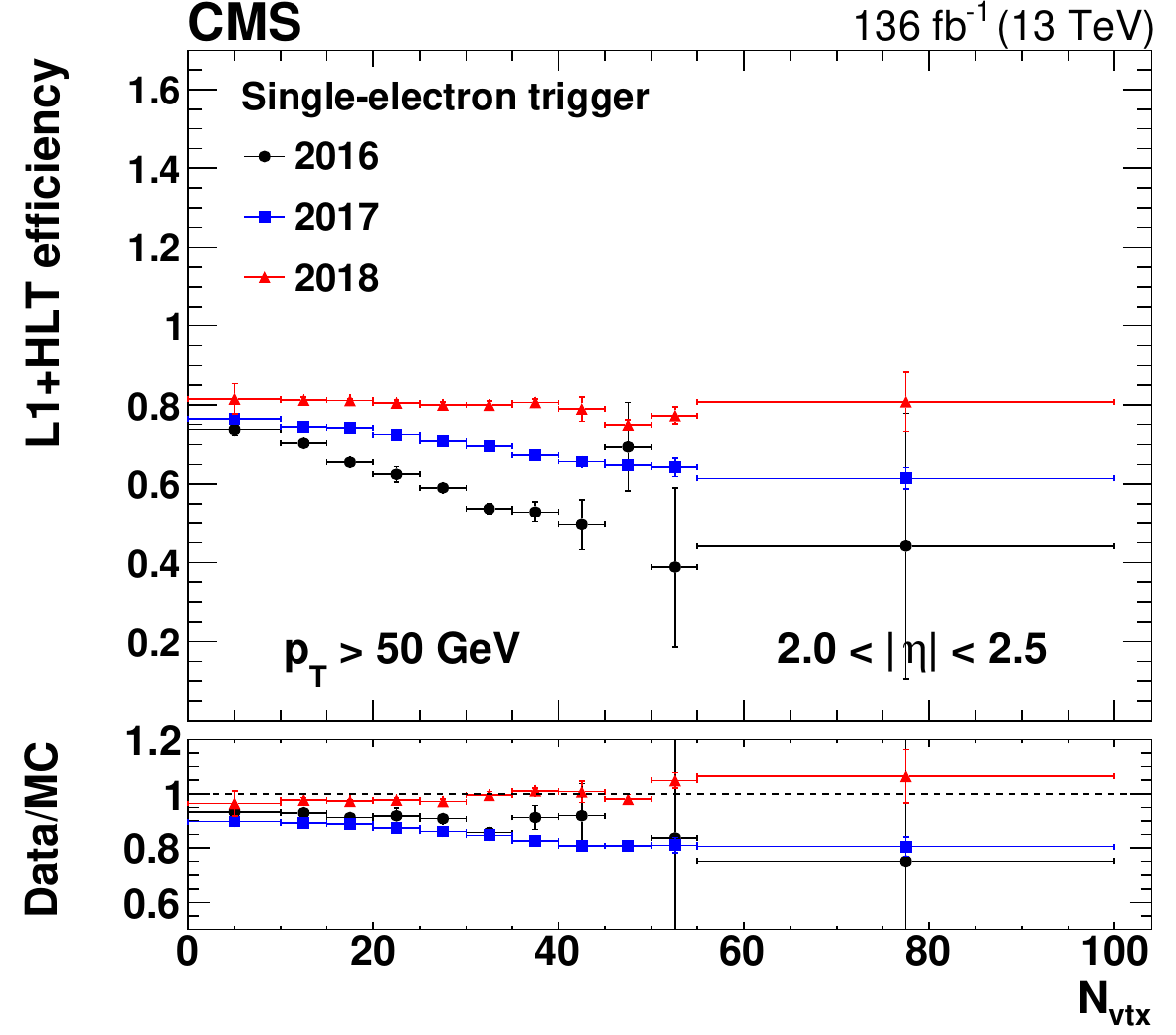}
  \caption{The L1+HLT efficiency of the single-electron HLT path
    with $\pt>27$ (32)\GeV in 2016 (2017 and 2018)
    with respect to an offline-reconstructed electron as
    a function of $N_{\text{vtx}}$, obtained for
    $0 < \abs{\eta} < 1.44$ (upper left),
    $1.57 < \abs{\eta} <  2.0$ (upper right),
    and $2.0 < \abs{\eta} <  2.5$ (lower).
    The electron \pt is required to be
    above 50\GeV.
    The lower panel of each plot shows the ratio of data to MC simulation.
The vertical bars on the markers represent combined statistical and systematic uncertainties.
  }
  \label{fig:SingleEle_npv}
\end{figure}

\begin{figure}[hbtp]
 \centering
   \includegraphics[width=0.48\textwidth]{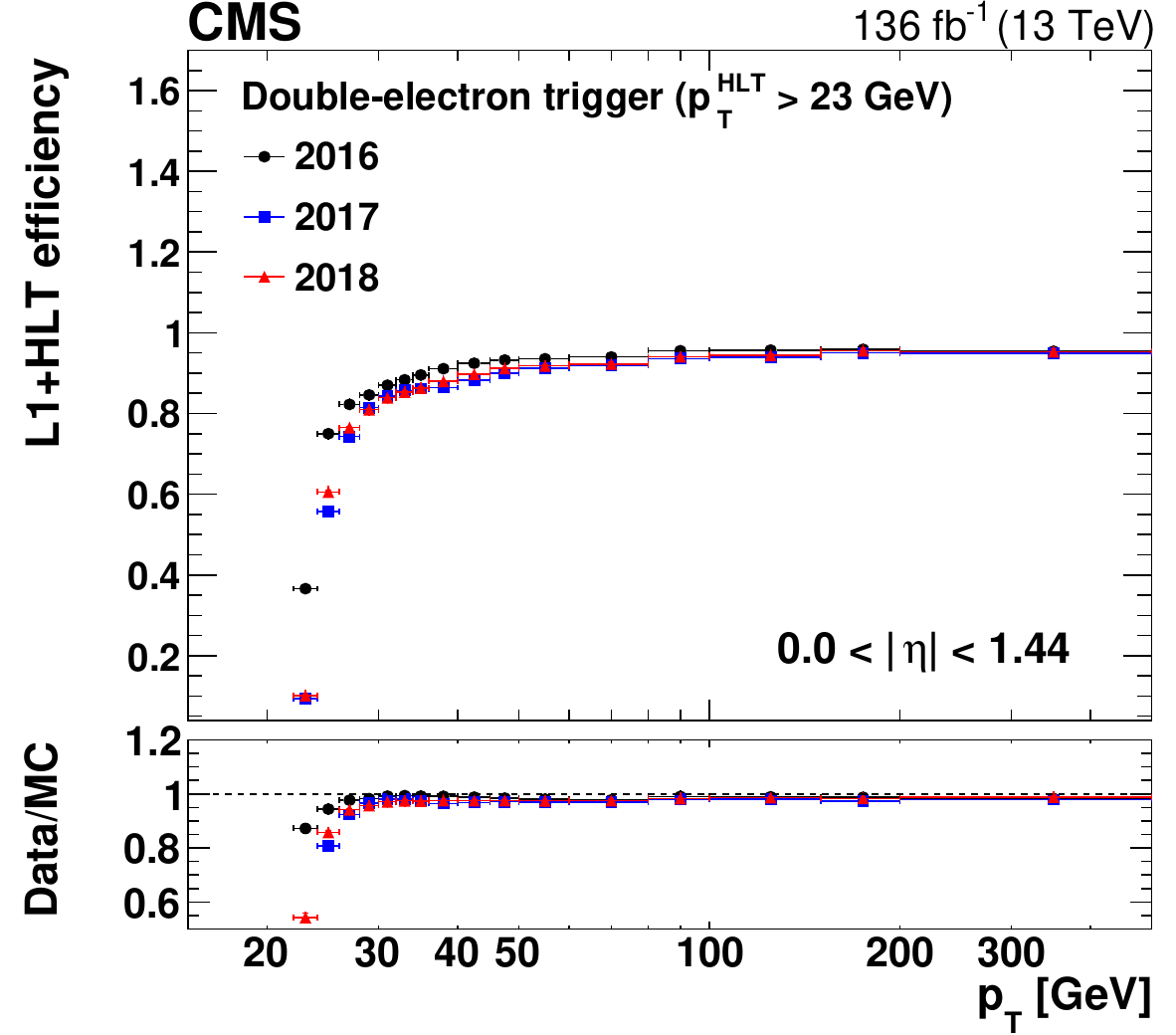}
   \includegraphics[width=0.48\textwidth]{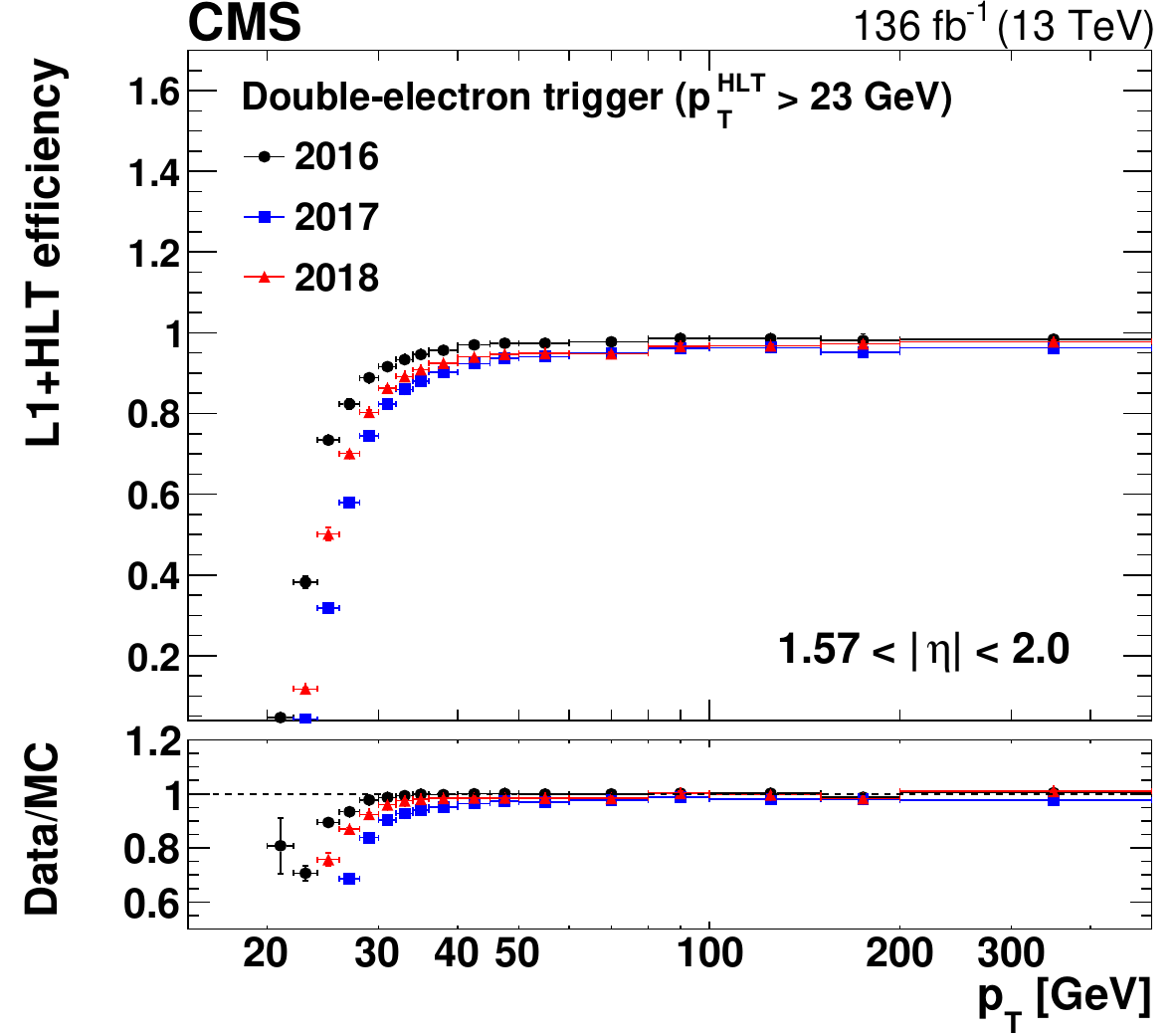}
   \includegraphics[width=0.48\textwidth]{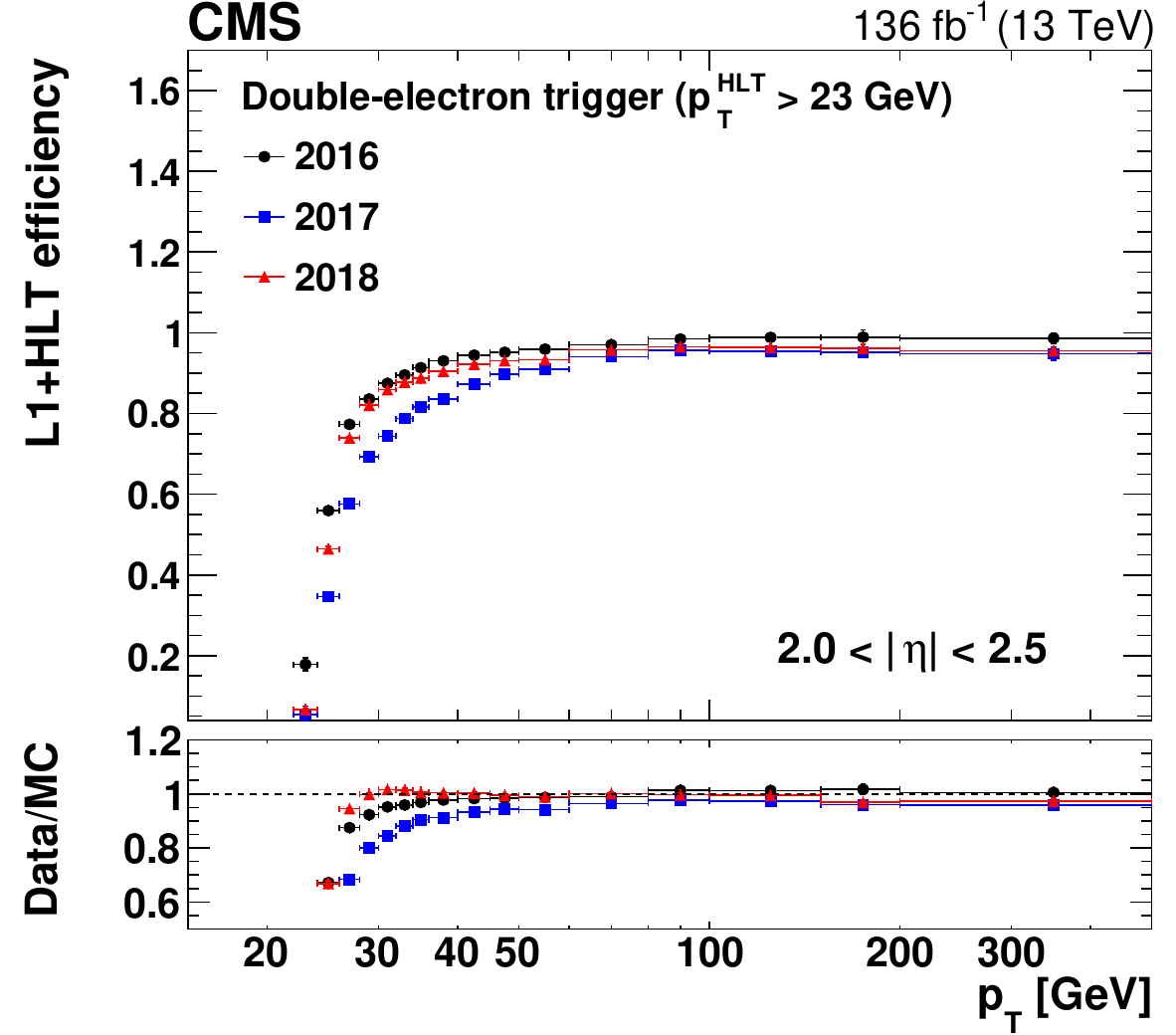}
 \caption{The L1+HLT efficiency of the $\pt>23\GeV$ leg of the
   double-electron trigger  with respect to an offline-reconstructed electron as a function of the electron
   \pt, obtained for
   $0 < \abs{\eta} <  1.44$ (upper left),
   $1.57 < \abs{\eta} <  2.0$ (upper right),
   and $2.0 < \abs{\eta} < 2.5$ (lower).
   The lower panel of each plot shows the ratio of data to MC simulation.
   The vertical bars on the markers represent combined statistical and systematic uncertainties.
 }
 \label{fig:Leg1_pt}
\end{figure}

\begin{figure}[hbtp]
  \centering
    \includegraphics[width=0.48\textwidth]{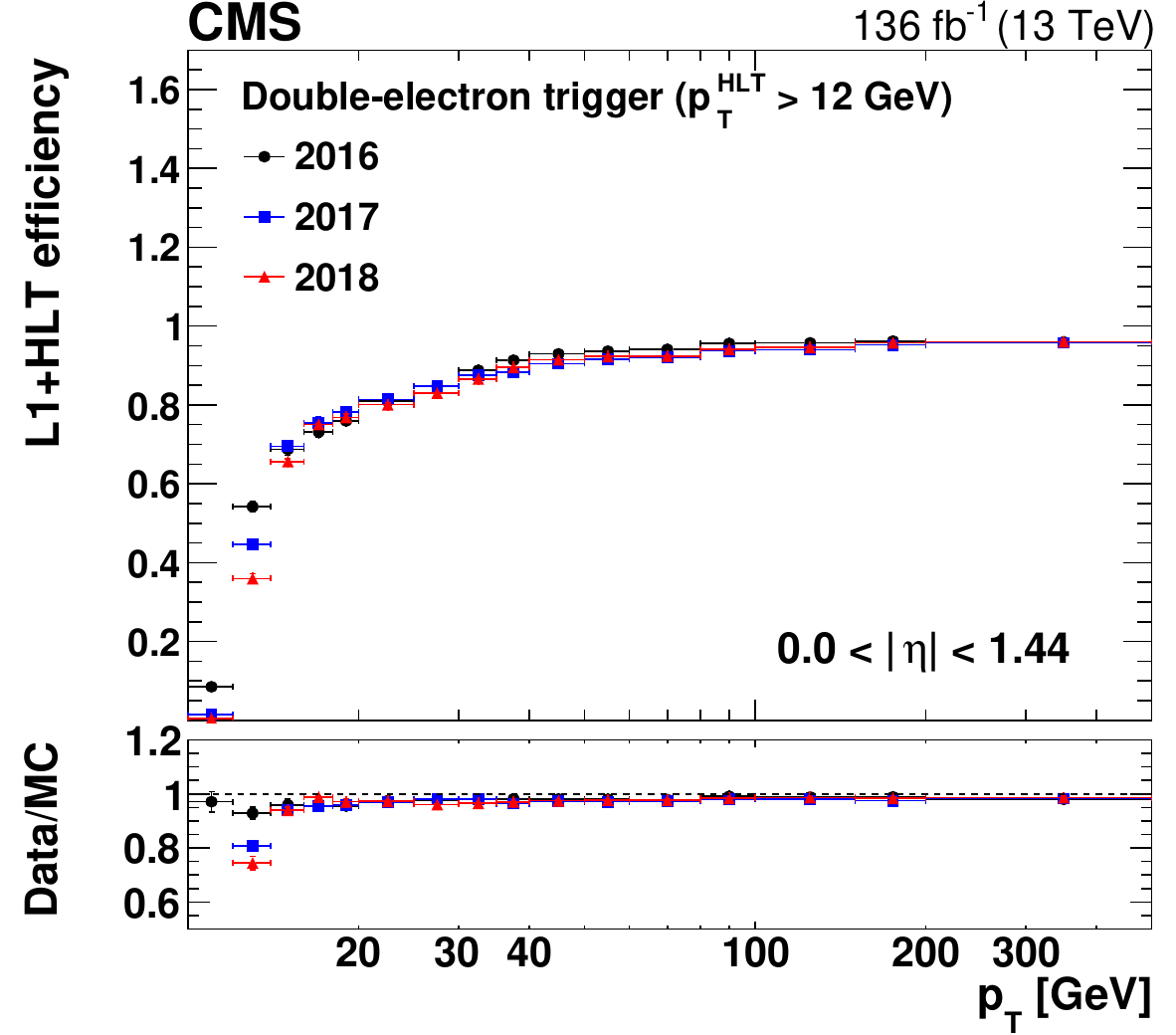}
    \includegraphics[width=0.48\textwidth]{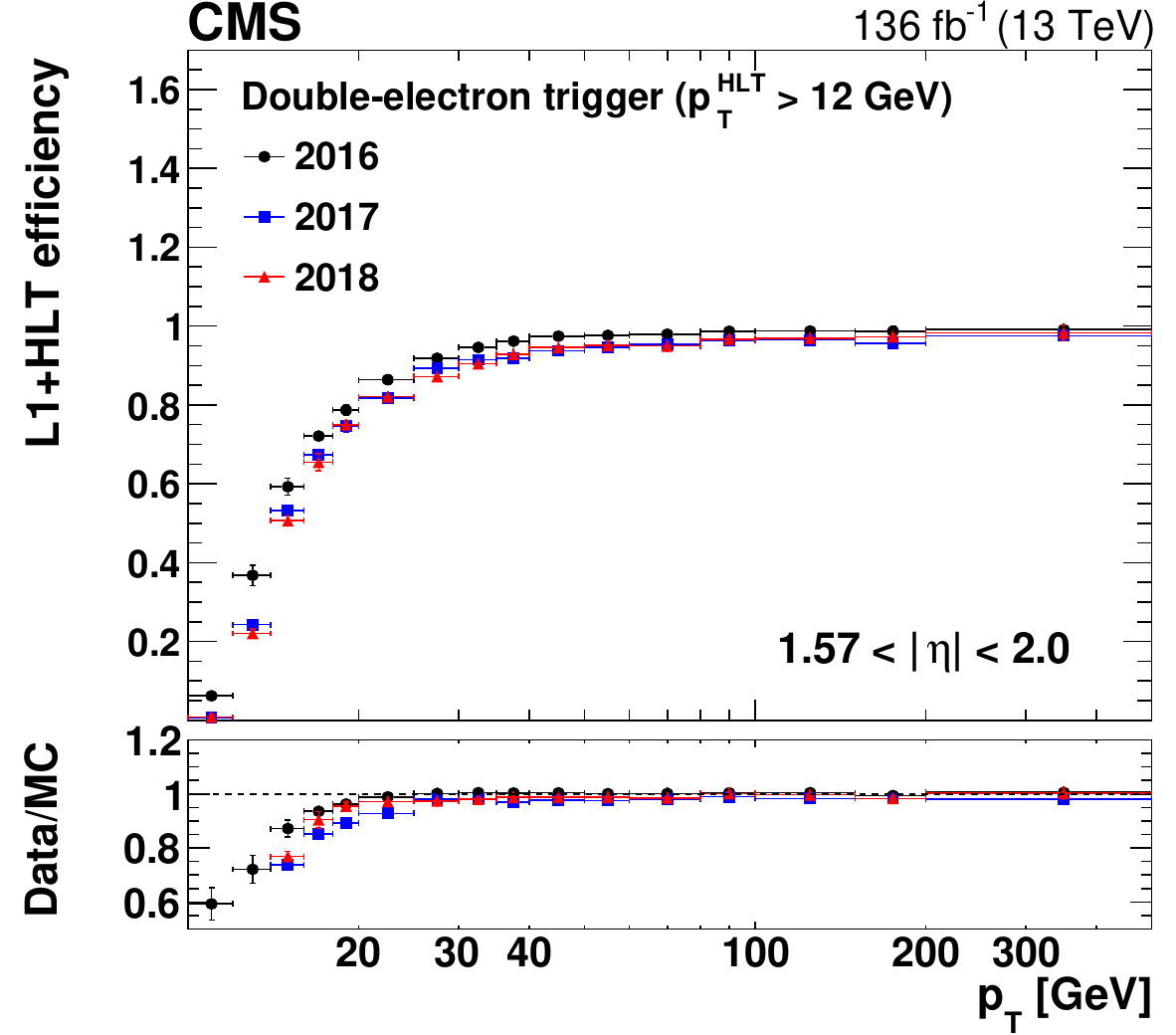}
    \includegraphics[width=0.48\textwidth]{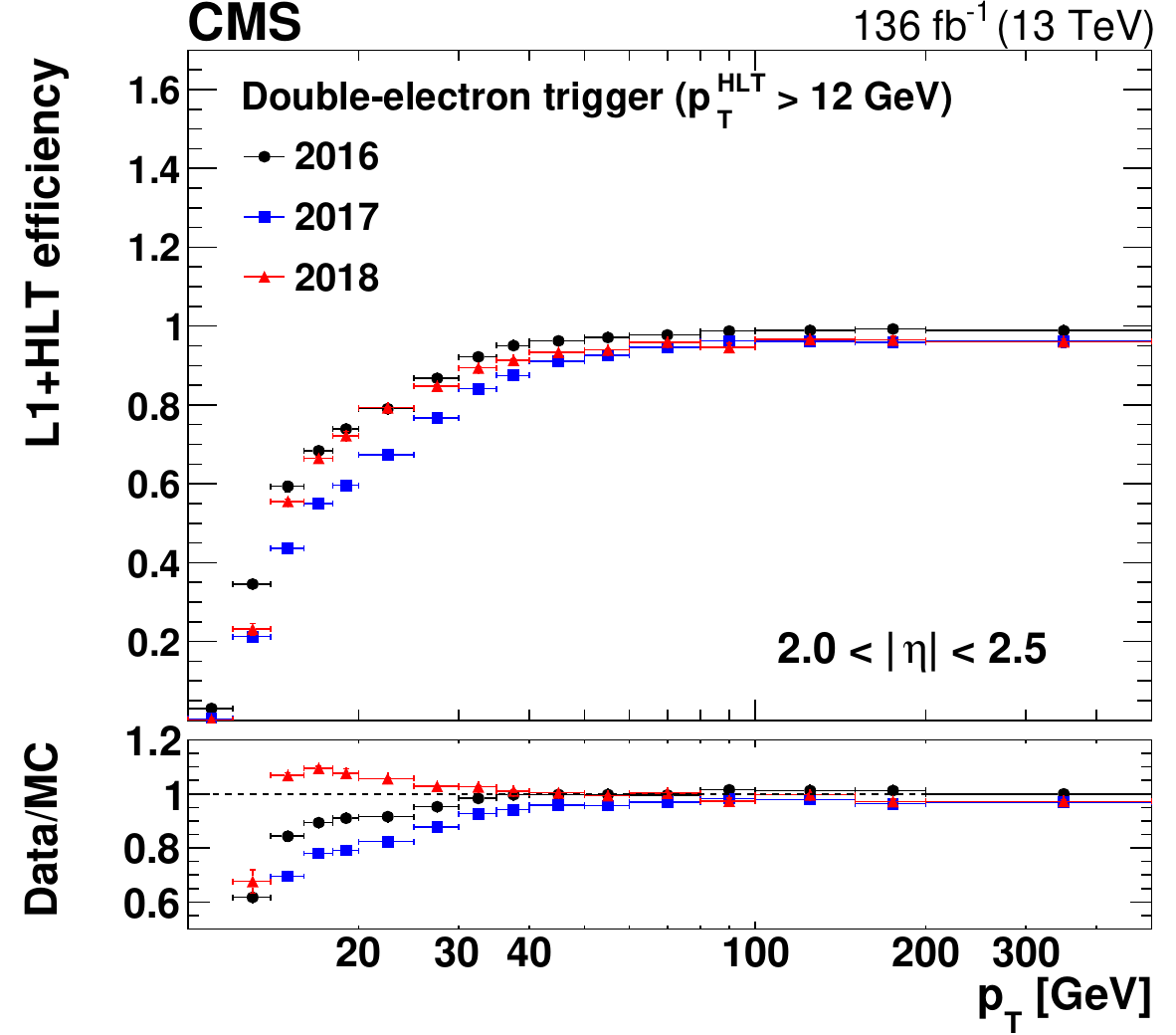}
  \caption{The L1+HLT efficiency of the $\pt>12\GeV$ leg of the
    double-electron trigger  with respect to an offline-reconstructed electron as a function of the electron
    \pt, obtained for
    $0 < \abs{\eta} <  1.44$ (upper left),
    $1.57 < \abs{\eta} <  2.0$ (upper right),
    and $2.0 < \abs{\eta} <  2.5$ (lower).
    The lower panel of each plot shows the ratio of data to MC simulation.
    The vertical bars on the markers represent combined statistical and systematic uncertainties.
  }
  \label{fig:Leg2_pt}
\end{figure}

\begin{figure}[hbtp]
  \centering
    \includegraphics[width=0.48\textwidth]{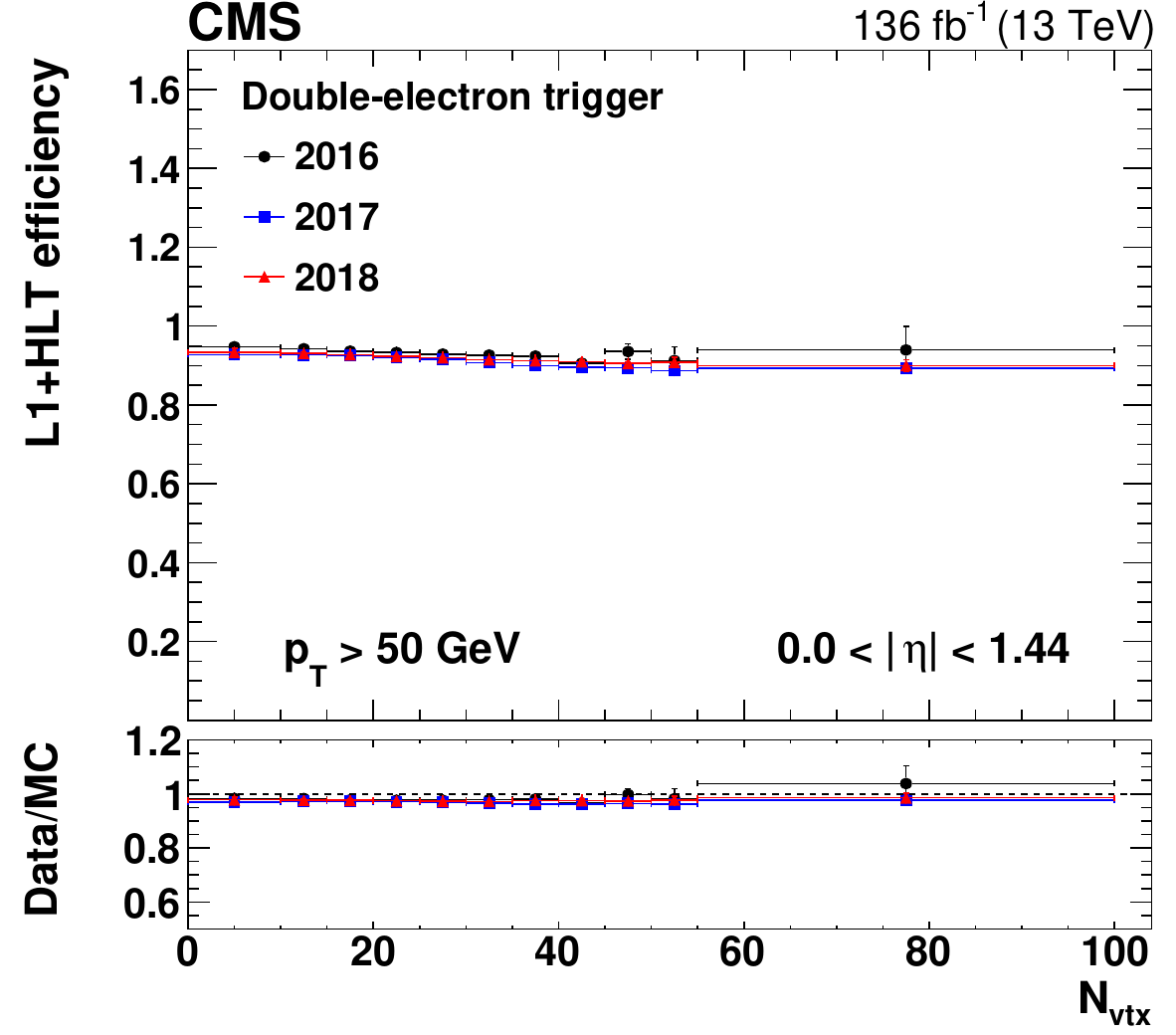}
    \includegraphics[width=0.48\textwidth]{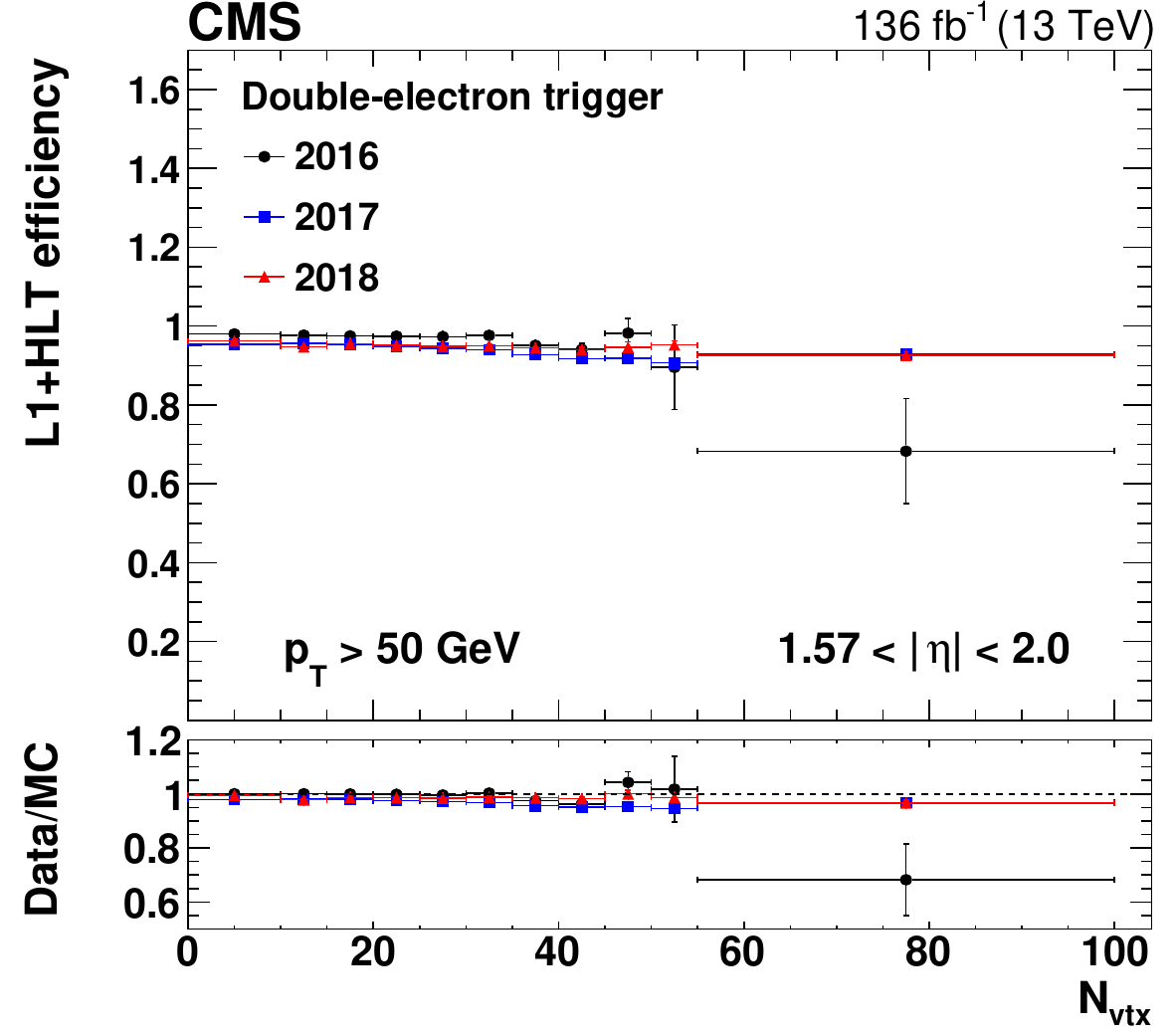}
    \includegraphics[width=0.48\textwidth]{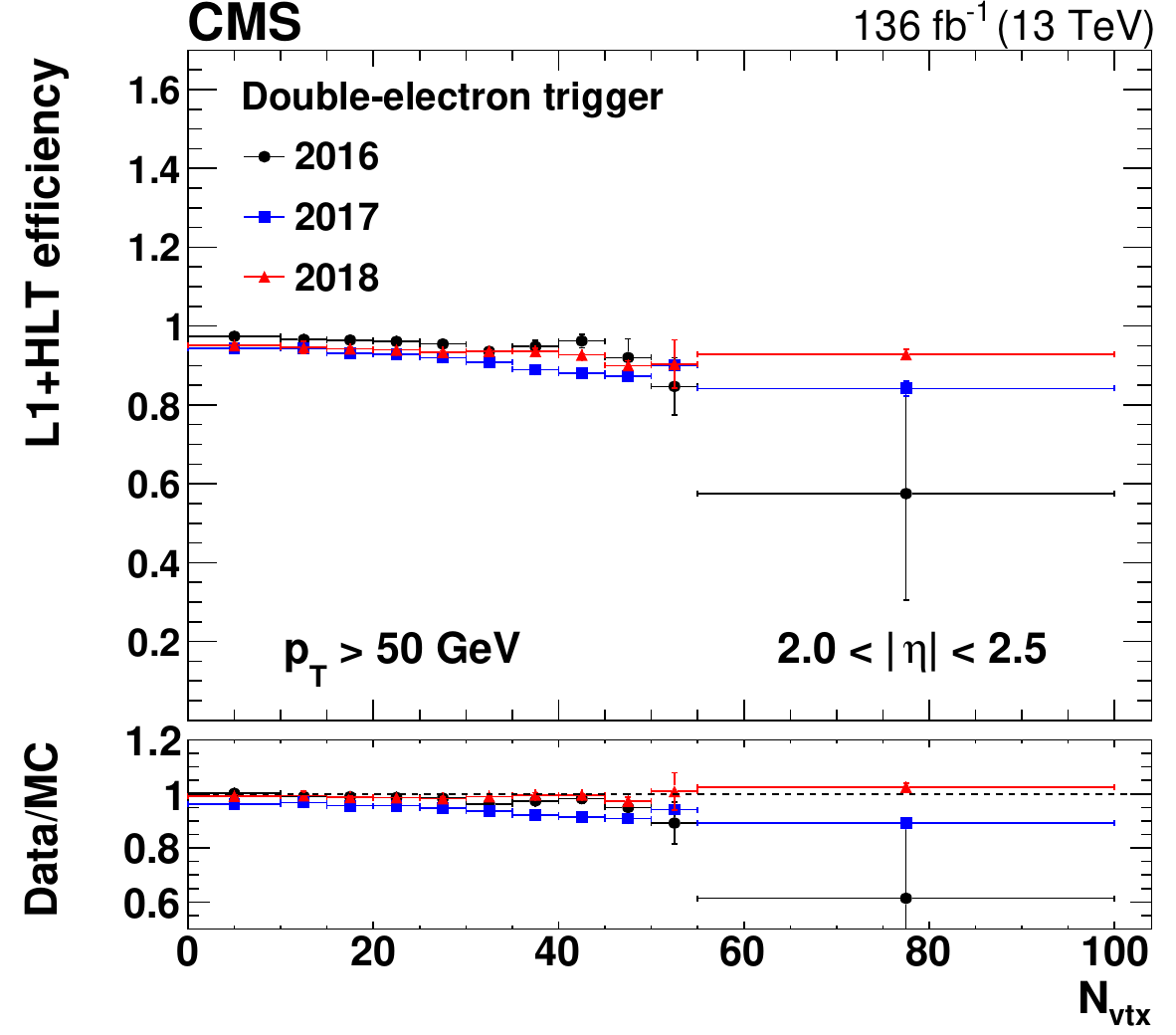}
  \caption{The L1+HLT efficiency of either leg of the double-electron
    trigger with respect to an offline-reconstructed electron as a function of $N_{\text{vtx}}$, obtained for 
    $0 < \abs{\eta} < 1.44$ (upper left),
    $1.57 < \abs{\eta} <  2.0$ (upper right), and
    $2.0 < \abs{\eta} <  2.5$ (lower).
    The electron \pt is required to be above 50\GeV.
    The lower panel of each plot shows the ratio of data to MC simulation.
    The vertical bars on the markers represent combined statistical and systematic uncertainties.
  }
  \label{fig:Leg1_npv}
\end{figure}

The identification requirements are based on the shower shapes in ECAL
and HCAL, and the isolation requirements on energy and momentum sums
in a cone around the electron.
Figures~\ref{fig:SingleEle_pt}--\ref{fig:Leg1_npv} show the
L1+HLT efficiency of these two electron triggers with respect to an offline-reconstructed electron as
a function of the electron \pt and $N_{\text{vtx}}$, for different $\eta$ regions of the
supercluster. The offline-reconstructed electron efficiency is typically above 95\% for electrons with $\pt>20\GeV$~\cite{CMS:2020uim}.
The lower panel of each plot shows the
ratio of the efficiency for data to MC simulation.
The data/MC discrepancy in the turn-on at low \pt,
seen for all years and $\eta$ values, mainly comes from the small differences
that exist between the online and offline ECAL response corrections~\cite{CMS:2020uim}.  
Small inefficiencies that arose during 2017 from L1 energy clusters
misassigned to the previous bunch crossing at high \pt~\cite{TRG-17-001}
primarily affect higher $\abs{\eta}$ than reported here.

The single-electron trigger performance reported in
Figs.~\ref{fig:SingleEle_pt} and~\ref{fig:SingleEle_npv}
is affected by a change in the strict identification and isolation
selections required in this HLT path, which together are known as the 
tight working point, whose target is a signal efficiency of about 80\%.
These criteria were retuned in 2017, and
some requirements in the endcap were loosened. Consequently, the
single-electron trigger efficiency is higher for 2017 and 2018 with
respect to 2016, in particular at high $\eta$ values.
The different shape as a function of \pt in 2016 with respect to 2017 and 2018
arises mainly from the different energy threshold, namely, 27 instead of
32\GeV. 
In 2017, the CMS pixel detector was upgraded by introducing extra
layers in the barrel and forward regions, and a commissioning period
at the beginning of the year led to a slightly reduced efficiency. As
a consequence of the upgraded detector, the algorithm used to reconstruct
electrons matching ECAL superclusters to pixel detector tracks was revised,
causing a significant rate reduction for a minimal performance
loss. However, problems with the pixel detector DC-DC converters
(discussed in Section~\ref{sec:tracking}) led to
a gradual efficiency reduction towards the end of the year.
Moreover, the majority of the high PU data in 2017 also came toward the
end of that year. Thus, 
for these reasons, the single-electron trigger performance in 2017 is slightly worse than in 2018.

The efficiency of the double-electron trigger, shown in
Figs.~\ref{fig:Leg1_pt}--\ref{fig:Leg1_npv}, is
in general higher in the turn-on region in 2016 compared with 2017 and 2018. This is because
the \ET thresholds of the lowest unprescaled L1 seed requiring two
electrons, which seeds this path, increased across the years. The
effect is especially evident at low \pt. Moreover, the
2017 trigger performance is slightly worse than the other years
because of the issues related to the pixel detector and PU described in the
previous paragraph. More details are reported in
Ref.~\cite{CMS:2020uim}.

\subsection{Jets}

Jets are reconstructed at the HLT using the anti-\kt clustering algorithm~\cite{antikt}  with a nominal distance parameter of 0.4, and 0.8 for wide jets used in Lorentz-boosted topologies and multijet triggers.
The inputs for the jet algorithm can be either calorimeter towers or
reconstructed objects from the PF algorithm.
Most HLT jet paths use the PF inputs (``PF-jets''), whereas
calorimeter jets (``Calo-jets'') are used as a first step to identify jet signatures and initiate the PF reconstruction.
To account for detector and collision conditions, several corrections are applied to the estimated PF hadron energies, average PU energy, and jet energy scale.
The performance of the jet triggers is measured in terms of their efficiency to select events that have an offline-reconstructed jet.
For this purpose, an unbiased set of $\Pp\Pp$ collision events collected with an isolated single-muon trigger with a $\pt > 27\GeV$ requirement is used.
The events are required to have exactly one loosely identified offline muon with $\pt > 10 \GeV$ within $\abs{\eta} < 2.4$, which has a relative isolation value less than 0.4 in a $\DR$ cone of radius 0.4 to match the trigger criteria.
Any events with additional loosely identified electrons having $\pt > 10 \GeV$ within $\abs{\eta} < 2.5$ are rejected to ensure that the chosen data have high purity of hadronic jets in the event.
The offline-reconstructed PF jets used in the measurement are clustered using the anti-\kt algorithm with radius 0.4; have $\pt > 18 \GeV$ within $\abs{\eta} < 2.4$; and pass selection criteria based on the charged-hadron fraction, number of constituents, \etc that are able to reject a good fraction of leptons misidentified as jets.
Events are selected requiring at least one such offline jet that is well separated from the offline muon by a $\DR$ of at least 0.4, so that the muon lies outside the reconstructed jet radius.

\begin{figure}[!htbp]
  \centering
  \includegraphics[width=0.45\textwidth]{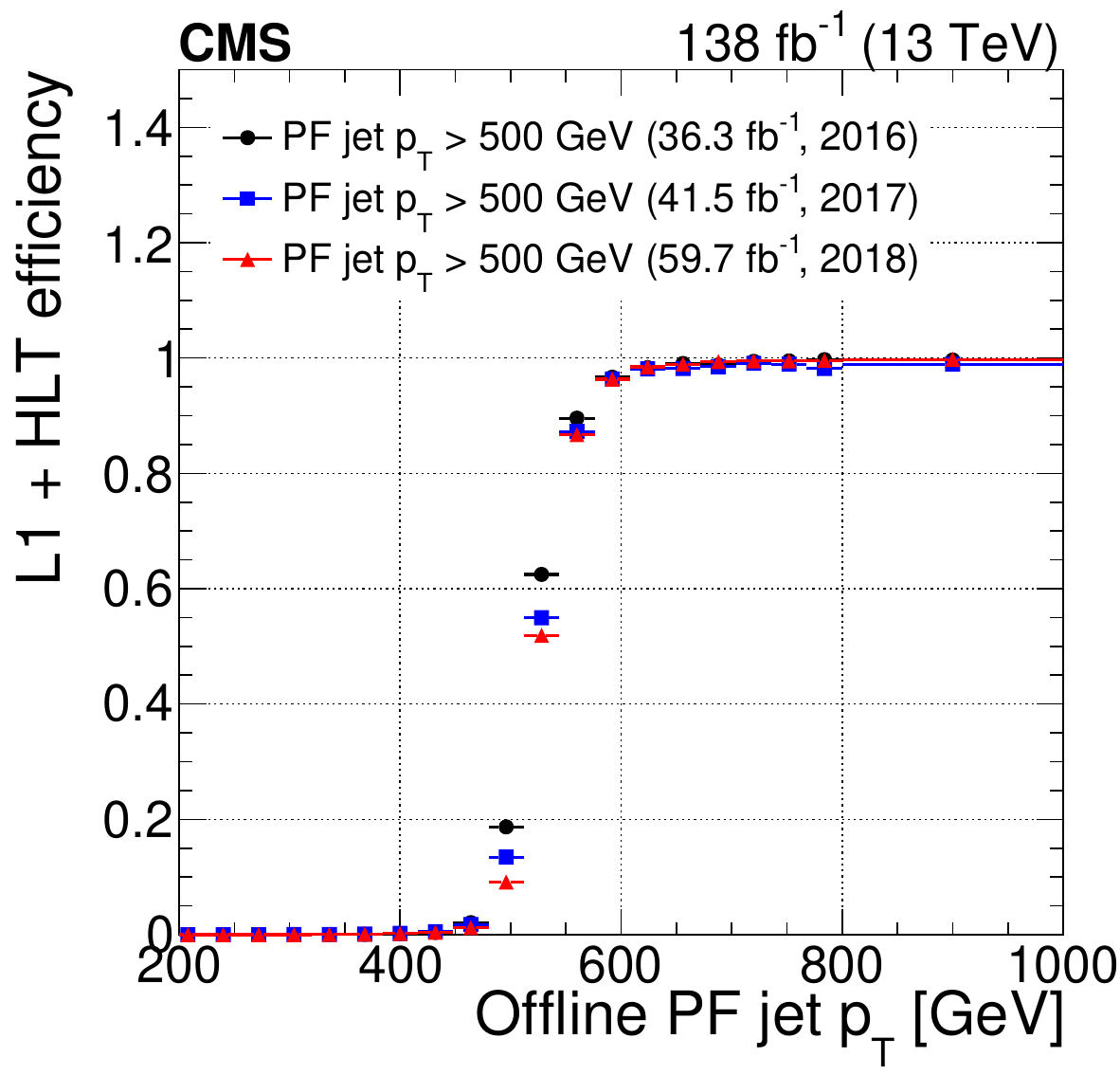}
  \caption{
    The L1+HLT efficiency of the unprescaled single PF jet trigger having an online \pt threshold of 500\GeV, measured with respect to the offline reconstructed PF jet \pt, for the data collected using an unbiased single-muon trigger during 2016, 2017, and 2018. The slight variation in turn-on curve is caused by differences in offline jet energy scale calibrations.
The vertical bars on the markers represent statistical uncertainties.
  }
  \label{fig:hlt_jet_performance}
\end{figure}

The efficiency is defined as the ratio of the number of events that have an HLT PF jet that passes the trigger threshold and matches the highest-\pt offline PF jet within  $\DR < 0.2$, to the total number of events with a reconstructed offline jet.
The efficiency for the lowest threshold unprescaled single PF jet trigger as a function of the offline PF jet \pt is shown separately for each data-taking year in Fig.~\ref{fig:hlt_jet_performance}.
Results are shown for the total integrated luminosities collected in each year during 2016, 2017, and 2018.
For a trigger threshold of 500\GeV, the efficiency reaches 100\% at
about 600\GeV in the offline reconstructed jet \pt for all three years. 
The jet trigger efficiency, measured as a function of the offline reconstructed jet \pt,  is affected by the calibration of the offline-reconstructed jets. The offline jet energy corrections were recalculated multiple times during \Runtwo, whereas only one set of online calibrations were used.
Hence, depending on the energy scale and resolution of the offline-reconstructed jets, the turn-on of the efficiency curve can be shifted and become slightly faster or slower.

\subsection{Scalar energy sums}

The global \HT energy sum is based on the scalar sum of jet
\pt, and is sensitive to multijet signatures.
The same set of unbiased events triggered by an isolated muon, described in the previous section, is used to measure the efficiency of \HT triggers.
The event preselection requirements based on leptons are also the same.
To suppress the effects from PU, the \HT reconstructed at the HLT is calculated using HLT PF jets having $\pt > 30 \GeV$ within $\abs{\eta} < 2.4$.
The same \pt and $\eta$ requirements are also applied to the offline-reconstructed PF jets to calculate the offline \HT, in addition to passing the jet identification criteria.
The offline jets are required to be separated from the offline muon by a $\DR$ of at least 0.4.

\begin{figure}[!htbp]
  \centering
  \includegraphics[width=0.45\textwidth]{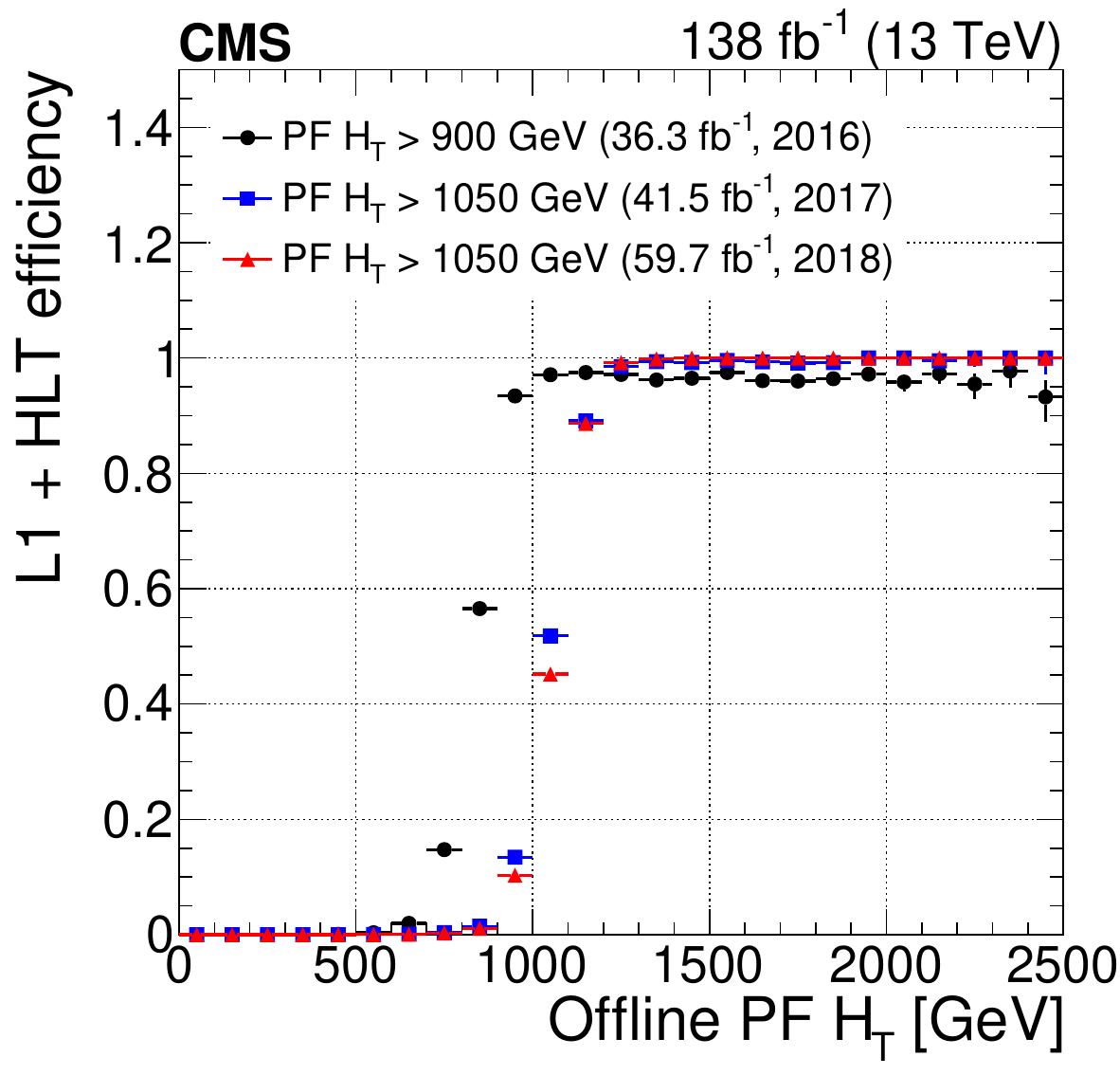}
  \caption{
    The L1+HLT efficiency of the unprescaled \HT triggers with the lowest thresholds, measured with respect to the offline-reconstructed \HT, for the data collected during 2016, 2017, and 2018. The inefficiency for 2016 is caused by an effect in the L1 trigger seed firmware implementation.
The vertical bars on the markers represent statistical uncertainties.
  }
  \label{fig:hlt_ht_performance}
\end{figure}

The trigger efficiency is defined as the ratio of the number of events where the \HT at the HLT passes the applied threshold to the total number of events selected by the offline \HT algorithm with the same threshold.
The performance of the unprescaled \HT triggers with the lowest thresholds, as a function of the offline \HT, is shown separately in Fig.~\ref{fig:hlt_ht_performance} for the total integrated luminosities collected during 2016, 2017, and 2018.
During 2016, a lower threshold of 900\GeV was applied online and
was increased in subsequent years to maintain a similar
total trigger rate as the \Linst increased.
This is because \HT is highly sensitive to PU events, causing a nonlinear increase in the trigger rate.

For an online threshold of 1050\GeV, the efficiency reaches 100\%
at about 1300\GeV in the offline-calculated \HT.
The \HT trigger efficiency for 2016 was lower because of an effect in the L1 trigger seed firmware implementation that limited the plateau to ${<}100\%$.

\subsection{Missing transverse momentum}

At the HLT, the missing transverse momentum is defined as the negative
vector sum of the \pt of all the PF candidates in an
event, and its magnitude is denoted as \ptmiss.
It is crucial to account for the instrumental effects of noise and beam-induced
backgrounds to keep the rates of these
triggers within reasonable limits. Additional filtering algorithms are applied
during reconstruction to achieve lower rates for \ptmiss
triggers. Calorimeter deposits consistent with noise signature or beam
halo are removed from the energy sum computation at the HLT.

The performance of the  \ptmiss triggers is measured with respect to the offline-reconstructed \ptmiss also based on PF candidates and including jet energy corrections, referred to as corrected \ptmiss.
An unbiased sample of events, triggered by a single isolated electron of $\pt > 32 \GeV$, is used for this measurement.
To match the online requirement, the events must contain exactly one well-identified and isolated offline electron, having $\pt > 35 \GeV$ within $\abs{\eta} < 2.5$ and passing the electron identification criteria based on track quality and electromagnetic shower shape variables.
Events with any additional loosely identified electrons or with muons with $\pt > 10 \GeV$ are excluded.

\begin{figure}[!htbp]
  \centering
  \includegraphics[width=0.45\textwidth]{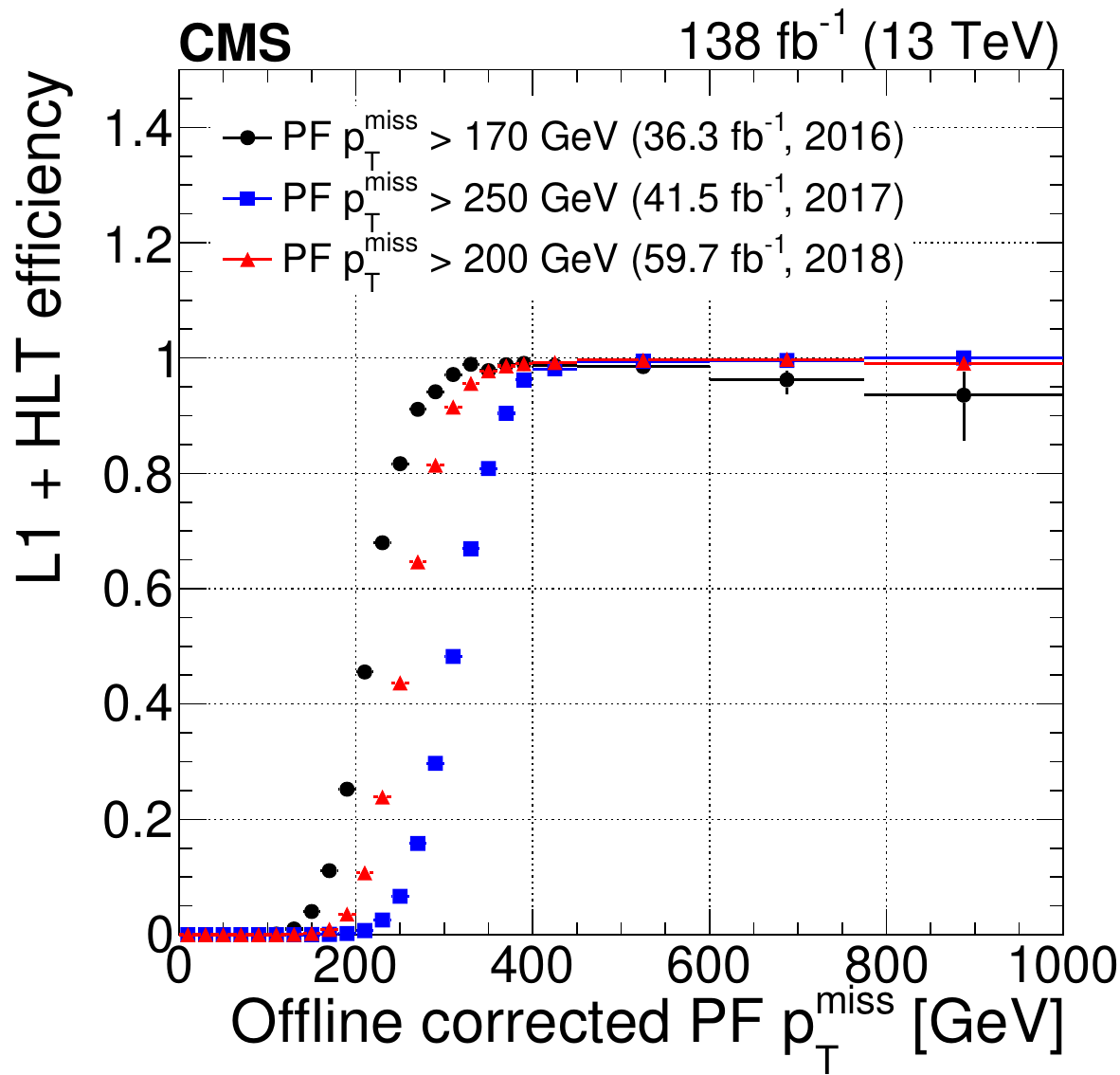}
  \caption{
    The L1+HLT efficiencies of the unprescaled  \ptmiss triggers with the lowest thresholds, measured with respect to the offline-reconstructed corrected  \ptmiss, for the data collected during 2016, 2017, and 2018.
The vertical bars on the markers represent statistical uncertainties.
  }
  \label{fig:hlt_met_performance}
\end{figure}

The trigger efficiency is defined as the ratio of the number of events that satisfy a given online \ptmiss threshold requirement to the total number of events selected by the offline \ptmiss algorithm with the same threshold.
The performance of the unprescaled triggers with the lowest thresholds  using the total integrated luminosities collected during 2016, 2017, and 2018, is shown in Fig.~\ref{fig:hlt_met_performance}.
Since \ptmiss triggers are also susceptible to PU effects similar to \HT triggers, the thresholds are different in the three years because of variations in the \Linst.
For a trigger of threshold of 170\GeV, the efficiency reaches 100\%
for an offline \ptmiss of about 350\GeV in 2016, with a
similar performance seen in later years for slightly shifted thresholds.

\begin{figure}[!htbp]
  \centering
  \includegraphics[width=0.45\textwidth]{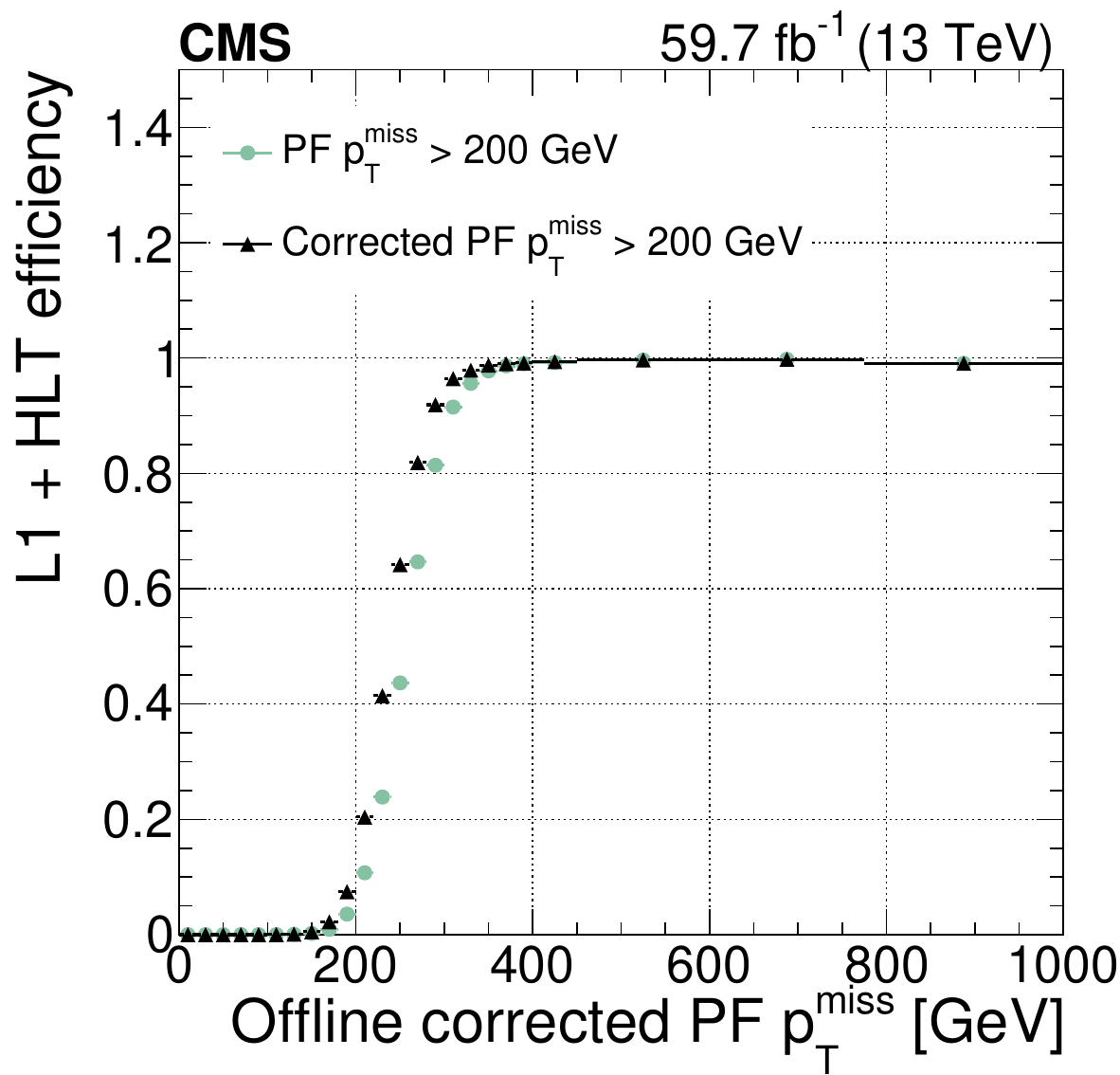}
  \includegraphics[width=0.45\textwidth]{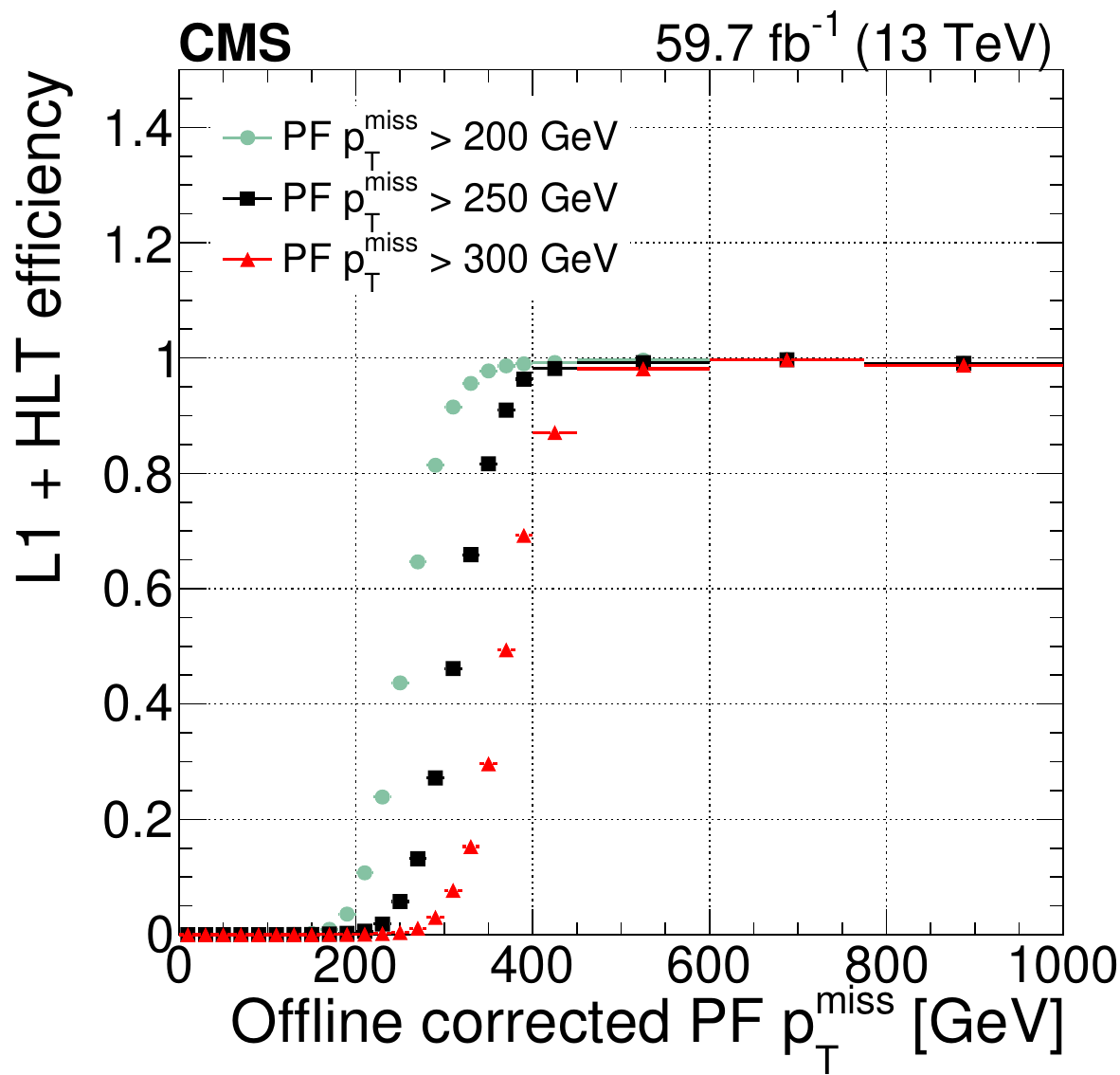}
  \caption{
    Left: comparison of the L1+HLT efficiencies of the corrected
    vs. nominal  \ptmiss triggers of the same
    threshold, for the data collected during 2018. Right:
    nominal \ptmiss trigger L1+HLT efficiencies using
    different thresholds.
  }
  \label{fig:hlt_met_comparison}
\end{figure}

At the HLT, online jet energy corrections can also be propagated to the calculation of the \ptmiss similar to that performed offline.
Figure~\ref{fig:hlt_met_comparison} (left) compares the performance of the nominal and corrected  \ptmiss at the HLT for the 2018 data-taking year.
The trigger with corrected  \ptmiss has a slightly faster turn-on compared with that of the nominal \ptmiss trigger having the same threshold.
However, the rate of the corrected \ptmiss trigger is also observed to increase by about 20\% compared with that of the nominal \ptmiss trigger.
Figure~\ref{fig:hlt_met_comparison} (right) shows the performance of different unprescaled thresholds on online \ptmiss against offline-corrected \ptmiss, again using 2018 data.
The turn-on curves are observed to behave consistently with increasing thresholds.

\begin{figure}[!htbp]
  \centering
  \includegraphics[width=0.45\textwidth]{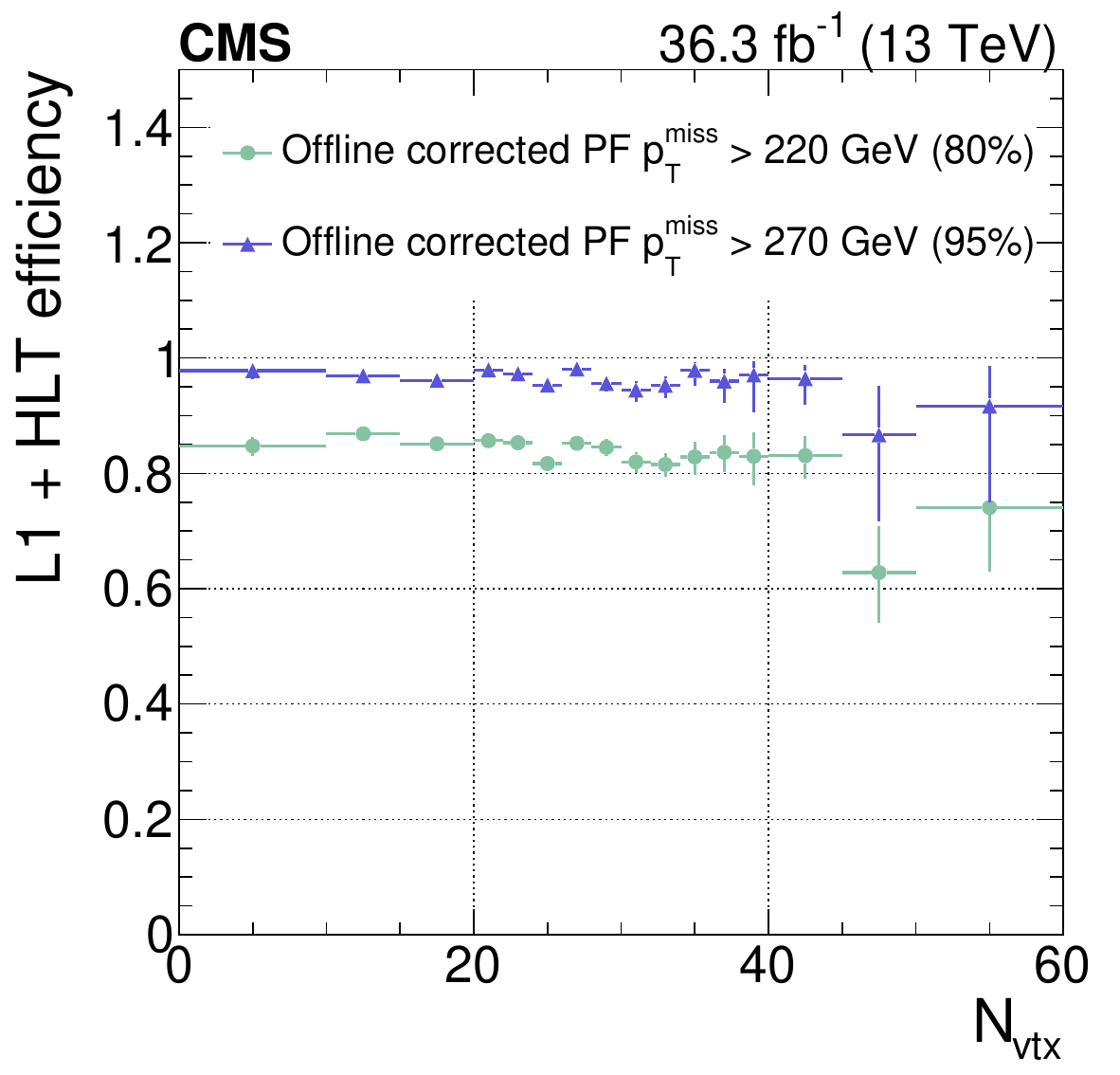} \\
  \includegraphics[width=0.45\textwidth]{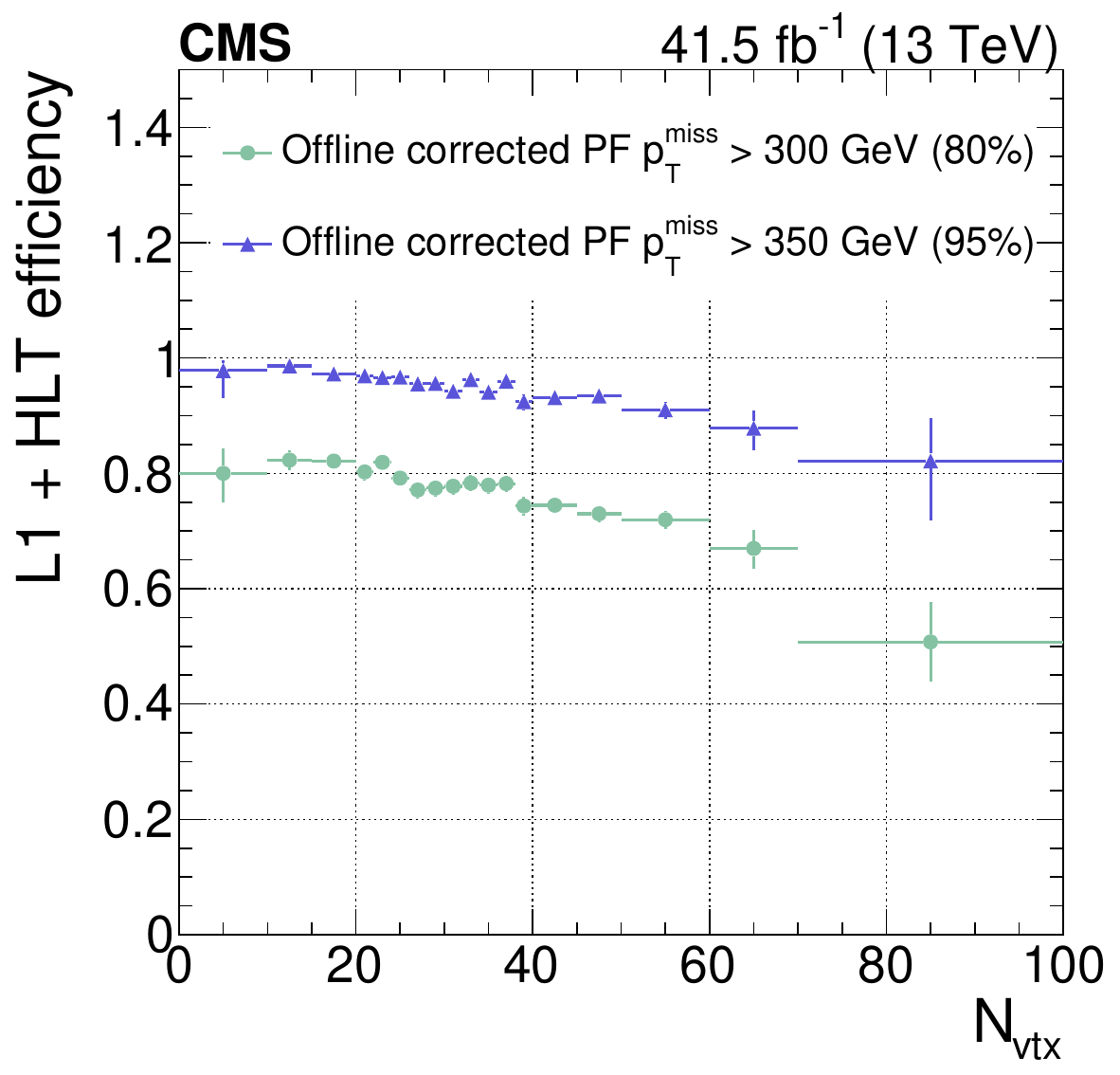}
  \includegraphics[width=0.45\textwidth]{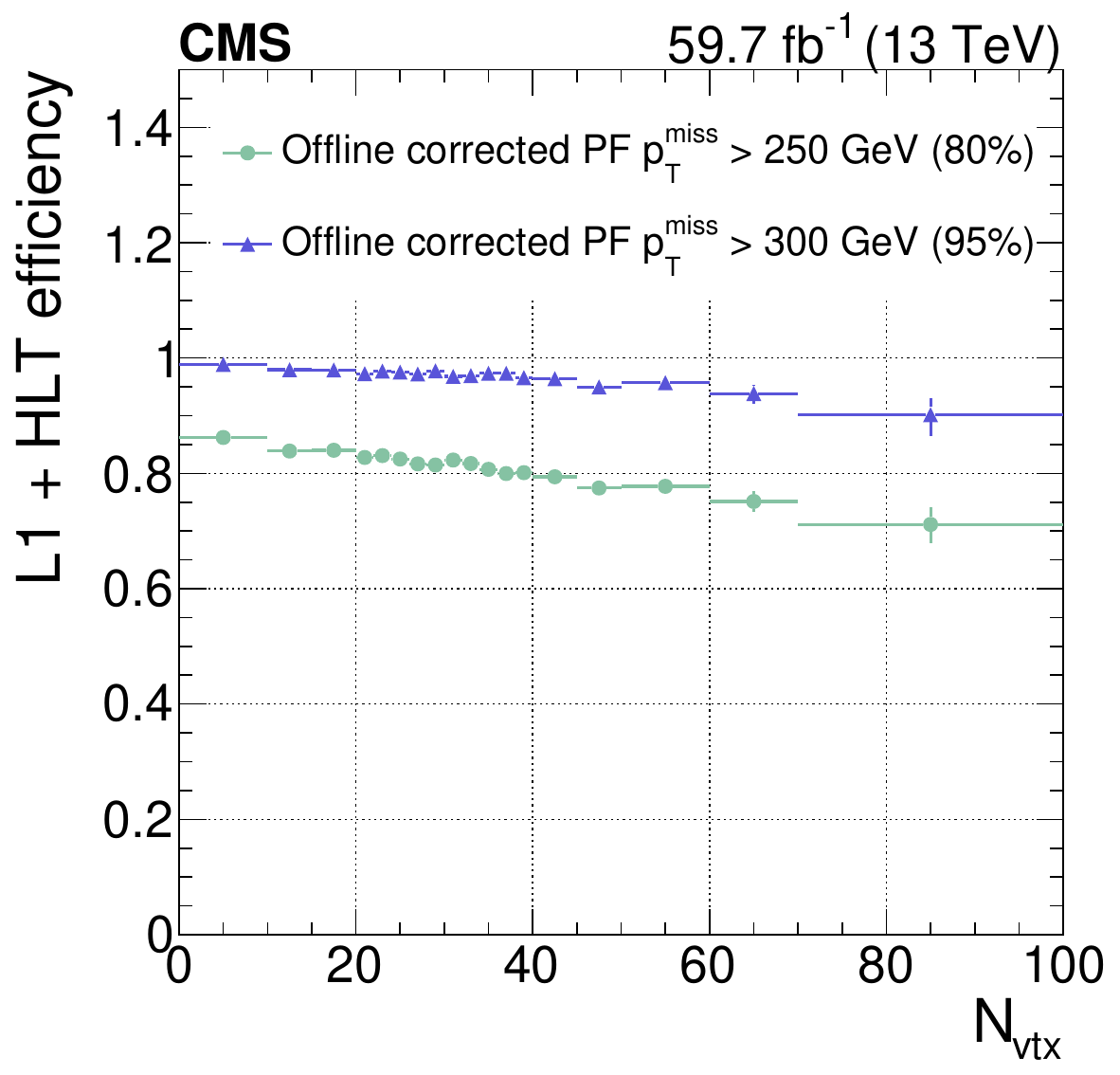}
  \caption{
    Event selection efficiencies as a function of $N_{\text{vtx}}$ for fixed L1+HLT efficiency values of 80 and 95\% of the unprescaled \ptmiss triggers with the lowest \ptmiss thresholds in 2016 (upper), 2017 (lower left), and 2018 (lower right).
    The PU is considerably lower in 2016,
    which allowed to lower thresholds for all triggers
    involving jets and \ptmiss.
The vertical bars on the markers represent statistical uncertainties.
  }
  \label{fig:npvRun2}
\end{figure}

The event selection efficiencies as a function of $N_{\text{vtx}}$ are shown in
Fig.~\ref{fig:npvRun2} for different data-taking years.
The offline thresholds are chosen at the fixed L1+HLT efficiency values of 80 and 95\% for each year, as determined from Fig.~\ref{fig:hlt_met_performance}.
The thresholds applied online are 170, 250, and 220\GeV in 2016, 2017, and 2018, respectively.
The efficiencies decrease in events with a larger $N_{\text{vtx}}$, which is expected since particle tracks from only a limited number of vertices are reconstructed at the HLT and \ptmiss is underestimated.

\begin{figure}[!htbp]
  \centering
  \includegraphics[width=0.45\textwidth]{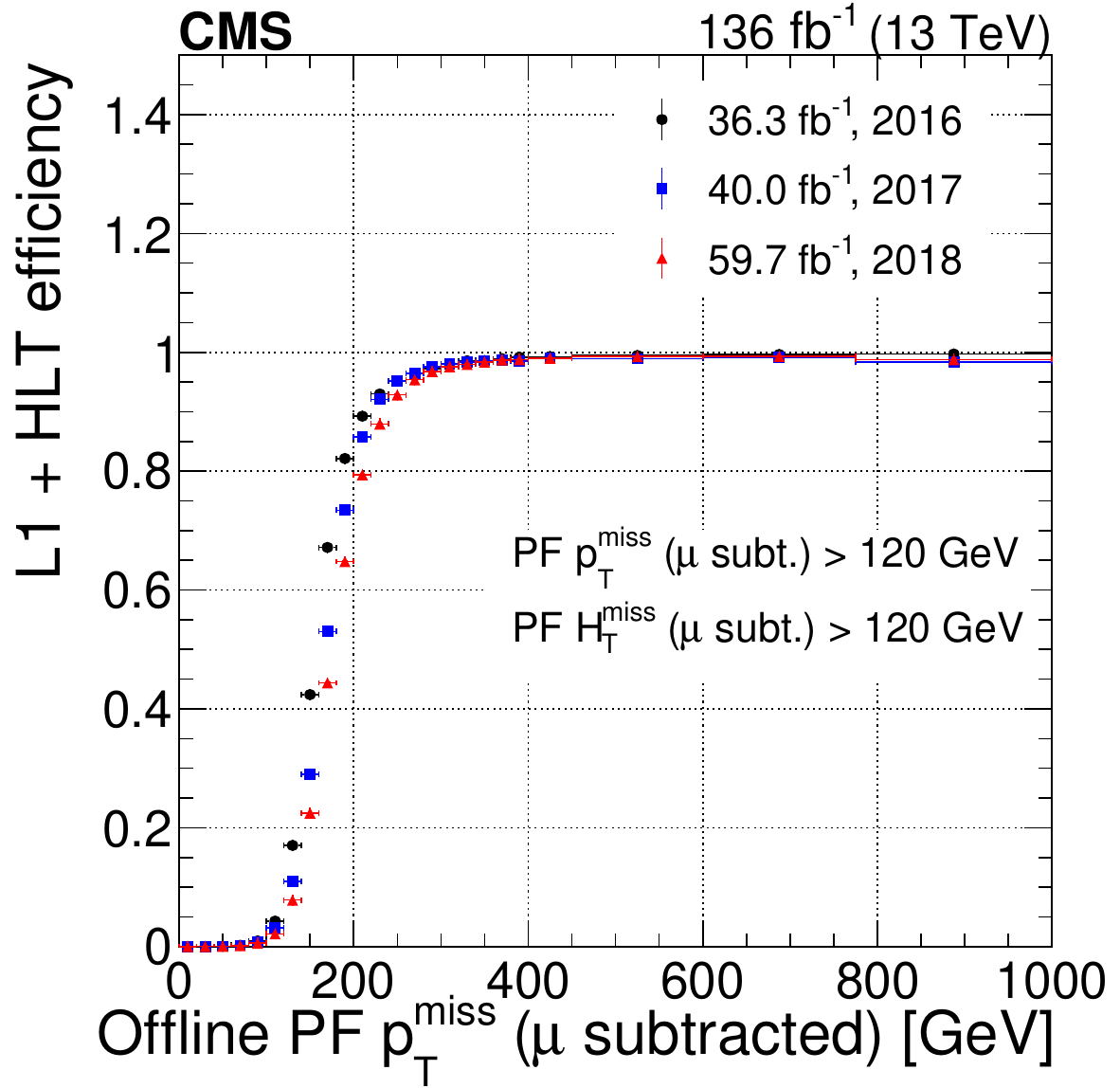}
  \caption{
    The L1+HLT efficiency of the $\PGm$-subtracted  \ptmiss trigger with a threshold of 120\GeV on both \ptmiss  and  missing \HT  measured with respect
    to the offline reconstructed $\PGm$-subtracted \ptmiss,
    shown separately for data collected during 2016, 2017, and 2018.
The vertical bars on the markers represent statistical uncertainties.
  }
  \label{fig:pfmet_nomu_mht}
\end{figure}

An alternative \ptmiss trigger is based on a
calculation that uses all the reconstructed PF objects except for
muons, leading to the ``$\PGm$-subtracted'' trigger paths. 
Therefore, in this approach, events with high-\pt muons are also
assigned large online \ptmiss,
whereas for events with no reconstructed muons, the two calculations
coincide.
An unprescaled trigger path that selects $\PGm$-subtracted \ptmiss
and missing \HT (similarly without muons) both ${>}120\GeV$  was available during the majority of \Runtwo.
The main uses for this path are searches for new physics in final states with only jets and \ptmiss,
which require the lowest \ptmiss thresholds
possible.
Figure~\ref{fig:pfmet_nomu_mht} shows the
performance of this path with respect to the offline $\PGm$-subtracted
\ptmiss, during 2016, 2017, and 2018.
A trigger efficiency above 95\% is reached for $\PGm$-subtracted
$\ptmiss > 250 \GeV$.

\subsection{\texorpdfstring{\PQb}{b} quark jets}

\label{sec:bTagging}
The identification of \PQb quark jets at the trigger level is essential to
collect events that do not pass standard lepton, jet, or \ptmiss
triggers, and to increase the purity of the recorded sample
for analyses requiring \PQb quark jets in the final state. The L1 trigger uses
information from the calorimeters and muon detectors to reconstruct
objects, such as charged leptons and jets. Sophisticated identification of \PQb quark jets similar to the one performed offline is
not possible at that stage as it relies on the reconstructed tracks
from charged particles available only at the HLT. In this section, we
describe \PQb quark jet identification at the HLT.

Because of latency constraints at the HLT, it is not feasible to
reconstruct the tracks and primary vertex with the algorithms used for
offline reconstruction. The time needed for track finding can be
significantly reduced if the position of the primary vertex is
known. Although the position in the transverse plane is defined with a
precision of 20\mum, its position along the beam line is not known.
However, it is possible to obtain a rough estimate of the primary
vertex position along the beam line by projecting
the position of the silicon pixel tracker hits compatible with the
jets onto the $z$ direction. A pixel tracker hit in the barrel (endcap) is compatible with a
jet when the difference in $\phi$ between the hit and the jet
is less than 0.21 (0.14). The region along the beam line with the
highest number of projected pixel detector hits is most likely to
correspond to the position of the primary vertex.  

This fast primary vertex finding algorithm is sensitive to pixel
detector hits from PU interactions. Therefore, a number of
selection requirements based on the shape of the charge deposition
clusters associated with the pixel detector hits are applied to select
those that most likely correspond to a particle with a large
\pt. In addition, only pixel detector hits compatible with up to
four highest-\pt jets with $\pt>30\GeV$ and $\abs{\eta}<2.4$ are
used. Finally, each pixel detector hit is assigned a weight reflecting
the probability that it corresponds to a track in one of the
considered jets. The weight is obtained by using information related
to the shape of the charge deposition cluster, the $\phi$
between the jet and the cluster, and the jet \pt. Since the spread
of projected hits from the primary vertex is proportional to the
distance from the beam line, a larger weight is assigned to pixel
detector hits closer to the beam line. 

Since \PQb tagging relies on the precise measurement of the displaced
tracks with respect to the primary vertex, it is crucial to use tracks
that use the information of both the pixel and the silicon strip
tracker to improve the spatial and momentum resolutions. To reduce the
HLT algorithm processing time, these tracks are reconstructed only
when originating near the primary vertex and if they are close to the
direction of the highest-\pt jets, sorted according to decreasing jet
\pt. Up to eight jets with $\pt>30\GeV$ and $\abs{\eta}<2.4$ are
considered in an event. In the first step, the trajectories of charged
particles are reconstructed from the pixel detector hits. To reduce
the reconstruction time, tracks are only reconstructed when they have
$d_{xy} < 15\unit{mm}$, $d_z < 2\unit{mm}$, and are
compatible in angle with the direction of one of the jets. The tracks are
reconstructed using the information from the pixel and strip
detectors. An iterative procedure is applied that is similar to the
offline track reconstruction except for the number of iterations and
the seeds used for track finding in each iteration. 

The reconstructed tracks and the primary vertex are then used to
reconstruct secondary vertices with the inclusive vertex finder reconstruction
algorithm~\cite{CMS:2011yuk,CMS:2017wtu}. These vertices and tracks are then used as input for the \PQb tagging algorithms.

Usually a loose \PQb tagging selection based on Calo-jets is used as an intermediate filter
to select the events for which the full PF-jet reconstruction
will be executed and from which the final PF-jet \PQb tagging
selection is applied. However, Calo-jet \PQb tagging was sufficient
for some physics triggers.

\subsubsection{Description of the algorithm}

During the 2016 data-taking period, \textsc{CSVv2} was the recommended
algorithm for \PQb tagging at the HLT.
The tagger and its performance are described in Ref.~\cite{CMS:2017wtu}.
After 2016, CMS developed a new tagger dedicated to identification
of \PQb quark jets reconstructed with the anti-\kt jet clustering algorithm with a
distance parameter equal to 0.4 (AK4-jets). This new algorithm, \DeepCSV~\cite{CMS:2017wtu},
relies on a multiclassifier neural network structure made of four fully
connected layers with 100 neurons each. The input variables list
comprises PF jet properties, tracks, and vertex related
variables. The \DeepCSV output values range from zero to
one and are interpreted as the probability for a given jet to
originate from the hadronization of a \PQb quark, a \PQc quark, or a gluon or
light quark.
The \DeepCSV tagger was deployed at the HLT at the start of the 2017 data-taking period,
and it was the recommended online \PQb tagging algorithm until the end of \Runtwo in 2018.
The training of the \DeepCSV algorithm has been
carried out using the input variable collections obtained from
MC simulated events after the trigger selection and the full
CMS event reconstruction. 

An important figure of merit for \PQb tagging algorithms is
the \PQb quark jet identification efficiency
versus the gluon or light-quark jet misidentification rate evaluated
in simulated events in the form of receiver operating characteristic
curves (ROC curves); this provides a
direct comparison of the performance of different taggers.
The flavor of offline-reconstructed jets in simulation is identified using the so-called ``ghost-matching'' technique~\cite{antikt}.
In this method, only the directional information of the four-momentum of the generator-level (ghost) hadron is used to prevent any modification to the four-momentum of the reconstructed jet.
Jets containing at least one \PQb hadron are assigned \PQb quark jets.
Similarly, labels are defined for jets originating from \PQc hadrons
and from gluons \Pg or light-flavor (\PQu, \PQd, \PQs) quarks (light jets).
Preference is given to jets with \PQb hadrons over \PQc hadrons.
Online jets are matched to offline jets if their direction agrees within a cone of $\DR<0.4$, and their flavor is assigned using the offline jet.

\begin{figure}[!htb]
  \centering
  \includegraphics[width=0.80\textwidth]{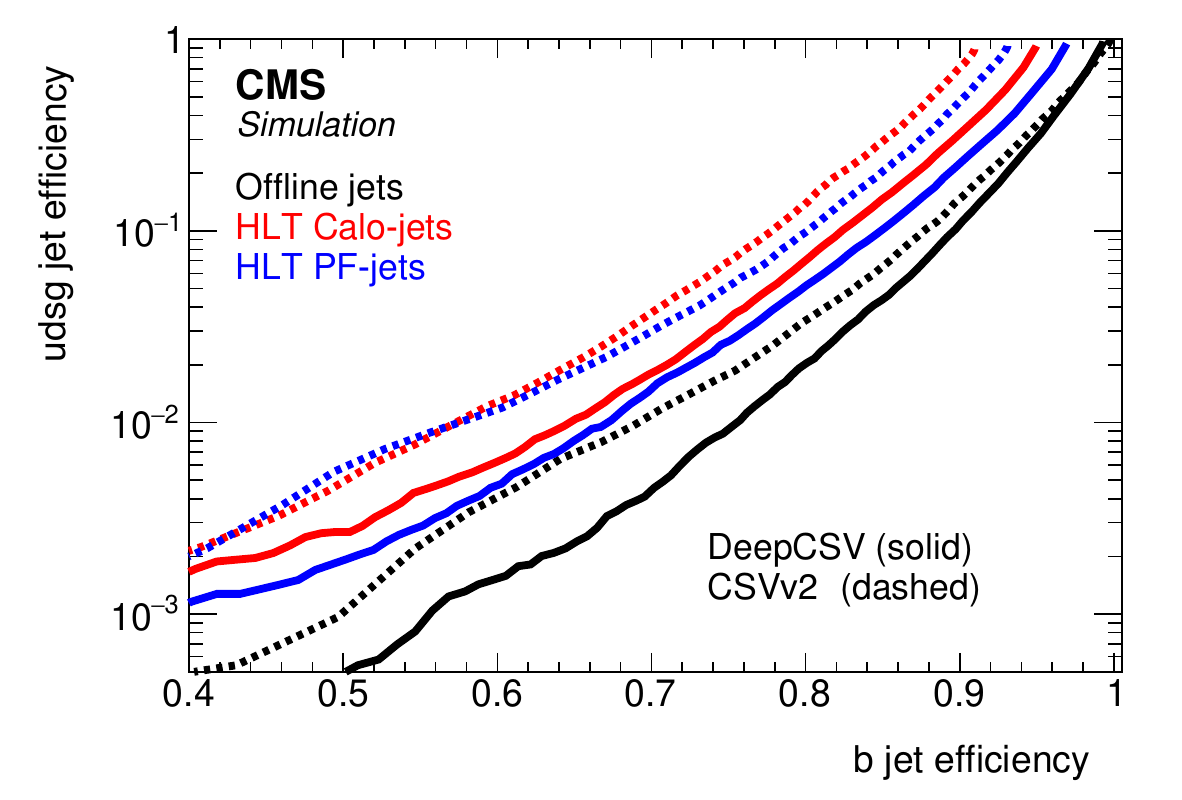}
  \caption{Performance of the online (red and blue) and offline (black)
    \PQb quark jet identification algorithms demonstrated as the probability for a
    light jet to be misidentified as a \PQb quark jet as a function of the
    efficiency to correctly identify a \PQb quark jet. The performance of the
    \textsc{CSVv2} (dashed) and \DeepCSV (solid) algorithms are shown. The curves
    are obtained for online and offline jets with $\pt>30\GeV$ and
    $\abs{\eta}<2.4$ in simulated \ttbar events. The plot is obtained using the
    2017 detector conditions.}
  \label{fig:ROC_MC}
\end{figure}

In Fig.~\ref{fig:ROC_MC}, the ROC curves obtained for the two \PQb
tagging algorithms, \DeepCSV and \textsc{CSVv2}, are compared using two
different sets of HLT input variables for the jet algorithm: full
event PF reconstruction (in blue) and
reconstruction based on information in regions around jets from the
CMS calorimeters (in red); the offline performance is shown
for reference in black. The conclusion of this study is that \DeepCSV
outperforms \textsc{CSVv2}, with an improvement of the \PQb quark jet tagging
efficiency for a fixed gluon or light-quark misidentification rate
of 5--15\%.

\subsubsection{Performance measurement in data and simulation}
The performance is evaluated using $\Pp\Pp$ collision data
collected at $\sqrt{s}=13\TeV$ in 2017 and 2018 during LHC \Runtwo, corresponding to
an integrated luminosity of about 30 and 48\fbinv, respectively. The performance is
assessed for events consistent with the \ttbar process. Events are selected at the HLT
using a combination of trigger paths that require the
presence of at least one muon and one electron.  For the offline analysis,
events are selected that contain one isolated electron with
$\pt>30\GeV$ and one isolated muon with $\pt>20\GeV$.  In addition, at
least two jets with $\pt>30\GeV$ are required.  This event selection
is enriched with \ttbar events and ensures an unbiased selection of
\PQb quark jets with only a small contribution of $\PQt\PW$ events.

\begin{figure}[!htb]
  \centering
  \includegraphics[width=0.49\textwidth]{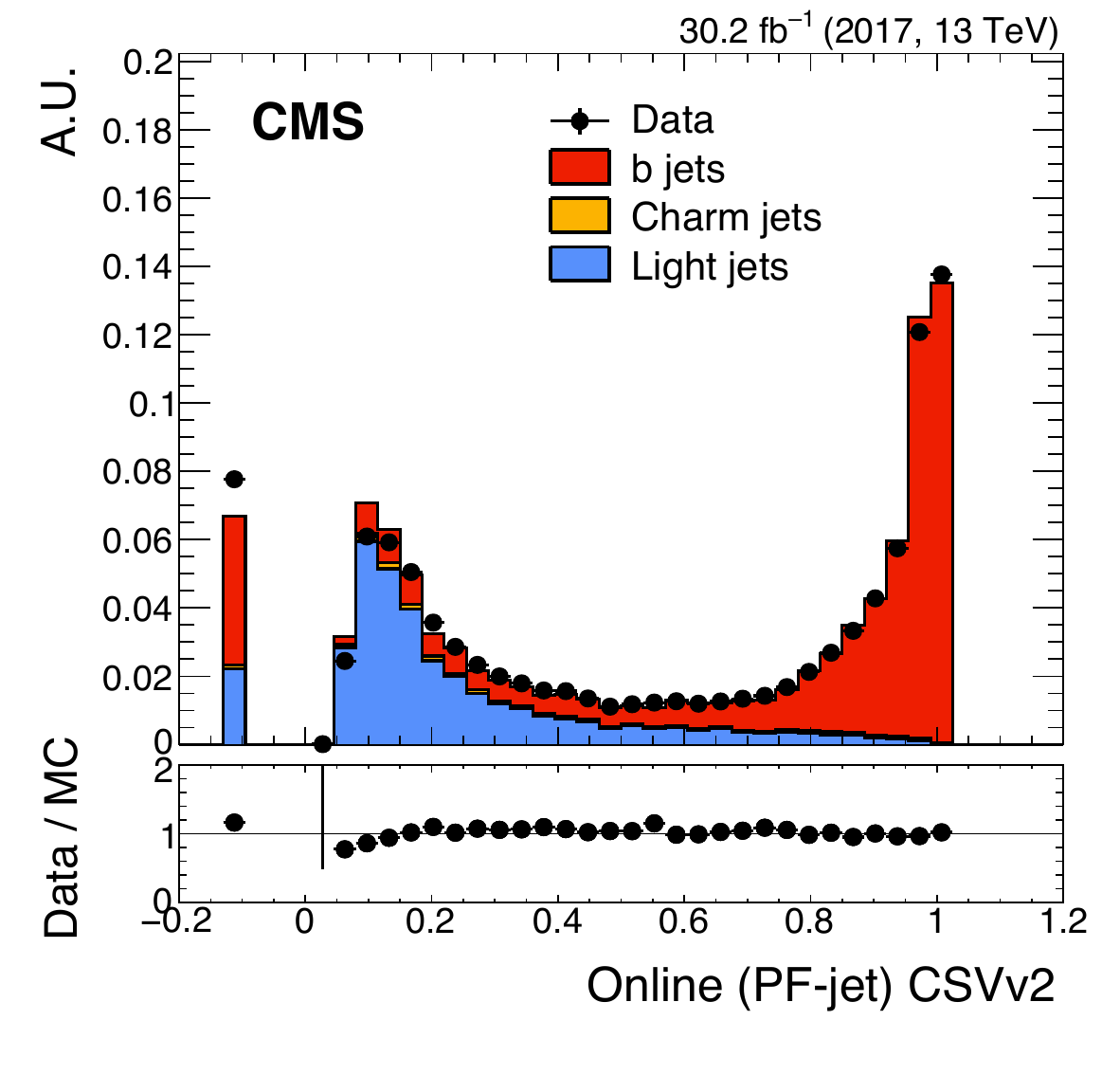}
  \includegraphics[width=0.49\textwidth]{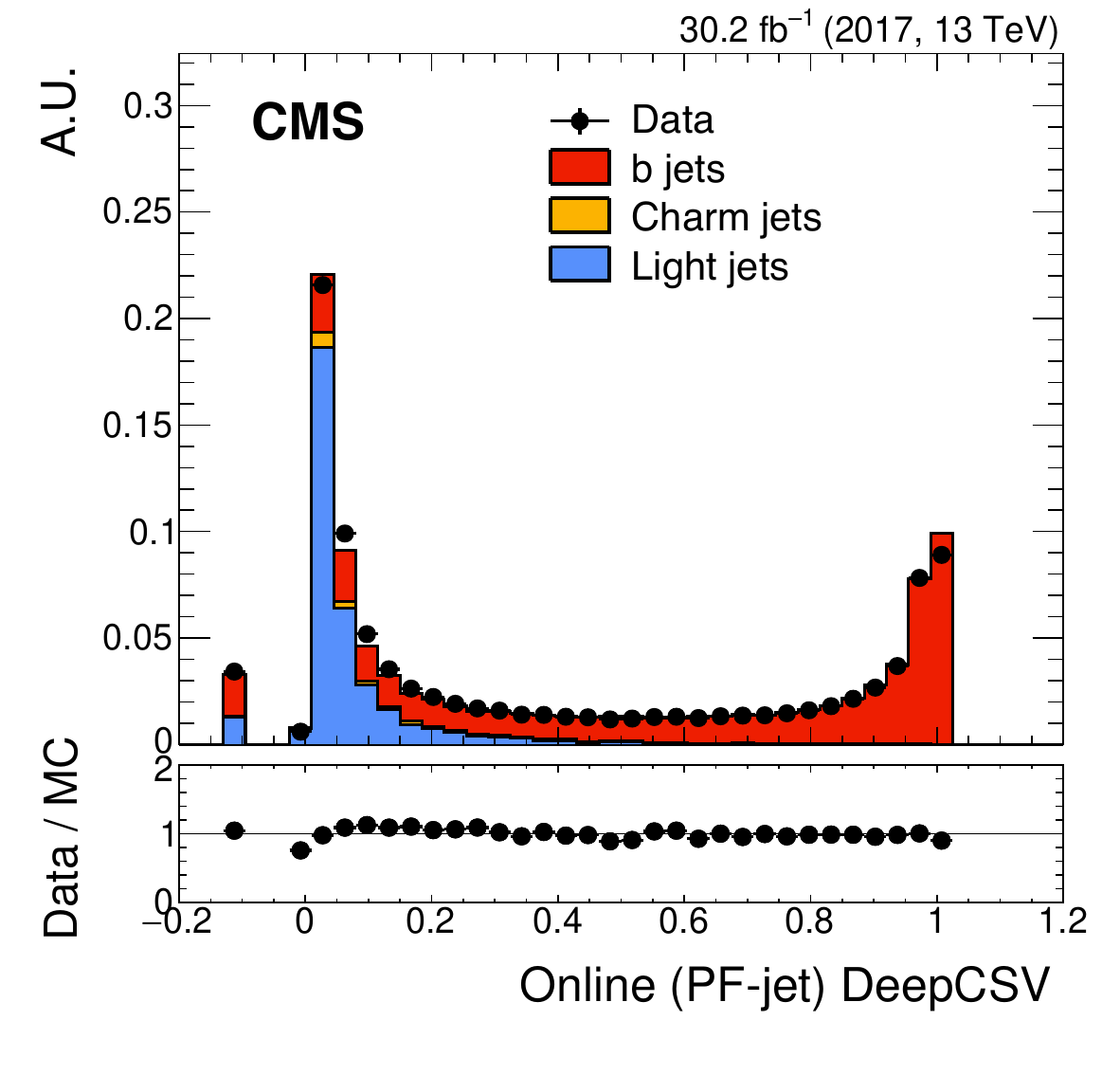}
  \caption{Left: Online (PF-Jets) \textsc{CSVv2} discriminator distribution, normalized to unity for both data and the summed simulation. Different colors show the contributions
    from simulation of different jet flavors. The plot is
    obtained using the 2017 detector conditions. A negative value indicates that the tracking
    preselection has failed and the discriminator is not evaluated.
    Right: Same, but for the \DeepCSV discriminator. The plot is obtained using 2017 data and
    MC simulation using the \DeepCSV algorithm as it was run in
    2018.
    The lower panel of each plot shows the ratio of data to MC simulation.
The vertical bars on the markers represent statistical uncertainties.
  }
  \label{fig:DeepCSV_data_MC}
\end{figure}

Efficiencies are measured by selecting events that contain at least
one offline-reconstructed jet passing a working point that
corresponds to a light jet mistag rate of 1\%.
Figure~\ref{fig:DeepCSV_data_MC} shows the online PF-jet \DeepCSV and \textsc{CSVv2} discriminator score. The left (right) plot was obtained using 2017 (2018) data. As described earlier, the output scores range from $0$ to $1$, and a negative value is assigned if the tracking preselection has failed and the discriminator was not evaluated.
For both tagging algorithms, the data agree well with the MC simulation predictions.
The distribution for true \PQb quark jets peaks at unity, while a peak at low values in the \PQb tag score is observed for light jets.
For \DeepCSV, the separation between signal and background is observed
to be much larger.

\begin{figure}[!htb]
  \centering
  \includegraphics[width=0.49\textwidth]{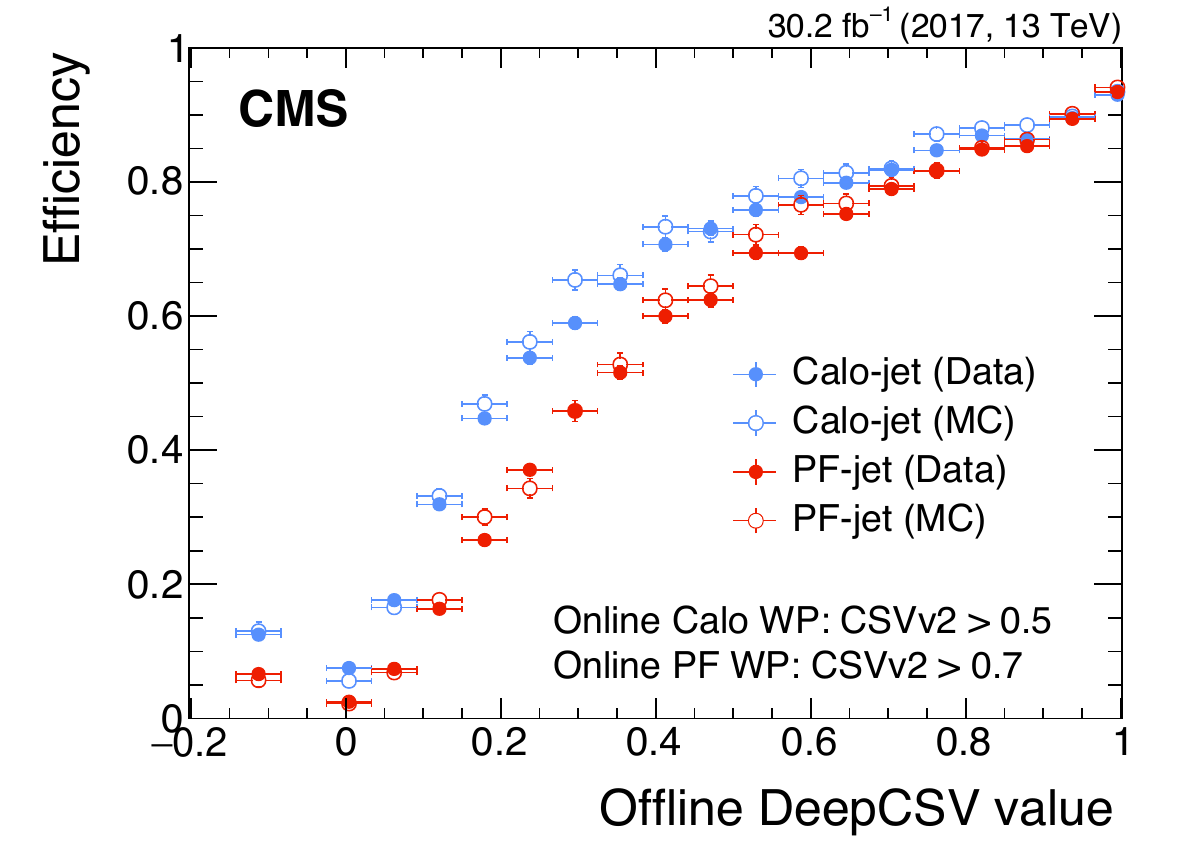}
  \includegraphics[width=0.49\textwidth]{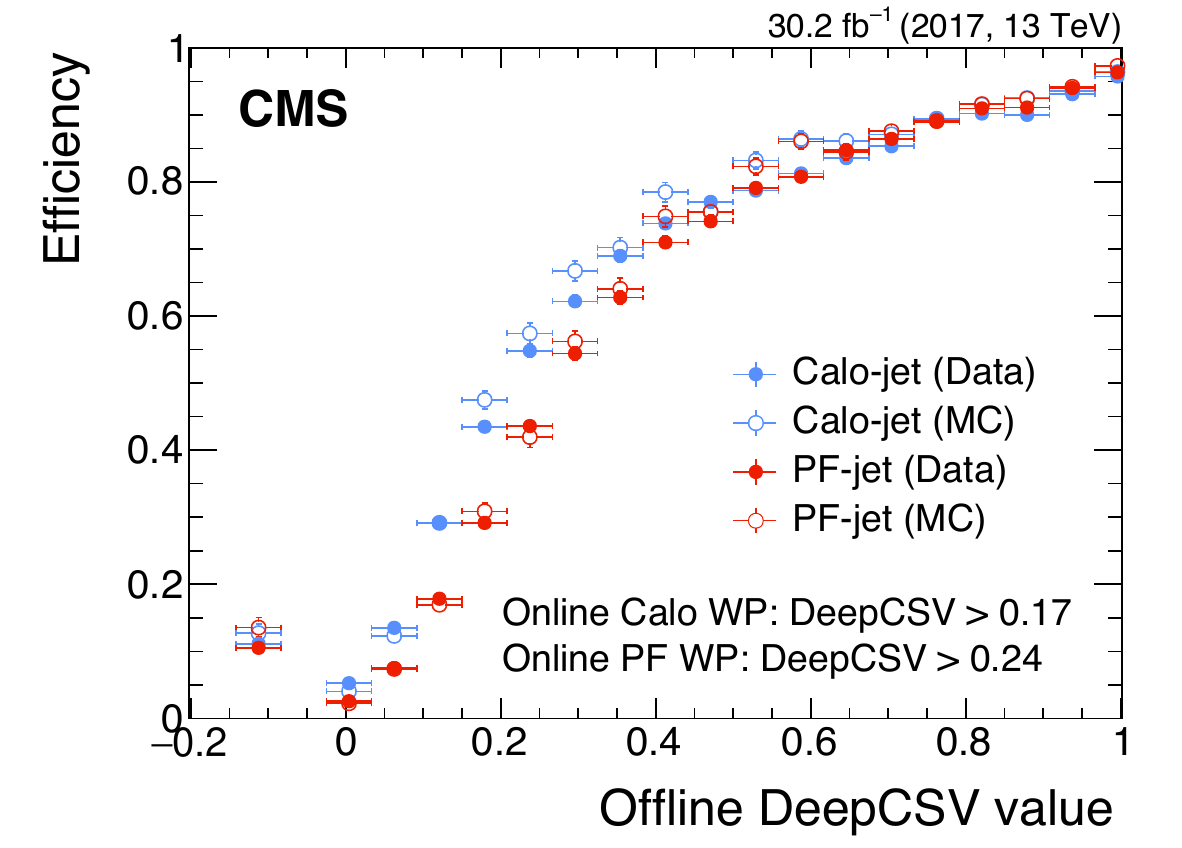}
  \caption{Efficiency to pass the online \textsc{CSVv2} (left) and \DeepCSV (right) working points as a
    function of the corresponding offline \DeepCSV value. Data collected
    in 2017 are shown in closed circles; the result of the simulation is shown
    in open circles. The turn-on with respect to the online Calo-jets is
    shown in blue. The turn-on with respect to the online PF-jets is
    shown in red. The right plot is obtained using 2017 data and MC simulation using
    the \DeepCSV algorithm as it was run in 2018. A negative value
    indicates that the tracking preselection has failed and the
    discriminator is not evaluated.
The vertical bars on the markers represent statistical uncertainties.
  }
  \label{figBTagEfficiency1}
\end{figure}

The efficiencies at the HLT are displayed in Fig.~\ref{figBTagEfficiency1}.
The efficiency is defined as the fraction of jets selected by the HLT with respect to
the number of jets selected by the offline \DeepCSV algorithm, with representative working points as indicated in the plots.
The study is performed using data and simulated samples for the conditions of 2017.
The performances of Calo- and PF-jets are shown individually.
Good agreement between data and simulation is observed.
The rise of the efficiency using \DeepCSV both online and offline is
much steeper because of the larger correlation between the scores.
In these plots, the choice of a looser working point, and thus higher efficiency,
for Calo-jet \PQb tagging compared to PF-jet \PQb tagging reflects
how these algorithms are applied in physics triggers.

\begin{figure}[!htb]
  \centering
  \includegraphics[width=0.49\textwidth]{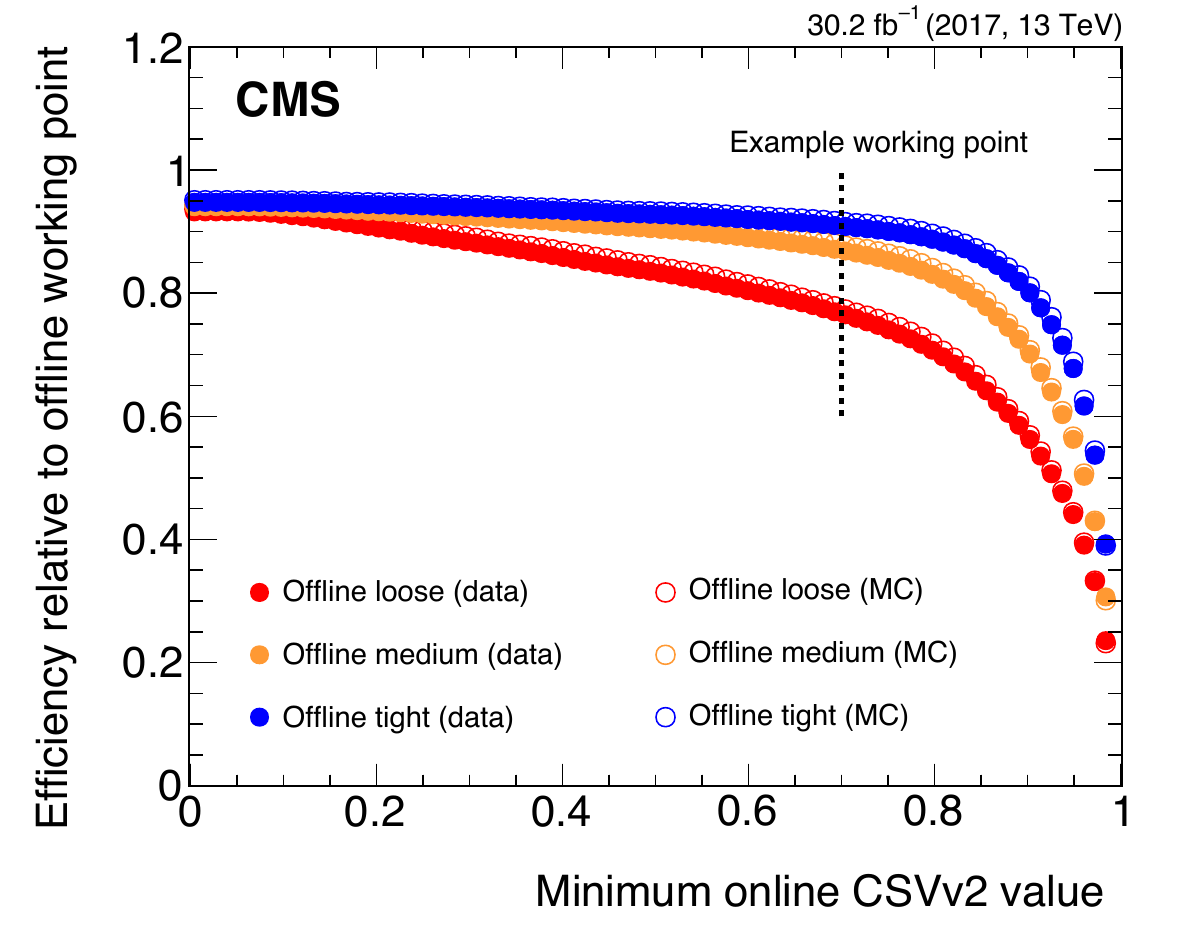}
  \includegraphics[width=0.49\textwidth]{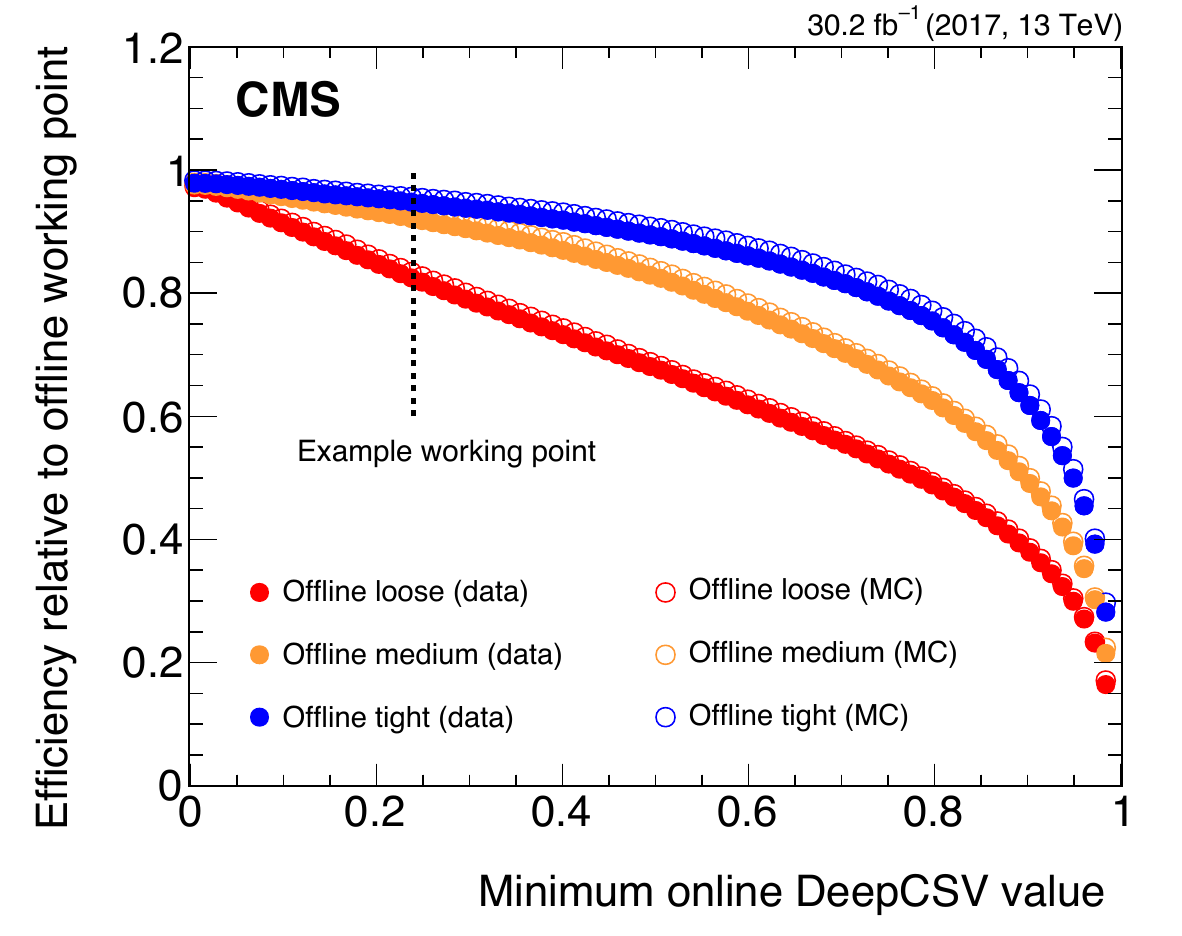}
  \caption{Efficiency of jets \PQb tagged offline to pass the online
    \textsc{CSVv2} (left) and \DeepCSV (right) \PQb tagging requirement, as a function of the online requirement.
    Three offline selections are shown: Loose (red), Medium
    (orange), and Tight (blue). Data are shown in closed circles; the
    result of the simulation is shown in open circles. The right plot is
    obtained using 2017 data and MC simulation but using the \DeepCSV algorithm as
    it was run in 2018.}
  \label{fig:BTagEfficiency2}
\end{figure}

For trigger paths using \PQb quark jet tagging, the online efficiency with respect to the offline performance is usually an important figure of merit.
This quantity can be evaluated for a fixed offline \PQb tagging efficiency corresponding to a light-jet efficiency of 0.1, 1, or 10\%, namely, the tight, medium, and loose working points, respectively. The relative \textsc{CSVv2} and \DeepCSV efficiency is shown in Fig.~\ref{fig:BTagEfficiency2}.
The efficiencies, as obtained from data and simulation, are in good agreement.
Using these curves, the online selection is tuned to reach a fixed efficiency given a fixed offline selection.
The \textsc{CSVv2} efficiency is smaller than that of \DeepCSV at high score values because the offline \DeepCSV discrimination is much better and the jets selected online are all also selected by the offline algorithm.

\subsection{Tau leptons}

Tau leptons have a relatively short lifetime and decay before reaching the
beampipe. In 64.8\% of cases, they decay hadronically into one or
three charged hadrons and mostly accompanied by neutral pions.
Neutral pions decay
promptly into two photons, which may convert into $\Pep\Pem$ pairs
while traversing the material of the tracker. As a result of the large
magnetic field of the CMS solenoid, the $\Pep\Pem$ pairs are
separated in the $(\phi, \eta)$ plane. Thus, neutral pions are
reconstructed from photons and electrons. The aim of the HLT tau
reconstruction is to
reconstruct hadronic decays of tau leptons ($\tauh$).

\subsubsection{Reconstruction of $\tauh$ trigger paths}

\begin{figure}[ht!]
  \centering
  \includegraphics[width=\textwidth]{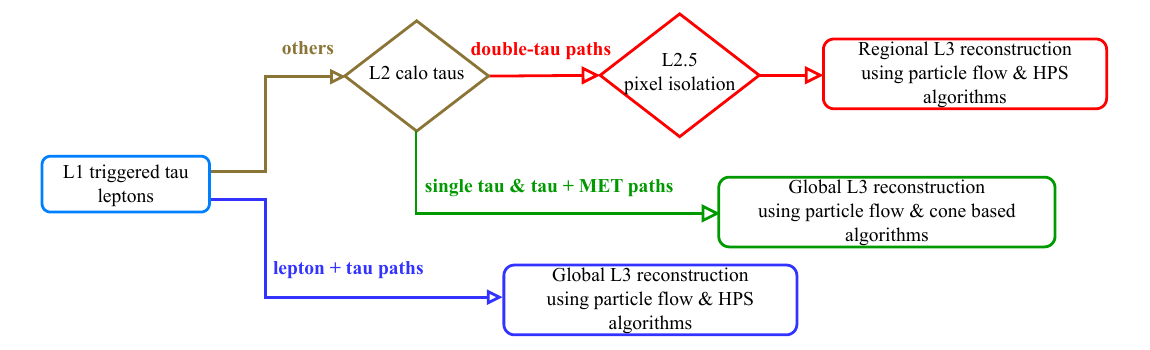}
  \caption{Flow chart for \tauh-candidate reconstruction at the HLT.}
  \label{fig:flowchart}
\end{figure}

The reconstruction of tau leptons at the HLT is performed in different
steps depending on the associated particle. A flow chart of the 
reconstruction steps is  summarized in Fig.~\ref{fig:flowchart}. The $\tauh$ triggers
used in the \Runtwo data-taking period and the corresponding
L1 and HLT conditions are listed in Table~\ref{tab:TauHLTConditions}.

\subsubsection* {Double-tau ($\tauh \tauh$) trigger paths}

A double-tau trigger path is formed when a tau lepton is associated with another tau lepton that also decays hadronically.
In these double-tau triggers, the reconstruction is performed in three steps.
The first step is called L2,
where reconstruction starts with the L1 trigger $\tauh$
candidates. The energy depositions in the calorimeter towers around
the seeded L1 $\tauh$ candidates within a cone of radius $0.8$ are
clustered, and L2 $\tauh$ candidates are reconstructed by using the
anti-\kt algorithm~\cite{antikt}
with $\DR = 0.2$.
In 2016 and 2017, the L2 candidates were reconstructed for all L1 $\tauh$ candidates,
including those with very low \pt that
did not contribute to any of the relevant seeds in the L1 menu.
In 2018, the reconstruction of the L2 candidates was updated to be performed around only those L1 $\tauh$ candidates
that satisfy the \pt and isolation criteria of the L1 seeds
that contributed to the event selection at L1.

In the second step, known as L2.5, charged-particle isolation based
on the information in the pixel detector is implemented. 
Reconstructed L2 $\tauh$ jets with $\pt > 20\GeV$ and $\abs{\eta} < 2.5$
are selected, and pixel detector tracks are reconstructed around the direction
of the selected L2 $\tauh$ candidates in a region of $\Delta \eta \times
\Delta \phi = 0.5 \times 0.5$ from the hits in the pixel detector. 
The reconstructed tracks are clustered, and vertices having at least
two tracks with $\pt > 1\GeV$ are formed by using the divisive
vertex finder~\cite{CMS:2014pgm}.
If no vertices are found, a $\tauh$ candidate is considered
perfectly isolated and the reconstruction continues with the next
step. If more than one vertex is reconstructed, the one with the
highest $\sum \pt^{2}$ of its constituent tracks is chosen as the primary vertex
of the hard-scattering event. Tracks that have at least three hits and
have a trajectory in a cone of $0.15 < \DR < 0.4$
centered around the L2 $\tauh$ candidates and originating from
the primary vertex are considered for the isolation requirement. These tracks are required to have
$d_{xy} < 0.2\unit{cm}$. An L2 $\tauh$ candidate is considered to be isolated if the
scalar sum of the \pt of the associated pixel detector tracks is less than
4.5\GeV. 

The final step is referred to as the L3 reconstruction, where
the full tracking information is included through the use of the online PF
reconstruction.
Instead of
reconstructing all tracks, the reconstruction is performed
regionally around the L2 $\tauh$ candidates with $\pt >20\GeV$ and
$\abs{\eta}<2.5$.
The L3 reconstruction was performed using a cone-based algorithm
until mid-2018, after which it was upgraded to
the hadron-plus-strips (HPS) algorithm 
that had already been in use for the offline
reconstruction of tau leptons~\cite{Sirunyan:2018pgf}. The HPS algorithm allows for the
exclusive reconstruction of specific hadronic decay modes, 
which is not possible with a cone-based
algorithm. Both algorithms start with PF jets reconstructed by the
anti-\kt algorithm with a distance parameter of $0.4$, and a maximum of one
tau lepton is reconstructed for each PF jet at the end of the
algorithm. The HPS-based algorithm, described later, will be the main
focus here, since the cone-based algorithm has already been described in
Ref.~\cite{Sirunyan:2018pgf}. 

The $\tauh\tauh$ triggers used in the \Runtwo data-taking period require a pair of isolated
L1 $\tauh$ candidates with variable \pt thresholds, as listed in
Table~\ref{tab:TauHLTConditions}.
The $\tauh$ candidates are required to
have $\pt>35\GeV$ at the HLT and to pass the medium working point of the combined
isolation, described below. The isolation is relaxed by 5\%/GeV for $\pt^{\tauh} > 100\GeV$
(\ie, the threshold on the \pt sum is increased by 5\% for each GeV in $\pt^{\tauh}$ above 100\GeV).
The $\tauh$ candidates must be separated by $\DR > 0.5$.

\subsubsection*{Lepton+\tauh trigger paths}

So-called cross triggers select events with at least one tau lepton and
another object, such as an electron or muon. For these triggers, the reconstruction
is performed in fewer steps than for double-tau trigger paths. These
triggers require the existence of an L1 trigger seeded by a muon or 
an electron together with a $\tauh$ lepton. Additionally,
they perform the reconstruction and selection of a muon or
electron before the $\tauh$ reconstruction starts, which
reduces the number of events for which the $\tauh$ reconstruction
is run. Because of that, the L3 reconstruction is run directly without
the L2 and L2.5 prefilters and thus over the entire CMS
detector acceptance. The L3 reconstruction is performed using the
HPS algorithm.

The lepton+$\tauh$ triggers require a muon or an electromagnetic object passing the
L1 \pt thresholds given in Table~\ref{tab:TauHLTConditions}.
A $\tauh$ candidate is also required to pass a given HLT \pt threshold and loose charged isolation.
In some cases, several L1
triggers with different \pt thresholds are used to keep the
L1 rate constant by dynamically selecting the unprescaled L1 seed based on \Linst.
The isolation is relaxed by 5\%/GeV for $\pt^{\tauh} >110\GeV$.
Finally, the lepton and $\tauh$ candidate must be separated
from each other by requiring $\DR > 0.3$.

\subsubsection*{The $\tauh$+$\ptmiss$ and single-$\tauh$ trigger paths}

Tau leptons can also be reconstructed in association with \ptmiss, which is also called a cross trigger.
The reconstruction in such triggers occurs in two steps. These
triggers require the existence of an L1 trigger seeded by \ptmiss together with
a $\tauh$ lepton, as with the lepton+$\tauh$ triggers. The reconstruction and
selection of \ptmiss before the $\tauh$ reconstruction is started
reduces the rate of events for which the $\tauh$ reconstruction
is run. Hence, the L3 reconstruction can be run without
the L2.5 prefilter and globally over the entire coverage of the CMS
detector.  In the case of single-$\tauh$ triggers, because of the high \pt requirement,
global L3 reconstruction is also affordable.
In both $\tauh$+$\ptmiss$ and single-$\tauh$ triggers, however, the L2 filters are used
to reduce the CPU processing time.
In contrast to the $\PGm\tauh$, $\Pe\tauh$, and $\tauh\tauh$ triggers, the
$\tauh$+$\ptmiss$ and single-$\tauh$ triggers reconstruct the tau
leptons using the cone-based algorithm as described in
Ref.~\cite{Sirunyan:2018pgf}.

\begin{table}[ht!]
  \centering
  \topcaption{List of $\tauh$ triggers used in 2016, 2017, and 2018 data-taking periods including both the L1 and HLT conditions.
    The $\Pe\tauh$ triggers evolved in 2016 data taking, with the labels ${(1)}$, ${(2)}$, and ${(3)}$ indicating use for the first 7.4\fbinv, the next 10.2\fbinv, and the last 18.3\fbinv of data, respectively.}
  \cmsTable{
  \begin{tabular}{lcll}  \hline
    Year & Trigger & HLT condition  & L1 condition \\ \hline
    {2016} && & \\
    & $\PGm \tauh$ & $\pt^{\PGm} > 19\GeV$ (isolated) ~  $\pt^{\tauh} > 20\GeV$ (unseeded) &  $\pt^{\PGm} > 18\GeV$ \\ [3pt]
    & {$\Pe \tauh$} & $\pt^{\Pe}>24\GeV$,  $\pt^{\tauh} > 20\GeV$ (unseeded) $^{(1)}$ &  $\pt^{\Pe} > 22\GeV$\\ [3pt]
    &    & $\pt^{\Pe} > 24\GeV$, $\pt^{\tauh} > 20\GeV$ (seeded \& nonisolated)  $^{(2)}$ &  $\pt^{\Pe} > 22\GeV$,\\
    & & & \,\,$\pt^{\tauh} > 20\GeV$\\ [3pt]
    &    & $\pt^{\Pe} > 24\GeV$,  $\pt^{\tauh} > 30\GeV$ (seeded \& isolated)$^{(3)}$ &   $\pt^{\Pe} > 22\GeV$,\\
    & & & \,\,$\pt^{\tauh} > 26\GeV$ \\ [3pt]
    
    & $\tauh \tauh$ &  $\pt^{\tauh} > 35\GeV$ (seeded \& isolated) &  $\pt^{\tauh} > 28$--36\GeV \\ [3pt]
    
    & $\tauh$+$\ptmiss$ & $\ptmiss > 90\GeV$,  $\pt^{\tauh} > 50\GeV$, $\pt^{h^{\pm}} >30\GeV$ (unseeded) &  $\ptmiss > 80$--100\GeV \\ [3pt]
    
    & Single $\tauh$ & $\pt^{\tauh} > 140\GeV$, $\pt^{h^{\pm}} >50\GeV$ (seeded) &  $\pt^{\tauh} > 120\GeV$ \\ [3pt]
    \\
    \multicolumn{2}{l}{{2017 \& 2018}}  & & \\
    &  $\PGm \tauh$ & $\pt^{\PGm}>20\GeV$ (isolated), ~  $\pt^{\tauh} > 27\GeV$ (seeded \& nonisolated) &   $\pt^{\PGm} > 18\GeV$,\\
    & & & \,\,$\pt^{\tauh} > 24/26\GeV$  \\ [3pt]
    
    & \multirow{1}{*}{$\Pe \tauh$}   & $\pt^{\Pe}>24\GeV$ ,  $\pt^{\tauh} > 30\GeV$ (seeded \& isolated) &  $\pt^{\Pe} > 22/24\GeV$,\\
    & & & \,\,$\pt^{\tauh} > 26/27\GeV$\\ [3pt]
    & $\tauh \tauh$ &  $\pt^{\tauh}> 35\GeV$ (seeded \& isolated) &   $\pt^{\tauh} > 32$--36\GeV \\ [3pt]
    
    & $\tauh$+$\ptmiss$ & $\ptmiss > 100\GeV$,  $\pt^{\tauh} > 50\GeV$, $\pt^{h^{\pm}} >30\GeV$ (seeded) &  $\ptmiss > 80$--110\GeV,\\
    & & & \,\,$\pt^{\tauh} > 40\GeV$\\ [3pt]
    
    & Single $\tauh$ & $\pt^{\tauh} > 180\GeV$, $\pt^{h^{\pm}} >50$\GeV (seeded) &  $\pt^{\tauh} > 120$--130\GeV \\ [3pt]
    \hline
  \end{tabular}
  \label{tab:TauHLTConditions}
  }
\end{table}

The $\tauh$+$\ptmiss$ triggers are seeded by
an L1 cross trigger as listed in Table~\ref{tab:TauHLTConditions}. The
$\tauh$+$\ptmiss$ triggers require high
\ptmiss and a tau lepton with $\pt^{\tauh} > 50\GeV$ passing
the medium working point of the charged isolation by rejecting the tracks with low
\pt. The isolation is relaxed by 5\%/GeV for $\pt^{\tauh} >120\GeV$.
The high \pt single-$\tauh$ trigger is seeded by a single
L1 tau lepton. Tau leptons are reconstructed by using charged particle \PShpm tracks with
$\pt^{\PShpm} > 30\,(50)\GeV$ for the $\tauh$+$\ptmiss$ (single-$\tauh$) trigger,
and they are required to have large \pt and
to pass the medium working point of the combined isolation (discussed in the next section). The isolation is
relaxed for $\pt^{\tauh} > 300\GeV$ by 2\%/GeV and is
removed completely when $\pt^{\tauh} > 500\GeV$.

\subsubsection* {Hadron-plus-strips algorithm}

In the first step of the HPS algorithm, photons and electrons that exist in PF jets are
clustered into a ``strip'' around the highest-\pt photon or electron with a
$\Delta \eta \times \Delta \phi$ area of $0.05 \times 0.2$,
and the strip is assigned the \PGpz mass. To overcome any inefficiency
stemming from the misreconstruction of charged hadrons, PF neutral
hadrons in the signal cone are considered as a part of charged hadrons
in addition to PF charged hadrons. The signal cone that is used to reconstruct $\tauh$ candidates is defined as a
function of the \pt of the hadronic system by $R_{\text{sig}} =
3.0\GeV/\pt$, with the limits of the cone size to be in the range of
$0.05< R_{\text{sig}} < 0.10$.
We reconstruct $\tauh$ candidates in the following decay topology classes on the number of charged and neutral hadrons:
$(n_{\PShpm}, n_{\PGpz}) =$  (1,0), (1,1), (1,2), (2,0), (2,1), (3,0), and (3,1).  The
classes with two charged hadrons have been added to catch decays to
three charged hadrons where one track was not properly
reconstructed. 
A vertex is associated to each tau candidate,
selected to be the one closest in
$d_z$, to the track of the highest-\pt charged-hadron candidate. A tau
lepton is then selected by applying further requirements that
include the compatibility of the final states to given decay
modes. For this purpose, the reconstructed decay modes are required to
be within a mass window corresponding to either a $\PGrP{770}$ or
$\PaDoP{1260}$. The mass windows are optimized for the online
implementation of the HPS algorithm to reconstruct the online
tau leptons efficiently. The offline values of the mass windows
are reported in Ref.~\cite{Sirunyan:2018pgf}.

After this step, there are further cleaning steps applied. Soft tau
lepton candidates with two charged hadrons with $\pt < 5\GeV$ are rejected
in order to reduce the rate of tau leptons with one
charged hadron migrating to the decay mode with two charged hadrons.
Of the remaining candidates, the 
tau leptons with the largest \pt and largest strip multiplicity are
preferred by the HLT paths. The single-tau candidate with the lowest combined isolation,
associating its neutral components with the candidate, within the isolation cone
size of $\DR =0.5$, is selected. The combined isolation is
calculated as: 
\begin{equation}
 I_{\tauh}^{\text{L3}} = \sum \pt^{\text{charged}} +  \sum \pt^{\gamma},
\end{equation}
where $\sum \pt^{\text{charged}}$ and $ \sum \pt^{\gamma} $ are the
scalar sums of the \pt of charged hadrons and of photons,
respectively, that do not
belong to the $\tauh$ candidate. The value of the combined isolation
is relaxed in the HPS-based tau lepton reconstruction compared with the
one used in cone-based algorithm to achieve similar
efficiency, leading to the requirement that the isolation be smaller
than 3.9, 3.7, and 3.2\GeV for loose, medium, and tight working points,
respectively. Those values are further relaxed 
as a function of \pt to increase the reconstruction efficiency of
genuine $\tauh$ candidates at high \pt. This is only possible because
of the reduction in the
number of misidentified $\tauh$ candidates as a function of \pt,
which helps to control the trigger rates~\cite{Sirunyan:2018pgf}.

\subsubsection{Performance measurement in data and simulation}

The efficiencies of the $\tauh$ legs of the $\Pe\tauh$, $\PGm\tauh$,
and $\tauh\tauh$ triggers are estimated by using the tag-and-probe
method in $\PZ/\gamma^* \to \PGt \PGt \to \PGm \tauh$ events,
since the $\tauh$ purity is higher in $\PGm \tauh$ events than in $\Pe
\tauh$ and $\tauh \tauh$ events. Monitoring triggers based on $\PGm \tauh$
with the same isolation, identification, and \pt thresholds as in the $\Pe
\tauh$ or $\tauh \tauh$ triggers are used to measure the efficiency
of $\tauh$ leg in $\Pe \tauh$ and $\tauh \tauh$ triggers. The
trigger efficiency is calculated from the ratio of the number of events that
pass the baseline offline tag-and-probe selection, explained below, as well as
the given HLT path to the number of events that pass only the tag-and-probe
selection. The offline \tauh candidates are matched to the online \tauh
candidates for the numerator selection. The measured efficiencies always
depend on the offline selection.

Events passing an isolated single-muon L1
trigger with $\pt>27\GeV$ and $\abs{\eta} < 2.1$ are selected. Exactly one offline
muon passing loose identification criteria is required to suppress the contamination
from $\PZ \to \PGm \PGm$ events. An offline muon candidate is considered
matched with
an online object if $\DR < 0.5$. From this sample, a hadronically decaying
tau candidate with $\pt > 20\GeV$ and $\abs{\eta} < 2.1$ is selected as the
probe. The tau candidate is required to pass the medium working point
of the tau combined isolation to reject the events with misidentified tau leptons
reconstructed from the background from SM events composed uniquely of jets produced through the strong interaction, referred to as quantum chromodynamics  multijet events, with 70\% efficiency. To suppress the
misidentified muons and electrons, dedicated
discriminators are used. The $\tauh$ lepton is required to be
separated from the tagged $\PGm$. The selected $\PGm$ and $\tauh$
candidates are required to have opposite sign charges in both data and
simulation samples. The reconstructed offline tau leptons are matched with tau
leptons, electrons, or muons at the MC generator level to
suppress misidentified $\tauh$ particles from jets
in simulated
events. In data, the contribution from such events is subtracted using
events containing a muon and an hadronic tau lepton carrying the same
charge. To increase the purity of the $\PZ \to \PGt \PGt \to \PGm \tauh$ events,
offline selections on the transverse mass,
$\mT (\PGm, \ptmiss) < 30\GeV$, and visible mass, $40 < m_{\text{vis}} (\PGm \tauh) <80\GeV$
are applied. Furthermore, events with electrons
and \PQb-tagged jets are vetoed.

\begin{figure}[ht!]
  \centering
  \includegraphics[width=0.45\textwidth]{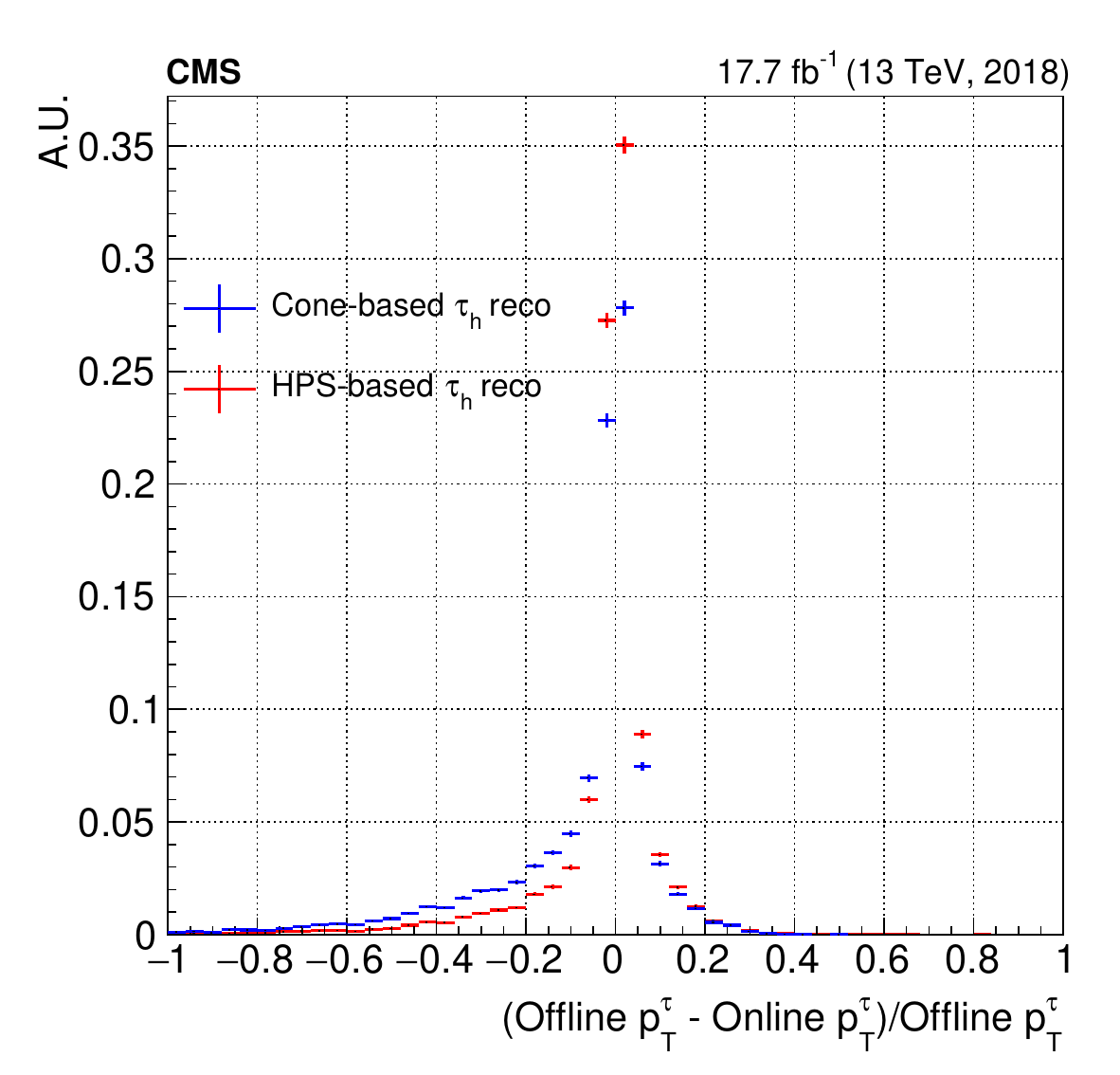}
  \caption{The \pt resolution for the $\PGm \tauh$ trigger for cone-based and HPS-based $\tauh$ reconstruction, calculated by using the first 17.7\fbinv of 2018 data taken with the cone-based tau reconstruction, where the trigger paths with HPS-based algorithm were included for the purpose of testing.
The vertical bars on the markers represent statistical uncertainties.
  }
  \label{fig:ptRes}
\end{figure}

\begin{figure}[htbp!]
  \centering
  \includegraphics[width=0.45\textwidth]{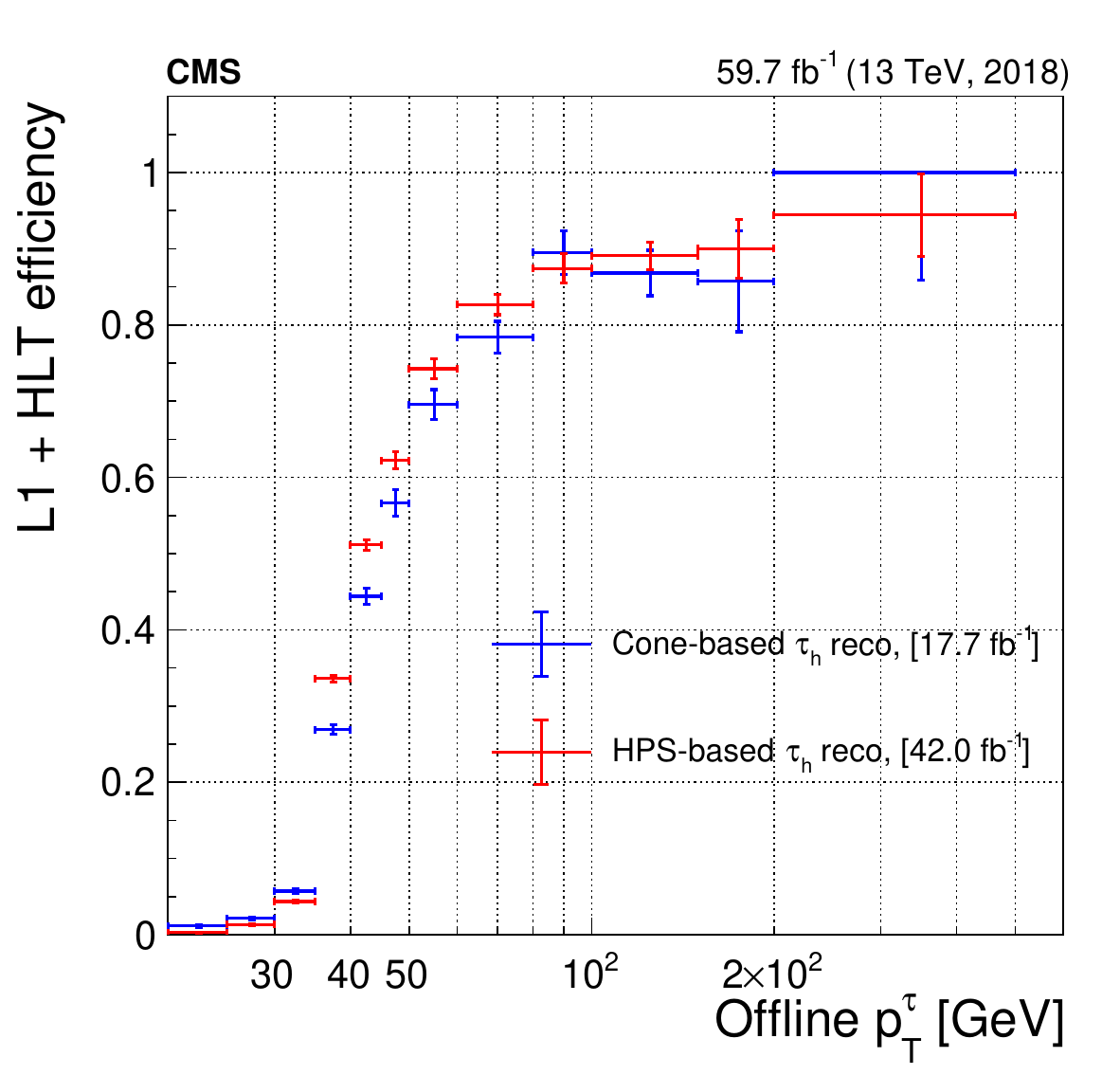}
  \includegraphics[width=0.45\textwidth]{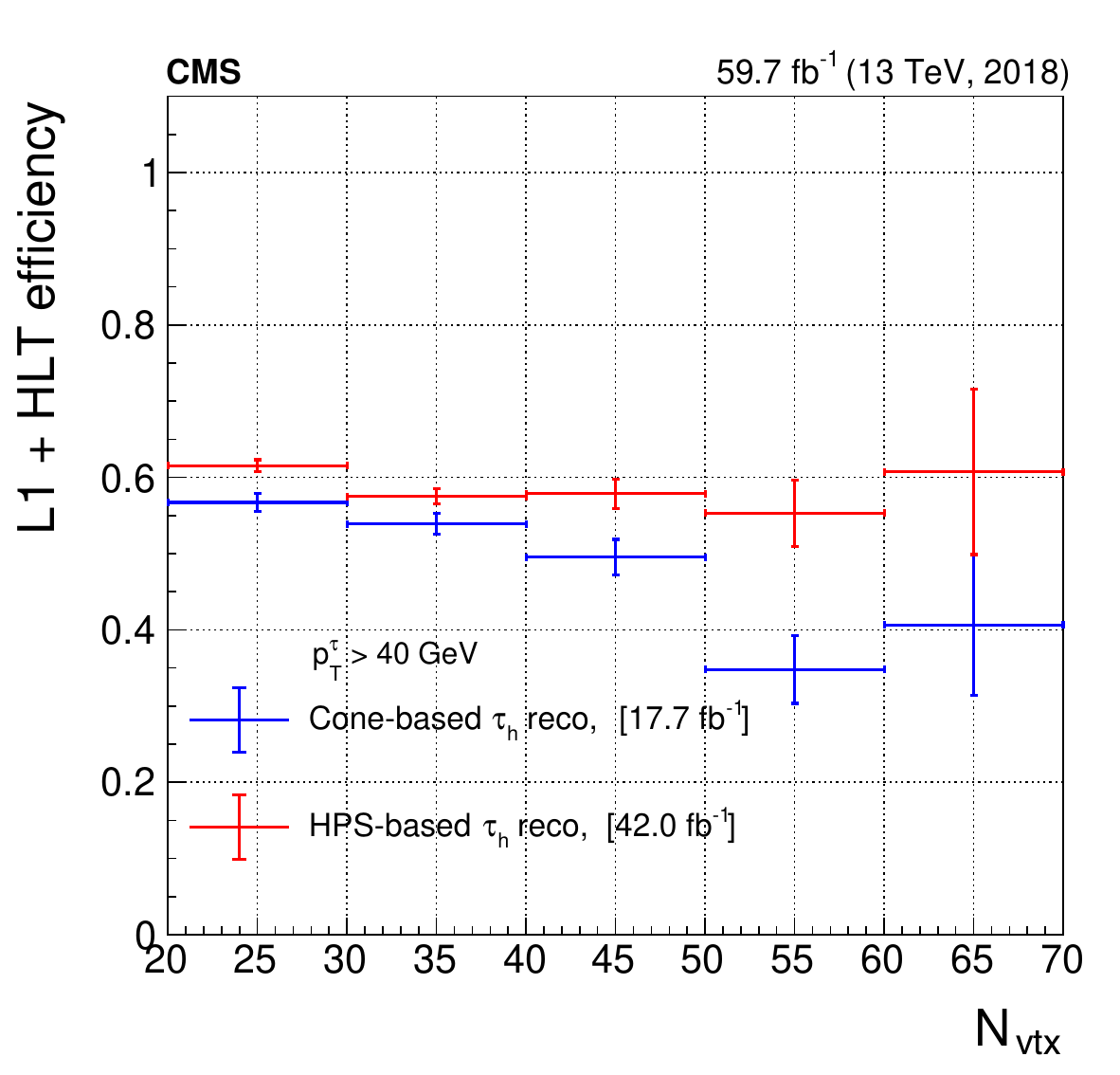}
  \caption{Combined L1 and HLT efficiency per leg of the $\tauh\tauh$ triggers with cone-based and HPS-based $\tauh$ reconstruction, using the first 17.7\fbinv and the next 42.0\fbinv of data in 2018. The figure shows the efficiency as a function of offline $\pt^{\PGt}$ (left) and $N_{\text{vtx}}$ (right).
The vertical bars on the markers represent statistical uncertainties.
  }
  \label{fig:tauReco1}
\end{figure}

The trigger efficiencies are measured for the full 2016, 2017, and 2018
data sets, corresponding to an integrated luminosity of 137.1\fbinv,
and in Drell--Yan simulated samples ($\PZ/\gamma^* \to \Pell \Pell$, where $\Pell =
\Pe, \PGm, \PGt$). The combined L1+HLT efficiency of the $\tauh$ triggers is presented
unless stated otherwise. Therefore,
generally, the results include the impact of L1 trigger selection
efficiency and the specific L1 seeding
efficiencies. Figure~\ref{fig:ptRes} compares the \pt resolution
of the two different algorithms used to reconstruct hadronically
decaying tau leptons from the same data set recorded from
the first 17.7\fbinv taken in 2018.
The figure shows that the
HPS-based tau reconstruction has a better \pt resolution compared with
the cone-based one. This reduces the fraction of misidentified
$\tauh$ candidates from low-\pt jets exceeding the nominal
\pt threshold, allowing lower \pt thresholds for the
same rate. The efficiency per leg of these two algorithms is presented in
Fig.~\ref{fig:tauReco1} for the $\tauh\tauh$ triggers, where one can see
that the HPS algorithm has
slightly higher efficiency in the turn-on region with \pt and has a
significant improvement in the region of high PU.

The implementation of the HPS $\tauh$ reconstruction reduced the HLT rate
of $\tauh$ by 10\% per $\tauh$ leg.
It is measured as $4.6$ and
$39\unit{Hz}$ for an average PU of approximately $50$
for $\PGm \tauh$ and $\tauh\tauh$ triggers, respectively,
while it was correspondingly $5.2$ and $50\unit{Hz}$ for the cone-based algorithm.
The
approximate processing time of a $\tauh \tauh$ trigger is around
50\unit{ms}, whereas it is around 10\unit{ms} for lepton+$\tauh$ triggers
for an
average PU of approximately 50.

\begin{figure}[htp!]
  \centering
  \includegraphics[width=0.49\textwidth]{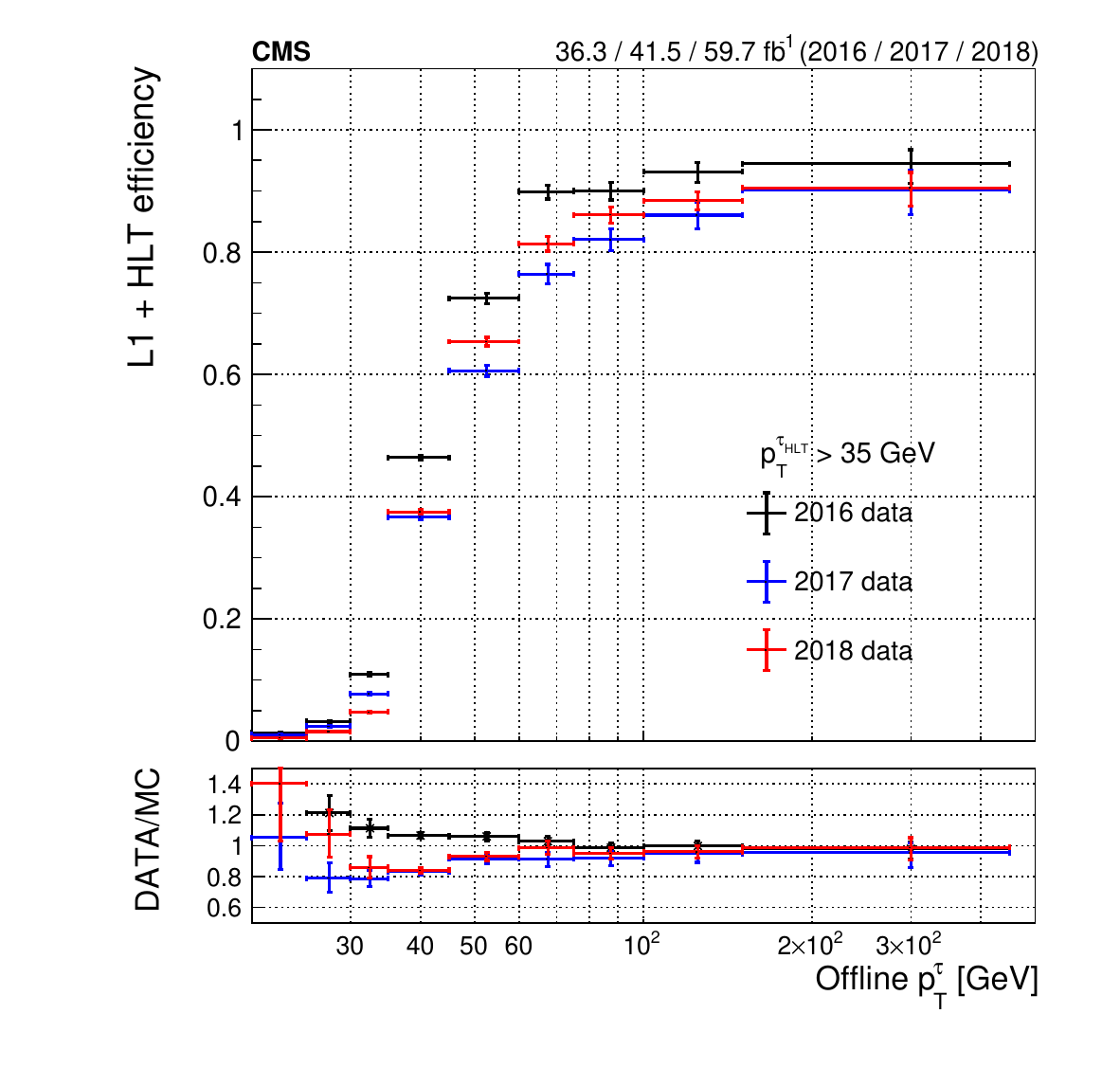}
  \includegraphics[width=0.49\textwidth]{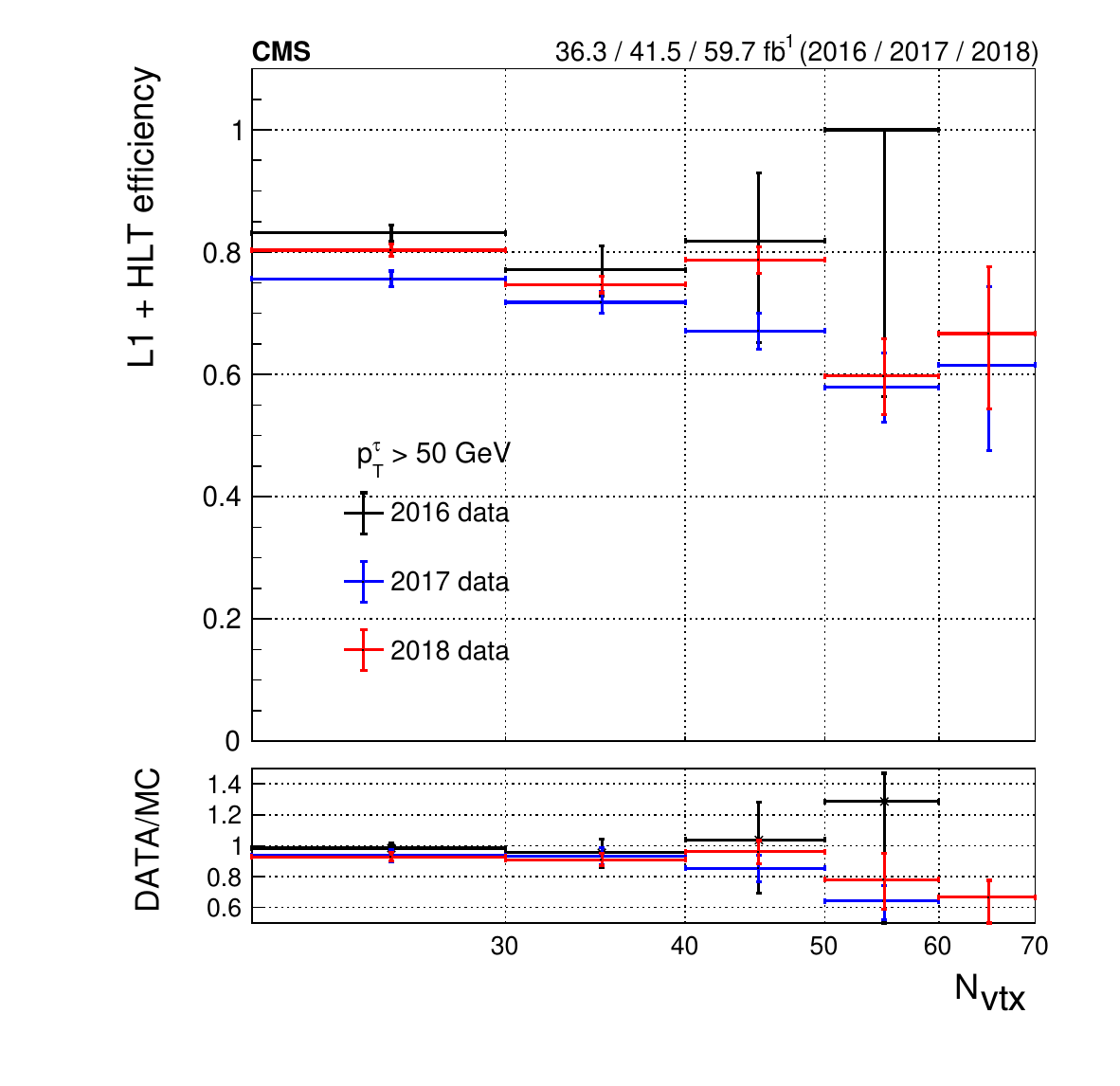}
  \caption{Combined L1 and HLT efficiency per leg of the $\tauh\tauh$ triggers for 2016, 2017, and 2018 data taking.
    The figure shows the trigger efficiency as a function of the offline $\pt^{\PGt}$ for a 35\GeV threshold (left) and as a function of $N_{\text{vtx}}$ (right), for which an offline requirement of $\pt^{\PGt}>50\GeV$ is applied.
    The lower panels show the ratio of data to MC simulation for each year.
The vertical bars on the markers represent statistical uncertainties.
  }
  \label{fig:ditauEffi}
\end{figure}

Figure~\ref{fig:ditauEffi} presents the trigger performance per
leg of the $\tauh \tauh$ triggers in 2016, 2017, and
2018. It shows that the 2016 data have a higher efficiency compared with 2017
and 2018 data, which is a consequence of the lower L1 \pt thresholds and
the lower PU, as presented in the right plot of the figure. The
HPS-based reconstruction algorithm deployed in the middle of 2018
data-taking results in good efficiency in general, compared with
the cone-based reconstruction algorithm that was used before. Since
there were no other significant differences in terms of \pt
thresholds, isolation, or L1 seeds between 2017 and 2018, one would
expect to see similar performance in 2017 and 2018. However, the
figure shows that the efficiency for 2017 data is lower than that in
2018
data. This was caused by the inactive pixel detector modules observed in the
beginning of 2017 data taking as well as the DC-DC converter issue (described in Section~\ref{sec:tracking})
  encountered at the end of
the 2017 data taking. The inefficiency coming from the inactive pixel detector modules
was recovered with the extra recovery iterations in the tracking, but the DC-DC
problem reduced the overall efficiency. In 2018,
the DC-DC pixel detector issue was mitigated, and this improved the tau
trigger efficiency.

\section{Summary}
\label{sec:summary}

The performance of the high-level trigger of CMS has been presented as
it evolved over the course of LHC \Runtwo in 2016--2018.
The algorithms of the high-level trigger were adapted to meet the
challenges of the increase of the LHC luminosity and pileup to twice their initial design
values as well as two to three times the previous \Runone values.
Imperfect detector effects that arose also were mitigated to minimize inefficiencies.
The trigger menu continuously evolved to meet the needs of the
experiment across a wide range of physics areas under these conditions
and within the CPU
capacity of the online computer farm.
The overall single isolated lepton
trigger efficiency was maintained at the level of 90\% for muons and
80\% for electrons with \pt thresholds of 24 and 32\GeV,
respectively.
Triggers based on jets and on energy sums such as
\ptmiss also were available.
Jets also could be tagged as arising from a \PQb quark jet with a
performance of the identification algorithm approaching that achieved offline
and using machine learning for the best discrimination.
Hadronically decaying tau leptons were also
reconstructed in the trigger and were typically used in combination
with other leptons, with minimum \pt thresholds from 20 to
35\GeV on the hadronic tau decay depending on the flavor of the other lepton and the
data-taking year. 
 
\begin{acknowledgments}
  We congratulate our colleagues in the CERN accelerator departments for the excellent performance of the LHC and thank the technical and administrative staffs at CERN and at other CMS institutes for their contributions to the success of the CMS effort. In addition, we gratefully acknowledge the computing centers and personnel of the Worldwide LHC Computing Grid and other centers for delivering so effectively the computing infrastructure essential to our analyses. Finally, we acknowledge the enduring support for the construction and operation of the LHC, the CMS detector, and the supporting computing infrastructure provided by the following funding agencies: SC (Armenia), BMBWF and FWF (Austria); FNRS and FWO (Belgium); CNPq, CAPES, FAPERJ, FAPERGS, and FAPESP (Brazil); MES and BNSF (Bulgaria); CERN; CAS, MoST, and NSFC (China); MINCIENCIAS (Colombia); MSES and CSF (Croatia); RIF (Cyprus); SENESCYT (Ecuador); ERC PRG, RVTT3 and MoER TK202 (Estonia); Academy of Finland, MEC, and HIP (Finland); CEA and CNRS/IN2P3 (France); SRNSF (Georgia); BMBF, DFG, and HGF (Germany); GSRI (Greece); NKFIH (Hungary); DAE and DST (India); IPM (Iran); SFI (Ireland); INFN (Italy); MSIP and NRF (Republic of Korea); MES (Latvia); LMTLT (Lithuania); MOE and UM (Malaysia); BUAP, CINVESTAV, CONACYT, LNS, SEP, and UASLP-FAI (Mexico); MOS (Montenegro); MBIE (New Zealand); PAEC (Pakistan); MES and NSC (Poland); FCT (Portugal);  MESTD (Serbia); MCIN/AEI and PCTI (Spain); MOSTR (Sri Lanka); Swiss Funding Agencies (Switzerland); MST (Taipei); MHESI and NSTDA (Thailand); TUBITAK and TENMAK (Turkey); NASU (Ukraine); STFC (United Kingdom); DOE and NSF (USA).

  \hyphenation{Rachada-pisek} Individuals have received support from the Marie-Curie program and the European Research Council and Horizon 2020 Grant, contract Nos.\ 675440, 724704, 752730, 758316, 765710, 824093, 101115353, 101002207, and COST Action CA16108 (European Union); the Leventis Foundation; the Alfred P.\ Sloan Foundation; the Alexander von Humboldt Foundation; the Science Committee, project no. 22rl-037 (Armenia); the Belgian Federal Science Policy Office; the Fonds pour la Formation \`a la Recherche dans l'Industrie et dans l'Agriculture (FRIA-Belgium); the F.R.S.-FNRS and FWO (Belgium) under the ``Excellence of Science -- EOS" -- be.h project n.\ 30820817; the Beijing Municipal Science \& Technology Commission, No. Z191100007219010 and Fundamental Research Funds for the Central Universities (China); the Ministry of Education, Youth and Sports (MEYS) of the Czech Republic; the Shota Rustaveli National Science Foundation, grant FR-22-985 (Georgia); the Deutsche Forschungsgemeinschaft (DFG), among others, under Germany's Excellence Strategy -- EXC 2121 ``Quantum Universe" -- 390833306, and under project number 400140256 - GRK2497; the Hellenic Foundation for Research and Innovation (HFRI), Project Number 2288 (Greece); the Hungarian Academy of Sciences, the New National Excellence Program - \'UNKP, the NKFIH research grants K 131991, K 133046, K 138136, K 143460, K 143477, K 146913, K 146914, K 147048, 2020-2.2.1-ED-2021-00181, and TKP2021-NKTA-64 (Hungary); the Council of Science and Industrial Research, India; ICSC -- National Research Center for High Performance Computing, Big Data and Quantum Computing and FAIR -- Future Artificial Intelligence Research, funded by the NextGenerationEU program (Italy); the Latvian Council of Science; the Ministry of Education and Science, project no. 2022/WK/14, and the National Science Center, contracts Opus 2021/41/B/ST2/01369 and 2021/43/B/ST2/01552 (Poland); the Funda\c{c}\~ao para a Ci\^encia e a Tecnologia, grant CEECIND/01334/2018 (Portugal); the National Priorities Research Program by Qatar National Research Fund; MCIN/AEI/10.13039/501100011033, ERDF ``a way of making Europe", and the Programa Estatal de Fomento de la Investigaci{\'o}n Cient{\'i}fica y T{\'e}cnica de Excelencia Mar\'{\i}a de Maeztu, grant MDM-2017-0765 and Programa Severo Ochoa del Principado de Asturias (Spain); the Chulalongkorn Academic into Its 2nd Century Project Advancement Project, and the National Science, Research and Innovation Fund via the Program Management Unit for Human Resources \& Institutional Development, Research and Innovation, grant B39G670016 (Thailand); the Kavli Foundation; the Nvidia Corporation; the SuperMicro Corporation; the Welch Foundation, contract C-1845; and the Weston Havens Foundation (USA).
\end{acknowledgments}

\bibliography{auto_generated}
\cleardoublepage \appendix\section{The CMS Collaboration \label{app:collab}}\begin{sloppypar}\hyphenpenalty=5000\widowpenalty=500\clubpenalty=5000
\cmsinstitute{Yerevan Physics Institute, Yerevan, Armenia}
{\tolerance=6000
A.~Hayrapetyan, A.~Tumasyan\cmsAuthorMark{1}\cmsorcid{0009-0000-0684-6742}
\par}
\cmsinstitute{Institut f\"{u}r Hochenergiephysik, Vienna, Austria}
{\tolerance=6000
W.~Adam\cmsorcid{0000-0001-9099-4341}, J.W.~Andrejkovic, L.~Benato\cmsorcid{0000-0001-5135-7489}, T.~Bergauer\cmsorcid{0000-0002-5786-0293}, S.~Chatterjee\cmsorcid{0000-0003-2660-0349}, K.~Damanakis\cmsorcid{0000-0001-5389-2872}, M.~Dragicevic\cmsorcid{0000-0003-1967-6783}, P.S.~Hussain\cmsorcid{0000-0002-4825-5278}, M.~Jeitler\cmsAuthorMark{2}\cmsorcid{0000-0002-5141-9560}, N.~Krammer\cmsorcid{0000-0002-0548-0985}, A.~Li\cmsorcid{0000-0002-4547-116X}, D.~Liko\cmsorcid{0000-0002-3380-473X}, I.~Mikulec\cmsorcid{0000-0003-0385-2746}, J.~Schieck\cmsAuthorMark{2}\cmsorcid{0000-0002-1058-8093}, R.~Sch\"{o}fbeck\cmsAuthorMark{2}\cmsorcid{0000-0002-2332-8784}, D.~Schwarz\cmsorcid{0000-0002-3821-7331}, M.~Sonawane\cmsorcid{0000-0003-0510-7010}, W.~Waltenberger\cmsorcid{0000-0002-6215-7228}, C.-E.~Wulz\cmsAuthorMark{2}\cmsorcid{0000-0001-9226-5812}
% ...
\par}
\vskip\cmsinstskip
\dag:~Deceased\\
$^{1}$Also at Yerevan State University, Yerevan, Armenia\\
$^{2}$Also at TU Wien, Vienna, Austria\\
% ...
\end{sloppypar}
\end{document}